\documentclass[a4paper,11pt]{article}
\usepackage{jheppub}
\usepackage{tikz-feynman}
\usepackage{multirow}
\usepackage{adjustbox}
\usepackage{tcolorbox}
\usepackage{multicol}
\usepackage{upgreek}
\usepackage{amsmath}
\usepackage{relsize}
\usepackage{hyperref}
\usepackage{cleveref}
\usepackage{appendix}       
\usepackage{mathtools,slashed}
\usepackage{subcaption}
\usepackage{tikz}
\usepackage{color, colortbl}
\usepackage{bigfoot}
\usepackage{soul}
\usepackage{orcidlink}
\usepackage[bottom]{footmisc}
\pdfoutput=1
\DeclareNewFootnote[para]{math}[fnsymbol]
\MakeSortedPerPage{footnotemath}
\definecolor{deepmagenta}{rgb}{0.8, 0.0, 0.8}
\definecolor{mediumtealblue}{rgb}{0.0, 0.33, 0.71}
\definecolor{warmblack}{rgb}{0.0, 0.26, 0.26}
\definecolor{bostonuniversityred}{rgb}{0.8, 0.0, 0.0}
\definecolor{junglegreen}{rgb}{0.16, 0.67, 0.53}
\definecolor{lightcornflowerblue}{rgb}{0.6, 0.81, 0.93}
\definecolor{mypink1}{rgb}{0.858, 0.188, 0.478}
\definecolor{mypink2}{RGB}{219, 48, 122}
\definecolor{mypink3}{cmyk}{0, 0.7808, 0.4429, 0.1412}
\definecolor{mygray}{gray}{0.2}
\definecolor{ForestGreen}{RGB}{34,139,34}

\newcommand{\eq}{eq\,.~}

\newcommand{\fig}{fig\,.~}

\newcommand{\GeV}{\text{GeV}}
\newcommand{\TeV}{\text{TeV}}

\newcommand{\mnone}{m_{\rm N_1^{}}^{}}
\newcommand{\mntwo}{m_{\rm N_2^{}}^{}}
\newcommand{\mnthree}{m_{\rm N_3^{}}^{}}
\newcommand{\metap}{m_{\eta^+}^{}}
\newcommand{\metaR}{m_{\eta_R^0}^{}}
\newcommand{\metaI}{m_{\eta_I^0}^{}}
\newcommand{\leta}{\lambda_{\eta}}
\newcommand{\letaH}{\lambda_{\eta\rm H}}
\newcommand{\letaHp}{\lambda_{\eta\rm H}^{\prime}}
\newcommand{\letaHpp}{\lambda_{\eta\rm H}^{\prime\prime}}
\newcommand{\deltam}{\delta m}
\newcommand{\Deltam}{\Delta m}
\usepackage[font={small},labelfont={bf}]{caption}
\usepackage{url}
\definecolor{MyDarkBlue}{rgb}{0.1, 0.1, 0.8}
\definecolor{SBlue}{rgb}{0.2, 0.4, 0.7} 
\definecolor{MyLightBlue}{rgb}{0.22,0.51,0.9}
\definecolor{MyGreen}{rgb}{0.0, 0.5, 0.0}
\definecolor{BrickRed}{rgb}{0.8, 0.25, 0.33}
\usepackage{hyperref}
\hypersetup{colorlinks, citecolor=BrickRed,linkcolor=MyDarkBlue, urlcolor=MyGreen}
\title{Vanilla Scotogenic Model at the future Muon Collider}
\author[a]{Niloy Mondal\orcidlink{0009-0006-5837-9772},}
\emailAdd{niloy18@iitg.ac.in}
\affiliation[a]{Department of Physics, Indian Institute of Technology Guwahati, North Guwahati, Assam-781039, India}
\author[b,c]{Dipankar Pradhan\orcidlink{0000-0002-2450-6677},}
\emailAdd{dipankar.pradhan@iopb.res.in}
\affiliation[b]{Institute of Physics, Sachivalaya Marg, Bhubaneswar, Odisha 751005, India}
\affiliation[c]{Homi Bhabha National Institute, BARC Training School Complex, Anushakti Nagar, Mumbai 400094, India}
\author[a,d]{Abhik Sarkar\orcidlink{0000-0003-1449-2934},}
\emailAdd{asarkar@lpsc.in2p3.fr}
\affiliation[d]{Laboratoire de Physique Subatomique et de Cosmologie, Université Grenoble-Alpes, CNRS/IN2P3, 53 Avenue des Martyrs, 38026 Grenoble, France}
\author[e]{Biplob Bhattacherjee\orcidlink{0000-0003-3668-8305},}
\emailAdd{biplob@iisc.ac.in}
\affiliation[e]{Centre for High Energy Physics, Indian Institute of Science, Bengaluru 560012, India}
\author[a]{and Subhaditya Bhattacharya\orcidlink{0000-0002-8841-603X}}
\emailAdd{subhab@iitg.ac.in}
\abstract{The vanilla scotogenic model (VSM) featuring three right-handed neutrinos and an inert scalar doublet, offers a unified explanation of neutrino masses, dark matter (DM), and the baryon asymmetry of the Universe through leptogenesis. The typical mass scale of such a scenario lies in the vicinity of $\sim$ 10 TeV and above, whereas enhanced second-generation Yukawa couplings lower the leptogenesis scale to $\sim 1$ \TeV, allowing the model to be probed at the future muon colliders. Adhering to DM, neutrino mass, Leptogenesis, and flavour dependent constraints, the charged scalar is likely to be a long-lived particle (LLP). We analyse both the prompt and displaced vertex signatures of the model to find that, prompt mono-photon signals remain challenging, but LLP searches offer a promising avenue to probe such a framework, subject to the timing resolutions, and displaced-vertex reconstruction efficiency.}
\begin{document} 
\begin{flushright}
IOP/BBSR/2026-08
\end{flushright}
\makeatletter
\gdef\@fpheader{}
\makeatother
\maketitle
\section{Introduction}
\label{sec:intro}
The existence of a non-luminous dark matter (DM) is supported from various astrophysical observations and Big Bang nucleosynthesis (BBN) \cite{Zwicky:1933gu, Rubin:1970zza,Rubin:1980zd,Tyson:1990yt,Bartelmann:1999yn,Clowe:2006eq,Planck:2018vyg,White:1993wm,Cyburt:2015mya}, although a discovery is still awaited. Direct \cite{Goodman:1984dc}, indirect \cite{Bergstrom:1989jr,Ellis:1988qp,Rudaz:1987ry}, and collider searches \cite{Fayet:1979sa,Fayet:1980rr,Fayet:1982ky,Goldberg:1983nd,Haber:1984rc,Jungman:1995df} have placed increasingly stringent upper limits on the relevant DM interaction cross sections.

Observed baryon asymmetry of the Universe (BAU) is another unsolved puzzle. Several mechanisms have been postulated \cite{Sakharov:1967dj,Trodden:1998ym, Cohen:1993nk, Affleck:1984fy, Bodeker:2020ghk, Riotto:1999yt, Cui:2015eba, Morrissey:2012db}, among which leptogenesis \cite{Fukugita:1986hr,Davidson:2008bu,Buchmuller:2005eh,Hugle:2018qbw} is a viable option, as explored here. Current CMB observations yield  $\eta_{\Delta B}^{\rm CMB} = (6.12 \pm 0.04)\times 10^{-10}$, \cite{Planck:2018vyg} in excellent agreement with the value inferred from BBN, using primordial light-element abundances \cite{Planck:2018vyg,ParticleDataGroup:2024cfk}. 

Absence of an active right-handed neutrino in the Standard Model (SM) of particle physics \cite{Goldhaber:1958nb,Wu:1957my,Feynman:1958ty,Sudarshan:1958vf,Gargamelle:1973,ALEPH:2005ab} primarily forbids neutrino mass generation using Higgs Yukawa interaction. However, various neutrino experiments \cite{Davis:1968cp, SNO:2001kpb, SNO:2011hxd, Super-Kamiokande:1998kpq, MACRO:1998ckv, KamLAND:2002uet, RENO:2012mkc, DayaBay:2012fng,K2K:2002icj, NOvA:2019cyt, T2K:2013ppw, MINOS:2011neo} indicate to neutrino flavour oscillation, 
and hence, non-zero neutrino masses. Several extensions beyond the SM have been proposed to address the same \cite{Weinberg:1979sa, Mohapatra:1979ia, Magg:1980ut, Foot:1988aq, Babu:1988ki, Zee:1980ai, Ma:2006km, Mohapatra:1986aw, Langacker:1998ut, Arkani-Hamed:1998wuz, Altarelli:2010gt, Mohapatra:2005wg, King:2003jb,Ardila-Tafurth:2025qdf}. 


We focus here on Vanilla Scotogenic model (VSM) \cite{Ma:2006km}, a SM extension comprising of an inert scalar doublet (ISD) and three right-handed neutrinos (RHNs), all odd under $\mathcal{Z}_2$ symmetry. This framework simultaneously accounts for neutrino masses, BAU, and DM, while remaining consistent with the current lepton flavor violation (LFV) constraints.
The simultaneous explanation of the observed DM relic abundance, neutrino masses, and BAU poses a few challenges for probing VSM: (\textbf{I}) the relic abundance of DM is satisfied only in a very narrow mass range around 550–650 GeV\footnote{
Satisfying DM relic abundance and direct-detection limits simultaneously in the high mass regime, a fine-tuned near-cancellation is required among the scalar couplings i.e. $(\letaH+\letaHp-\letaHpp) \to 0$. 
}, with the resonance region already excluded by collider searches \cite{Barbieri:2006dq,Hashemi:2015swh}; (\textbf{II}) the smallest RHN must be larger than $\sim$10 \TeV~\cite{Hugle:2018qbw} to account for correct BAU, thereby rendering the collider phenomenology of the model considerably less prominent and almost inaccessible through direct searches; (\textbf{III}) parameter space satisfying neutrino mass, BAU, and DM abundance are well below the current experimental sensitivities on radiative charged LFV decays; and (\textbf{IV}) DM production at colliders is largely unobservable due to the weak sensitivity of the relevant channels.

A few attempts have been made to make VSM phenomenologically viable \cite{Hugle:2018qbw, Racker:2024fpn}. The central strategy adopted here is to render second or third generation (neutrino-mass compatible) Yukawa couplings relatively large by an appropriate choice of the orthogonal matrix angle. Such couplings contribute sub-dominantly to neutrino mass generation and washout processes, while significantly enhancing the CP asymmetry required for successful baryogenesis. This also facilitates a collider search prospect of the model hitherto unexplored, which we elaborate upon. 


The compressed scalar spectrum naturally gives rise to a long-lived charged scalar, 
making the pair-production sensitive to both prompt and long-lived particle searches. At $\mu^{+}\mu^{-}$ colliders \cite{Accettura:2023ked,InternationalMuonCollider:2024jyv,Accettura:2025xgt}, charged scalar production receives contributions from both gauge- and Yukawa-mediated interactions. While the gauge contribution dominates at lower energies, the Yukawa-mediated channel becomes increasingly important at higher energies, providing sensitivity to the model-specific parameters. Furthermore, the Majorana nature of the RHNs gives rise to lepton number violating interactions that can be probed directly at the same-sign $\mu^{+}\mu^{+}$ collider, where the absence of gauge-mediated production channels allows the signal to depend more strongly on the underlying Yukawa interactions. Consequently, a physics program combining both $\mu^{+}\mu^{-}$ and $\mu^{+}\mu^{+}$ collider-runs offers a powerful and complementary strategy for exploring the parameter space of this model. 

We ensure consistency with direct detection experiments, in particular the LZ-2025 results \cite{LZ:2024zvo}, indirect detection limits from the combined analysis \cite{Fermi-LAT:2025gei} of Fermi-LAT, HAWC, H.E.S.S., MAGIC, and VERITAS. Additionally, we include the projected sensitivity to the DM-nucleon scattering cross section from the XLZD Collaboration \cite{XLZD:2024nsu}, and the SM Higgs decay constraint into di-photon final states \cite{Swiezewska:2012eh,Aiko:2023nqj,Braathen:2024ckk}.

This paper is organized as follows. In Sec.~\ref{sec:intro}, we briefly review the Scotogenic model and introduce the particle content and relevant interactions. The theoretical and experimental constraints imposed on the model parameter space are discussed in Sec.~\ref{sec:constraints}, including theoretical consistency conditions, neutrino masses and mixings, lepton flavor violation, $\mu$-$e$ conversion in nuclei, electric dipole moments, electroweak precision observables, and existing collider limits on long-lived particles. In Sec.~\ref{sec:pheno}, we investigate the phenomenological implications of the model, focusing on neutrino mass generation, lepton flavor violation, dark matter relic abundance and detection prospects, as well as leptogenesis and the generation of the observed matter-antimatter asymmetry. The prospects of probing the model at future muon colliders are presented in Sec.~\ref{sec:collider}, where both prompt and long-lived search strategies are explored. Finally, we summarize our findings and conclude in Sec.~\ref{sec:conclusion}.
\section{Vanilla Scotogenic model}
\label{sec:intro}
In VSM framework, we extend the SM by introducing an inert scalar doublet $\eta$ and three right-handed neutrinos $N_k$, all of which are odd under a discrete $\mathcal{Z}_2$ symmetry. The lightest $\mathcal{Z}_2$-odd particle, identified as the CP-odd component of the inert scalar doublet, is assumed to be a stable DM candidate. The charge assignments of the newly introduced fields are summarized in \cref{tab:fields}.
\begin{table}[!htb]
\centering
\begin{tabular}{ccccc}
\hline\hline
$\;\;$Field$\;\;$ & $\;\;$Spin$\;\;$ & Generations & {$\rm SU(3)_{\mathtt{C}} \otimes SU(2)_{\mathtt{L}} \otimes U(1)_{\mathtt{Y}}$} & $\;\;\mathcal{Z}_{2}\;\;$ \\
\hline\hline
 $\eta$ & $0$ & $1$ & $\left(1,\,2,\, \tfrac{1}{2}\right)$ & $-1$ \\ \hline
 $N_k$ & $\tfrac{1}{2}$ & $3$ & $(1,\,1,\,0)$ & $-1$ \\
\hline\hline
\end{tabular}
\caption{Particle content of VSM framework and their corresponding charge assignments, where $k \in \{1, 2, 3\}$ denotes the generation/family index.}
\label{tab:fields}
\end{table}
The VSM Lagrangian can be written as \cite{Ma:2006km},
\begin{align}
\mathcal{L}\supset i\sum_{k=1,2,3}\overline{N}_k\slashed{\partial}N_k+|\mathcal{D}_{\mu}\eta|^2+\mathcal{L}_{N}-\mathcal{V}\,(H, \eta)\,,
\label{eq:model}
\end{align}
where $H$ denotes the SM Higgs doublet, $\mathcal{D}_\mu$ is the electroweak covariant derivative, and $\eta$ is an inert scalar doublet of the form
$\begin{pmatrix}
\eta^{+} &~(\eta_{R}^0 + i\,\eta_{I}^0)/\sqrt{2}
\end{pmatrix}^T$ with hypercharge $Y=\frac{1}{2}$. The Lagrangian $\mathcal{L}_{N}$ contains the Yukawa interaction terms as well as the Majorana mass term for the RHNs, and is given by,
\begin{eqnarray}
\mathcal{L}_{\rm N} & \supset &- \left(h_{\alpha \beta}\overline{L}_{\alpha}\Tilde \eta N_{\beta} + h.c\right)-\dfrac{1}{2}\sum_{k=1,2,3}m_{N_k^{}}^{}\overline{N_{k}^{c}}N_{k}^{}\,,
\end{eqnarray}
where $\tilde{\eta} = i\sigma_2 \eta^\star$ denotes the charge-conjugate doublet, with $\sigma_2$ being the second Pauli matrix. The scalar potential $\mathcal{V}(H,\eta)$ term can be written as:
\begin{align}
\nonumber\mathcal{V} \,=\,& \mu_{\eta}^2\left(\eta^{\dagger}_{}\eta\right)-
\mu_{H}^2\left(H^{\dagger}_{}H\right)+\lambda_H\left|H^{\dagger}H\right|^{2}  + \lambda_\eta\left|\eta^{\dagger}\eta\right|^{2} + \lambda_{\eta H}\left(\eta^{\dagger}\eta\right) \left( H^{\dagger}H \right)\\ 
&+\lambda_{\eta H}^{\prime}\left(H^{\dagger}\eta\right) \left(\eta^{\dagger}_{} H\right)+
\dfrac{1}{2}\left[\lambda_{\eta H}^{\prime\prime}\left( \eta^{\dagger}H \right)\left( \eta^{\dagger}H \right)+ {\rm h.c.} \right] \,,
\label{eq:poten}
\end{align}
with $\mu_{H}>0$ and $\mu_{\eta}>0$, so that $H$ acquires a non-zero vacuum expectation value (VEV), while $\eta$ doesn't get one. After the electroweak symmetry breaking (EWSB) with $\langle H \rangle=v/\sqrt 2$, the masses of the physical scalars would be, 
\begin{align}
m_{\eta^+}^2&=\mu_{\eta}^2+\lambda_{\eta H}v^2\,,\\
m_{\eta^0_{R}}^2 &=m_{\eta^+}^2+\dfrac{1}{2}\left(\lambda_{\eta H}^{\prime}+\lambda_{\eta H}^{\prime\prime}\right)v^2\,,
\label{eq:metaR2}\\
m_{\eta^0_{I}}^2 &=m_{\eta^+}^2+\dfrac{1}{2}\left(\lambda_{\eta H}^{\prime}-\lambda_{\eta H}^{\prime\prime}\right)v^2\,.
\label{eq:metaI2}
\end{align}
The parameters defined as, 
\begin{align}
\begin{split}
\Deltam=\metaR-\metaI\,,\\
\deltam=\metap-\metaI\,,
\end{split}
\label{eq:dm-Dm}
\end{align}
are used as free parameters, rather than $\metap$ and $\metaR$, in the subsequent analysis of DM and related phenomenology.

\section{Model constraints}
\label{sec:constraints}
The New Physics (NP) model parameters ($\metaI,~\metaR,~\metap,~\leta,~\letaH$) are constrained by several theoretical and experimental constraints. The most relevant ones are discussed below.
\subsection{Theoretical constraints}
The co-positivity criteria for our model \cite{Lozano:2025tst}:
\begin{eqnarray}
\begin{split}
\lambda_{H}\geq 0\,, \hspace{0.25cm} \lambda_\eta\geq 0\,, \hspace{0.25cm} \lambda_{\eta H}^{\prime}+2\sqrt{\lambda_{H}\lambda_\eta}\geq 0\,, \\ \lambda_{\eta H}^{\prime}+\lambda_{\eta H}-\lambda_{\eta H}^{\prime\prime}+2\sqrt{\lambda_{H}\lambda_\eta}\geq 0\,.\hspace{0.35cm}
\end{split}
\end{eqnarray}
All the quartic couplings should be below $4\pi$ so that the perturbative expansion of the potential around its minimum is justified i.e., low-energy scalar potential is bounded from below. All the mass squared parameters should be positive so that $\eta$ is unable to get vev and prevents the breaking of $\mathcal{Z}_2$ symmetry. In addition, as all the portal couplings are related to physical mass eigenvalues, the perturbativity conditions are translated into the following constraints on masses \cite{Abada:2018zra}:
\begin{align}
|\letaH|\leq {\rm min}\left[\sqrt{4\pi}\,,m_{\eta^+}^2/\langle H\rangle\right]\,,~~\,\left|\dfrac{m^2_{\eta_R^0}+m^2_{\eta_I^0}-2m^2_{\eta^+}}{2\langle H\rangle^2}\right| \leq\sqrt{4\pi}\,,~~\, \left|\dfrac{m^2_{\eta_R^0}-m^2_{\eta_I^0}}{2\langle H\rangle^2}\right| \leq\sqrt{4\pi}\,.
\end{align}
\subsection{Neutrino masses and mixings}
Neutrino flavor oscillations, firmly established by solar, atmospheric, reactor, and accelerator neutrino experiments such as Super-Kamiokande, SNO, KamLAND, and more recently Hyper-Kamiokande \cite{Davis:1968cp, SNO:2001kpb, SNO:2011hxd, Super-Kamiokande:1998kpq, MACRO:1998ckv, KamLAND:2002uet, RENO:2012mkc, DayaBay:2012fng,K2K:2002icj, NOvA:2019cyt, T2K:2013ppw, MINOS:2011neo} provide unambiguous evidence for nonzero active neutrino masses via neutrino flavour oscillations. While the Dirac or Majorana nature of neutrinos and the absolute neutrino mass scale remain unknown, neutrino oscillation experiments are sensitive to the differences in the squared masses and to the parameters governing leptonic mixing. In particular, these experiments have measured the mass-squared differences ($\Delta m_{21}^2=m_{\nu^{}_{2}}^2-m_{\nu^{}_{1}}^2$, $|\Delta m_{31}^2|=|m_{\nu^{}_{3}}^2-m_{\nu^{}_{1}}^2|$, $\Delta m_{32}^2=m_{\nu^{}_{3}}^2-m_{\nu^{}_{2}}^2$), and the three mixing angles ($\theta_{12},\theta_{23},\theta_{13}$) of the Pontecorvo–Maki–Nakagawa–Sakata (PMNS) matrix. Possible CP-violating phase ($\delta_{\rm CP}$) in the lepton sector has been constrained \cite{Super-Kamiokande:1998kpq,SNO:2002tuh,KamLAND:2002uet,Esteban:2020cvm}.
Notably the measurement of the mod value of $|\Delta m_{31}^2|$ leads to the possibilities of normal ordering (NO) and inverted ordering (IO) given by the following equations, where the best fit values for the parameters are taken \cite{Esteban:2018azc, Esteban:2020cvm,Esteban:2024eli}: 
\begin{gather}
\begin{rcases}
m_{\nu^{}_{1}} \approx 0 \\
m_{\nu^{}_{2}} = \sqrt{m_{\nu^{}_{1}}^2 + \Delta m_{21}^2} \\
m_{\nu^{}_{3}} = \sqrt{m_{\nu^{}_{1}}^2 + \Delta m_{31}^2}
\end{rcases}
~
\begin{tabular}{l}
NO
\end{tabular}
~
\left\{
\begin{array}{l}
\theta_{12}/^{\circ}=33.68,\;
\theta_{23}/^{\circ}=48.5,\;
\theta_{13}/^{\circ}=8.52,\;
\delta^{\rm CP}_{\rm Dirac}/^{\circ}=177,\\[2mm]
\Delta m_{21}^2 = 7.49^{+0.19}_{-0.19}\times10^{-5}\,\text{eV}^2,\\
\Delta m_{31}^2 = 2.513^{+0.021}_{-0.019}\times10^{-3}\,\text{eV}^2
\end{array}
\right.
\label{eq:neutrino_mass1}
\end{gather}

\begin{align}
\begin{split}
\begin{rcases}
m_{\nu^{}_{1}} = \sqrt{m_{\nu^{}_{3}}^2 - \Delta m_{32}^2 - \Delta m_{21}^2} \\
m_{\nu^{}_{2}} = \sqrt{m_{\nu^{}_{3}}^2 - \Delta m_{32}^2} \\
m_{\nu^{}_{3}} \approx 0
\end{rcases}
~
\begin{tabular}{l}
IO
\end{tabular}
~
\left\{
\begin{array}{l}
\theta_{12}/^{\circ}=33.68,\;
\theta_{23}/^{\circ}=48.6,\;
\theta_{13}/^{\circ}=8.58,\;
\delta^{\rm CP}_{\rm Dirac}/^{\circ}=285,\\[2mm]
\Delta m_{21}^2 = 7.49^{+0.19}_{-0.19}\times10^{-5}\,\text{eV}^2,\\
\Delta m_{32}^2 = -2.484^{+0.020}_{-0.020}\times10^{-3}\,\text{eV}^2
\end{array}
\right.
\end{split}
\label{eq:neutrino_mass2}
\end{align}
In this work, the results in \eq\eqref{eq:neutrino_mass1} for NO are taken as inputs to determine the Yukawa matrix ($h$), using the Casas-Ibarra (CI) parametrization \cite{Casas:2001sr} (see Appendix~\ref{sec:numass}). The Yukawa matrix is then used in the analyses of leptogenesis and DM phenomenology. Throughout the subsequent analysis, the complex angles of the CI parametization are fixed to $z_{12}=0,~z_{13_R^{}}=z_{13_I^{}}=\sqrt{m_{\nu^{}_{1}}/(2 m_{\nu^{}_{3}})}~\text{and}~ z_{23}=0+7i\,$, while the RHN masses are chosen to follow the hierarchical pattern $m_{N_{3}^{}}^{}=\sqrt{10}~m_{N_{2}^{}}^{}=10~m_{N_{1}^{}}^{}$, with the additional condition $m_{N_{1}^{}}^{}\gtrsim\metaI\,$.

\subsection{Lepton flavor violation}
Different lepton flavour violating processes can occur due to VSM framework. 
Some of the future sensitivities of these measurements
can put significant bound on the model parameters, which we discuss below. 
\subsubsection{$\mu^+ \to e^+ \gamma$}
The MEG II experiment at PSI, searching for $\mu^+ \to e^+ \gamma$ decays, reports no excess signal over the expected background, yielding an upper limit on the branching ratio at 90\% confidence level (C.L.) \cite{MEGII:2025gzr},
\begin{gather}
{\mathcal {BR}} (\mu ^+ \rightarrow {\textrm{e}}^+ \gamma ) < 1.5 \times 10^{-13}\,.
\end{gather}
The target goal of this experiment is to reach a sensitivity of $\sim 6 \times 10^{-14}$ \cite{MEGII:2023ltw}.
%
\begin{figure}[htb!]
\centering
\begin{adjustbox}{width=\textwidth}
\begin{tcolorbox}[colback=gray!5, colframe=black!10, boxrule=1pt, arc=4mm, boxsep=0pt, left=0pt, right=0pt, top=0pt, bottom=0pt, width=1\textwidth, halign=center]
\begin{tikzpicture}[baseline={(current bounding box.center)},style={scale=0.9, transform shape}]
\begin{feynman}
\vertex (a){\(\color{black}{\ell_{\alpha}}\)};
\vertex [right=1.5cm of a] (b);
\vertex [above right=0.75cm and 0.75cm of b] (b1);
\vertex [below right=0cm and 1.5cm of b] (c1);
\vertex [above right=0.5cm and 1cm of b1] (b2);
\vertex [below right=0.0cm and 1cm of c1] (c2);
\diagram*{
(a)-- [line width=0.25mm,fermion, arrow size=1.2pt, style=bostonuniversityred,ultra thick] (b),
(c1) -- [line width=0.25mm, plain,arrow size=1.2pt, style=black,ultra thick, edge label={\(\color{black}{N_k}\)}] (b),
(b) -- [line width=0.25mm, charged scalar, arrow size=1.2pt, style=black, ultra thick, edge label={\(\color{black}{\eta^- }\)}] (b1),
(b1) -- [line width=0.25mm, charged scalar,arrow size=1.2pt, style=black, ultra thick, edge label={\(\color{black}{\eta^- }\)}] (c1),
(b1) -- [line width=0.25mm, boson,arrow size=1.2pt, style=mediumtealblue,ultra thick] (b2),
(c1) -- [line width=0.25mm, fermion,  arrow size=1.2pt, style=mediumtealblue,ultra thick] (c2)};
\vertex[above right=0.5cm and 1cm of b1]{\(\color{black}{\gamma }\)};
\vertex[below right=0.0cm and 1cm of c1]{\(\color{black}{\ell_{\beta}}\)};
\node at (b)[circle,fill,style=black,inner sep=1pt]{};
\node at (b1)[circle,fill,style=black,inner sep=1pt]{};
\node at (c1)[circle,fill,style=black,inner sep=1pt]{};
\end{feynman}
\end{tikzpicture}\hfill
\begin{tikzpicture}[baseline={(current bounding box.center)},style={scale=0.9, transform shape}]
\begin{feynman}
\vertex (a){\(\color{black}{\ell_{\alpha}}\)};
\vertex[right=1.5cm of a] (a1);
\vertex[right=1cm of a1] (a2);
\vertex[right=1cm of a2] (b);
\vertex[above right=1.0cm and 0.75cm of b] (b1);
\vertex[below right=1.0cm and 0.75cm of b] (b2);
\diagram*{
(a)-- [line width=0.25mm,fermion, arrow size=1.2pt, style=bostonuniversityred,ultra thick] (a1),
(a1) -- [line width=0.25mm, charged scalar, half left,arrow size=1.2pt, style=black,ultra thick, edge label={\(\color{black}{\eta^-}\)}] (a2),
(a1) -- [line width=0.25mm, plain,half right, arrow size=1.2pt, style=black, ultra thick, edge label'={\(\color{black}{N_k}\)}] (a2),
(a2) -- [line width=0.25mm, fermion, arrow size=1.2pt, style=gray, ultra thick, , edge label={\(\color{black}{\ell_{\beta}}\)}] (b),
(b) -- [line width=0.25mm, fermion,arrow size=1.2pt, style=mediumtealblue, ultra thick] (b2),
(b) -- [line width=0.25mm, boson,arrow size=1.2pt, style=mediumtealblue,ultra thick] (b1)};
\vertex[above right=1.0cm and 1cm of b] {\(\color{black}{\gamma }\)};
\vertex[below right=1.0cm and 1cm of b] {\(\color{black}{\ell_{\beta}}\)};
\node at (b)[circle,fill,style=black,inner sep=1pt]{};
\node at (a1)[circle,fill,style=black,inner sep=1pt]{};
\node at (a2)[circle,fill,style=black,inner sep=1pt]{};
\end{feynman}
\end{tikzpicture}\hfill
\begin{tikzpicture}[baseline={(current bounding box.center)},style={scale=0.9, transform shape}]
\begin{feynman}
\vertex (a){\(\color{black}{\ell_{\alpha}}\)};
\vertex [right=1.5cm of a] (b);
\vertex [above right=1.5cm and 1cm of b] (b4);
\vertex [below right=0.0cm and 1cm of b] (b1);
\vertex [below right=0.0cm and 2cm of b] (b2);
\vertex [below right=0.0cm and 1cm of b2] (b3);
\diagram*{
(a)-- [line width=0.25mm,fermion, arrow size=1.2pt, style=bostonuniversityred,ultra thick] (b),
(b) -- [line width=0.25mm, fermion, arrow size=1.2pt, style=gray,ultra thick, edge label'={\(\color{black}{\ell_{\alpha}}\)}] (b1),
(b1) -- [line width=0.25mm,half left, plain,arrow size=1.2pt, style=black,ultra thick, edge label={\(\color{black}{N_k}\)}] (b2),
(b1) -- [line width=0.25mm, charged scalar, half right,arrow size=1.2pt, style=black,ultra thick, edge label'={\(\color{black}{\eta^-}\)}] (b2),
(b2) -- [line width=0.25mm, fermion, arrow size=1.2pt, style=mediumtealblue, ultra thick] (b3),
(b) -- [line width=0.25mm, boson,arrow size=1.2pt, style=mediumtealblue, ultra thick] (b4)};
\vertex[above right=1.5cm and 1cm of b] {\(\color{black}{\gamma }\)};
\vertex[below right=0.0cm and 1cm of b2] {\(\color{black}{\ell_{\beta}}\)};
\node at (b)[circle,fill,style=black,inner sep=1pt]{};
\node at (b1)[circle,fill,style=black,inner sep=1pt]{};
\node at (b2)[circle,fill,style=black,inner sep=1pt]{};
\end{feynman}
\end{tikzpicture}
\end{tcolorbox}
\end{adjustbox}
\caption{One-loop Feynman diagrams related to $\ell_{\alpha}\to\ell_{\beta}\gamma$.}
\label{feyn:lfv}
\end{figure}
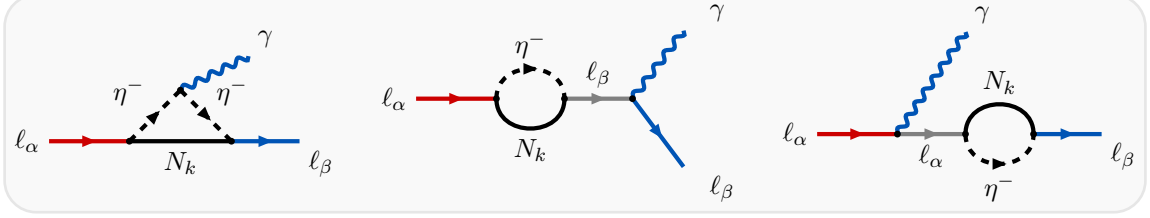

The branching ratio of $\ell_{\alpha}\to\ell_{\beta}\gamma$ decay via VSM (Feynman graphs are shown in \fig\ref{feyn:lfv}) is given 
by \cite{Toma:2013zsa,Vicente:2014wga,Valiente:2023blo,Hundi:2022iva,Chen:2019nud},
\begin{align}
\mathcal{BR}(\ell_{\alpha}\to\ell_{\beta}\gamma)=\dfrac{3(4\pi)^3\alpha_{\rm em}}{4 G_F^2}|F_D|^2~\mathcal{BR}(\ell_{\alpha}\to\ell_{\beta}\nu_{\alpha}\overline{\nu_{\beta}})\,,
\label{eq:mueg}
\end{align}
where $\alpha_{\rm em}$ and $G_F$ are the electromagnetic fine-structure constant and Fermi constant, respectively, and 
$F_D$ is the dipole form factor, provided in Appendix~\ref{sec:loop}. For SM leptonic decay branching, $\ell_{\alpha}\to\ell_{\beta}\nu_{\alpha}\overline{\nu_{\beta}}$, see \cite{ParticleDataGroup:2024cfk}.

\subsubsection{$\ell_\alpha \to 3\,\ell_\beta$}
Charged lepton flavor–violating decay $\ell_\alpha \to 3\,\ell_\beta$ (i.e., $\ell_\alpha \to \ell_\beta \,\bar{\ell}_\beta \,\ell_\beta$) is expected to provide a significant limit on NP with upcoming experimental sensitivities, particularly from the Mu3e experiment. 
The current exclusion and projection (proj) limit at $90\% \rm ~C.L.$ on $\mu \to 3e$ decay from the SINDRUM and Mu3e experiments are given as:
\begin{gather}
\mathcal{BR}(\mu \to 3e) < 1.0 \times 10^{-12} \quad(\text{SINDRUM~\cite{SINDRUM:1987nra})}\,,\\
\mathcal{BR}(\mu \to 3e) < 
2.0~(0.1) \times 10^{-15}  \quad\text{(Mu3e Phase-I~(II) proj.~\cite{Mu3e:2020gyw} ~(\cite{COMET:2025sdw}))}\,.
\end{gather}
At the one-loop level, the amplitude receives contributions from $\gamma$-penguin, $Z$-penguin, Higgs-penguin, and box diagrams. However, the Higgs-penguin contributions are neglected, as we focus on processes involving the first two generations of charged leptons, for which the smallness of the Yukawa couplings renders such effects negligible. This approximation does not hold for LFV processes involving $\tau$ leptons; however, the corresponding experimental constraints are presently less stringent than those for the first two generations.

For this model framework, the branching ratio for $\ell_\alpha \to 3\,\ell_\beta$ is given by \cite{Toma:2013zsa}:
\begin{align}\label{eq:mu3e}
\nonumber\mathcal{BR}(\ell_{\alpha}\to\ell_{\beta}\overline{\ell}_{\beta}\ell_{\beta})&=\dfrac{3(4\pi)^2\alpha_{\rm em}^2}{8\rm G_F^2}\left[|A_{ND}|^2+|F_D|^2\left(\dfrac{16}{3}\log\left(\dfrac{m_\alpha}{m_\beta}\right)-\dfrac{22}{3}\right)+\dfrac{1}{6}|B|^2\right.\\&\nonumber\left.
+\dfrac{1}{3}\left(2|F_{RR}|^2+|F_{RL}|^2\right)+\left(-2A_{ND}F_D^*+\dfrac{1}{3}A_{ND}B^*-\dfrac{2}{3}F_{D}B^*+{\rm h.c.}\right)\right]\\&\times\mathcal{BR}(\ell_{\alpha}\to\nu_{\alpha}\ell_{\beta}\overline{\nu}_{\beta})\,,
\end{align}
where $F_D,~A_{ND},~B,~F_{RR},~{\rm and }~F_{RL}$ are detailed in Appendix~\ref{sec:loop}. For this calculation, we have used $\mathcal{BR}(\mu\to\nu_\mu e\bar{\nu}_e)\sim100\%$ \cite{ParticleDataGroup:2024cfk}. Note however, $\mu\to 3e$ imposes a weaker constraint on the model compared to the $\mu\to e\gamma$, except the region where $x_i\ll 1$, as discussed in~\cite{Toma:2013zsa}. Despite relatively weak constraint, future measurements of this observable might provide an excellent probe of such models, since the projected sensitivity to the corresponding branching ratio is expected to reach the level of $\mathcal{O} (10^{-16})$ \cite{Mu3e:2020gyw}. For illustration, we quote the branching ratio of $\mu\to 3e$ at two characteristic benchmark points in \cref{tab:LFV} which satisfy all the constraints, and show the projected future sensitivities of $\mu\to 3e$ in \fig\ref{fig:mueg_neutrino} to lie close in probing our allowed model parameters. 
\subsubsection{$\mu \to e$ conversion}
%
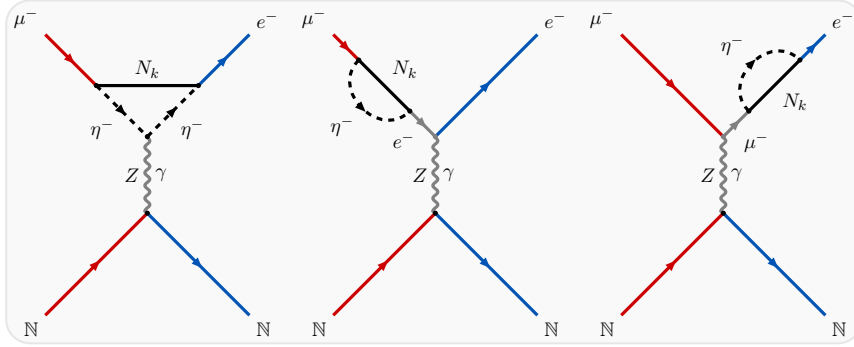
\begin{figure}[htb!]
\centering
\begin{adjustbox}{width=0.75\textwidth}
\begin{tcolorbox}[colback=gray!5, colframe=black!10, boxrule=1pt, arc=4mm, boxsep=0pt, left=0pt, right=0pt, top=0pt, bottom=0pt, width=1\textwidth, halign=center]
\begin{tikzpicture}[baseline={(current bounding box.center)},style={scale=0.9, transform shape}]
\begin{feynman}
\vertex (a);
\vertex[above right=1cm and 1.0cm of a] (a2);
\vertex[above right=2cm and 2.0cm of a] (a4);
\vertex[above left=1cm and 1.0cm of a] (a1);
\vertex[above left=2cm and 2.0cm of a] (a3);
\vertex[below=1.5cm of a] (b);
\vertex[below left=2cm and 2cm of b] (b1);
\vertex[below right=2cm and 2cm of b] (b2);
\diagram*{
(a3)-- [line width=0.25mm,fermion, arrow size=1.2pt, style=bostonuniversityred,ultra thick] (a1),
(a1) -- [line width=0.25mm, plain,arrow size=1.2pt, style=black,ultra thick, edge label={\(\color{black}{N_k}\)}] (a2),
(a2) -- [line width=0.25mm, fermion,arrow size=1.2pt, style=mediumtealblue,ultra thick] (a4),
(a1) -- [line width=0.25mm, charged scalar, arrow size=1.2pt, style=black, ultra thick, edge label'={\(\color{black}{\eta^- }\)}] (a),
(a) -- [line width=0.25mm, charged scalar,arrow size=1.2pt, style=black, ultra thick, edge label'={\(\color{black}{\eta^- }\)}] (a2),
(a) -- [line width=0.25mm, boson,  arrow size=1.2pt, style=gray,ultra thick, edge label={\(\color{black}{\gamma}\)}, edge label'={\(\color{black}{Z}\)}] (b),
(b1) -- [line width=0.25mm,fermion, arrow size=1.2pt, style=bostonuniversityred,ultra thick] (b),
(b) -- [line width=0.25mm, fermion,arrow size=1.2pt, style=mediumtealblue,ultra thick] (b2)};
\vertex[above left=2cm and 2.0cm of a]{\(\color{black}{\mu^- }\)};
\vertex[above right=2cm and 2.0cm of a]{\(\color{black}{e^-}\)};
\vertex[below left=2cm and 2cm of b]{\(\color{black}{\mathbb{N}}\)};
\vertex[below right=2cm and 2cm of b]{\(\color{black}{\mathbb{N}}\)};
\node at (a)[circle,fill,style=black,inner sep=1pt]{};
\node at (b)[circle,fill,style=black,inner sep=1pt]{};
\node at (a1)[circle,fill,style=black,inner sep=1pt]{};
\node at (a2)[circle,fill,style=black,inner sep=1pt]{};
\end{feynman}
\end{tikzpicture}\hfill
\begin{tikzpicture}[baseline={(current bounding box.center)},style={scale=0.9, transform shape}]
\begin{feynman}
\vertex (a);
\vertex[above right=1cm and 1.0cm of a] (a2);
\vertex[above right=0.5cm and 0.5cm of a] (a22);
\vertex[above right=2cm and 2.0cm of a] (a4);
\vertex[above right=1.5cm and 1.5cm of a] (a44);
\vertex[above left=1cm and 1.0cm of a] (a1);
\vertex[above left=0.5cm and 0.5cm of a] (a11);
\vertex[above left=2cm and 2.0cm of a] (a3);
\vertex[above left=1.5cm and 1.5cm of a] (a33);
\vertex[below=1.5cm of a] (b);
\vertex[below left=2cm and 2cm of b] (b1);
\vertex[below right=2cm and 2cm of b] (b2);
\diagram*{
(a3)-- [line width=0.25mm,fermion, arrow size=1.2pt, style=bostonuniversityred,ultra thick] (a33),
(a33) -- [line width=0.25mm, plain,arrow size=1.2pt, style=black,ultra thick,edge label={\(\color{black}{N_k}\)}] (a11),
(a33) -- [line width=0.25mm, half right,charged scalar,arrow size=1.2pt, style=black,ultra thick,edge label'={\(\color{black}{\eta^-}\)}] (a11),
(a11) -- [line width=0.25mm, fermion, arrow size=1.2pt, style=gray, ultra thick, edge label'={\(\color{black}{e^- }\)}] (a),
(a)-- [line width=0.25mm,fermion, arrow size=1.2pt, style=mediumtealblue,ultra thick] (a4),
(a) -- [line width=0.25mm, boson,  arrow size=1.2pt, style=gray,ultra thick, edge label={\(\color{black}{\gamma}\)}, edge label'={\(\color{black}{Z}\)}] (b),
(b1) -- [line width=0.25mm,fermion, arrow size=1.2pt, style=bostonuniversityred,ultra thick] (b),
(b) -- [line width=0.25mm, fermion,arrow size=1.2pt, style=mediumtealblue,ultra thick] (b2)};
\vertex[above left=2cm and 2.0cm of a]{\(\color{black}{\mu^- }\)};
\vertex[above right=2cm and 2.0cm of a]{\(\color{black}{e^-}\)};
\vertex[below left=2cm and 2cm of b]{\(\color{black}{\mathbb{N}}\)};
\vertex[below right=2cm and 2cm of b]{\(\color{black}{\mathbb{N}}\)};
\node at (a)[circle,fill,style=gray,inner sep=1pt]{};
\node at (b)[circle,fill,style=black,inner sep=1pt]{};
\node at (a11)[circle,fill,style=black,inner sep=1pt]{};
\node at (a33)[circle,fill,style=black,inner sep=1pt]{};
\end{feynman}
\end{tikzpicture}\hfill
\begin{tikzpicture}[baseline={(current bounding box.center)},style={scale=0.9, transform shape}]
\begin{feynman}
\vertex (a);
\vertex[above right=1cm and 1.0cm of a] (a2);
\vertex[above right=0.5cm and 0.5cm of a] (a22);
\vertex[above right=2cm and 2.0cm of a] (a4);
\vertex[above right=1.5cm and 1.5cm of a] (a44);
\vertex[above left=1cm and 1.0cm of a] (a1);
\vertex[above left=0.5cm and 0.5cm of a] (a11);
\vertex[above left=2cm and 2.0cm of a] (a3);
\vertex[above left=1.5cm and 1.5cm of a] (a33);
\vertex[below=1.5cm of a] (b);
\vertex[below left=2cm and 2cm of b] (b1);
\vertex[below right=2cm and 2cm of b] (b2);
\diagram*{
(a3)-- [line width=0.25mm,fermion, arrow size=1.2pt, style=bostonuniversityred,ultra thick] (a),
(a44)-- [line width=0.25mm,fermion, arrow size=1.2pt, style=mediumtealblue,ultra thick] (a4),
(a22) -- [line width=0.25mm, plain,arrow size=1.2pt, style=black,ultra thick,edge label'={\(\color{black}{N_k}\)}] (a44),
(a22) -- [line width=0.25mm, half left,charged scalar,arrow size=1.2pt, style=black,ultra thick,edge label={\(\color{black}{\eta^-}\)}] (a44),
(a) -- [line width=0.25mm, fermion, arrow size=1.2pt, style=gray, ultra thick, edge label'={\(\color{black}{\mu^- }\)}] (a22),
(a) -- [line width=0.25mm, boson,  arrow size=1.2pt, style=gray,ultra thick, edge label={\(\color{black}{\gamma}\)}, edge label'={\(\color{black}{Z}\)}] (b),
(b1) -- [line width=0.25mm,fermion, arrow size=1.2pt, style=bostonuniversityred,ultra thick] (b),
(b) -- [line width=0.25mm, fermion,arrow size=1.2pt, style=mediumtealblue,ultra thick] (b2)};
\vertex[above left=2cm and 2.0cm of a]{\(\color{black}{\mu^- }\)};
\vertex[above right=2cm and 2.0cm of a]{\(\color{black}{e^-}\)};
\vertex[below left=2cm and 2cm of b]{\(\color{black}{\mathbb{N}}\)};
\vertex[below right=2cm and 2cm of b]{\(\color{black}{\mathbb{N}}\)};
\node at (a)[circle,fill,style=gray,inner sep=1pt]{};
\node at (b)[circle,fill,style=black,inner sep=1pt]{};
\node at (a22)[circle,fill,style=black,inner sep=1pt]{};
\node at (a44)[circle,fill,style=black,inner sep=1pt]{};
\end{feynman}
\end{tikzpicture}
\end{tcolorbox}
\end{adjustbox}
\caption{1-loop Feynman diagrams related to $\mu~\mathbb{N}\to e~\mathbb{N}$.}
\label{feyn:lfv}
\end{figure}

Muon-to-electron ($\mu\to e$) conversion in muonic atoms serves as a major probe for LFV. This process originates when a muon, brought to rest in a target material, is captured by an atom to form a muonic state, and subsequently cascading into the $1s$ orbital. While the SM limits muon's fate to either decay via $\mu^- \to e^- \bar{\nu}_e \nu_\mu$ in orbit, or undergo standard \textit{nuclear muon capture} ($\mu^- + \mathbb{N}(A, Z) \to \nu_\mu + \mathbb{N}(A, Z - 1)$), many BSM frameworks, including ours (see Fig. \ref{feyn:lfv}), predict neutrinoless coherent conversion, $\mu^- + \mathbb{N}(A, Z) \to e^- + \mathbb{N}(A, Z)$. Although this transition violates individual lepton flavor numbers ($L_e$ and $L_\mu$) by one unit, it maintains the conservation of the total lepton number. As the conversion is a coherent process involving the entire nucleus, it offers a remarkably clean signature for detecting NP.

The branching ratio of $\mu\to e$ conversion can be written as \cite{Arganda:2007jw},
\begin{align}\label{eq:mute}
\mathcal{R}_{\mu\to e}^{\mathbb{N}} &= \frac{\Gamma(\mu^- + \mathbb{N}(A,Z) \to e^- + \mathbb{N}(A,Z))}{\Gamma(\mu^- + \mathbb{N}(A,Z) \to \nu_{\mu} + \mathbb{N}(A,Z-1))}\nonumber\\
&=
\frac{G_F^2~\alpha_{\rm em}^3 \,Z_{\rm eff}^4\,F_p^2\,p_e\,E_e \,m_\mu^3 
}{
8\pi^2 Z\,\Gamma_{\rm capt}
}\Bigg\{\left|
(Z+N)\left(g_{LV}^{(0)}+g_{LS}^{(0)}\right)
+(Z-N)\left(g_{LV}^{(1)}+g_{LS}^{(1)}\right)
\right|^2\nonumber\\
&\quad\quad\quad\quad\quad\quad\quad\quad+\left|
(Z+N)\left(g_{RV}^{(0)}+g_{RS}^{(0)}\right)+(Z-N)\left(g_{RV}^{(1)}+g_{RS}^{(1)}\right)
\right|^2
\Bigg\}\,,
\end{align}
following the notation in~\cite{Arganda:2007jw}. Z(N) denotes the proton(neutron) numbers of the target nucleus, $Z_{\rm eff}$ denotes effective atomic charge, $F_p$ corresponds to the nuclear form factor, while $\Gamma_{\rm capt}$ is the total muon capture rate in the nucleus. Apart, $p_e\simeq E_e\simeq m_\mu$ indicates the momentum and energy of the outgoing electron, and the $\alpha_{\rm em}$ denotes electromagnetic fine-structure constant. For details, see Appendix.~\ref{sec:loop}.

SINDRUM II experiment provides the most stringent limit for neutrinoless $(\mu\to e)$ conversion to date, $7 \times 10^{-13}$ for gold target. Next-generation facilities such as Mu2e and COMET are designed to improve this sensitivity by four orders of magnitude using aluminium targets. The exclusion limits and projected sensitivities at $90\%$ C.L. for these upcoming experiments are summarized below:
\begin{gather}
\mathcal{R}_{\mu\to e}^{\rm Au} < 7\times 10^{-13} \quad\text{(SINDRUM-II~\cite{SINDRUMII:2006dvw})}\,,\\
\mathcal{R}_{\mu\to e}^{\rm Al} < 
\begin{cases}
6.2~(0.3)\times 10^{-16}  \quad\text{(Mu2e Run-I~(II) proj.~\cite{Mu2e:2022ggl}~(\cite{Mu2e-II:2022blh}))}\,,\\
7.0~(0.02)\times 10^{-15}  \quad\text{(COMET Phase-I~(II) proj.~\cite{COMET:2018auw} ~(\cite{Moritsu:2022lem})}\,.
\end{cases}
\end{gather}
\begin{table}[htb!]
\centering
\begin{adjustbox}{max width=\textwidth}
\begin{tabular}{c c c c c c c c}
\hline \hline 
 $\metaI~(\text{GeV})$ &
$\mntwo~(\text{TeV})$ &
$\Deltam~(\text{GeV})$ & $|h_{\mu 2}|$
 &
$\mathcal{BR}(\mu\to e\gamma)$ &
$\mathcal{BR}(\mu\to 3e)$ &
$\mathcal{R}_{\mu\to e}^{\rm Au}$ &
$\mathcal{R}_{\mu\to e}^{\rm Al}$\\
\hline
\hline
\multirow{3}*{$580$} & \multirow{3}*{$5.2$} & $0.01$ & $0.55$& $6.6\times 10^{-13}$ & $5.2\times 10^{-15}$ & $4.2\times 10^{-14}$ &
$3.6\times10^{-14}$\\
 &  & $0.02$ & $0.41$ & $2.1\times 10^{-13}$ & $1.6\times 10^{-15}$ & $1.3\times 10^{-14}$ & $9.0\times 10^{-15}$\\
 &  & $0.03$ & $0.34$ & $9.3\times 10^{-14}$ & $7.3\times 10^{-16}$ & $5.9\times 10^{-15}$ & $4.0\times 10^{-15}$\\
 \hline
 \multirow{3}*{$1260$} & \multirow{3}*{$6.4$} & $0.01$ & $0.55$& $1.8\times 10^{-13}$ & $1.3\times 10^{-15}$ & $4.8\times 10^{-15}$ &
 $3.6\times10^{-16}$\\
 &  & $0.02$ & $0.37$ & $4.4\times 10^{-14}$ & $3.2\times 10^{-16}$ & $1.2\times 10^{-15}$ & $8.2\times10^{-16}$\\
 &  & $0.03$ & $0.30$ & $1.9\times 10^{-14}$ & $1.4\times 10^{-16}$ & $5.3\times 10^{-16}$ & $3.6\times10^{-16}$\\
 \hline \hline
\end{tabular}
\end{adjustbox}
\caption{ Different LFV observables for two characteristic benchmark points of our model are provided here with $\delta m=0.2~\mathrm{GeV}$, and smallest active neutrino mass $m_{\nu^{}_{1}}=10^{-5}$ eV.}
\label{tab:LFV}
\end{table}
We calculate the branching ratios of all these different LFV processes for two characteristic benchmark points of the model 
in \cref{tab:LFV}. 

\begin{figure}[htb!]
 \centering
 \includegraphics[width=0.45\linewidth]{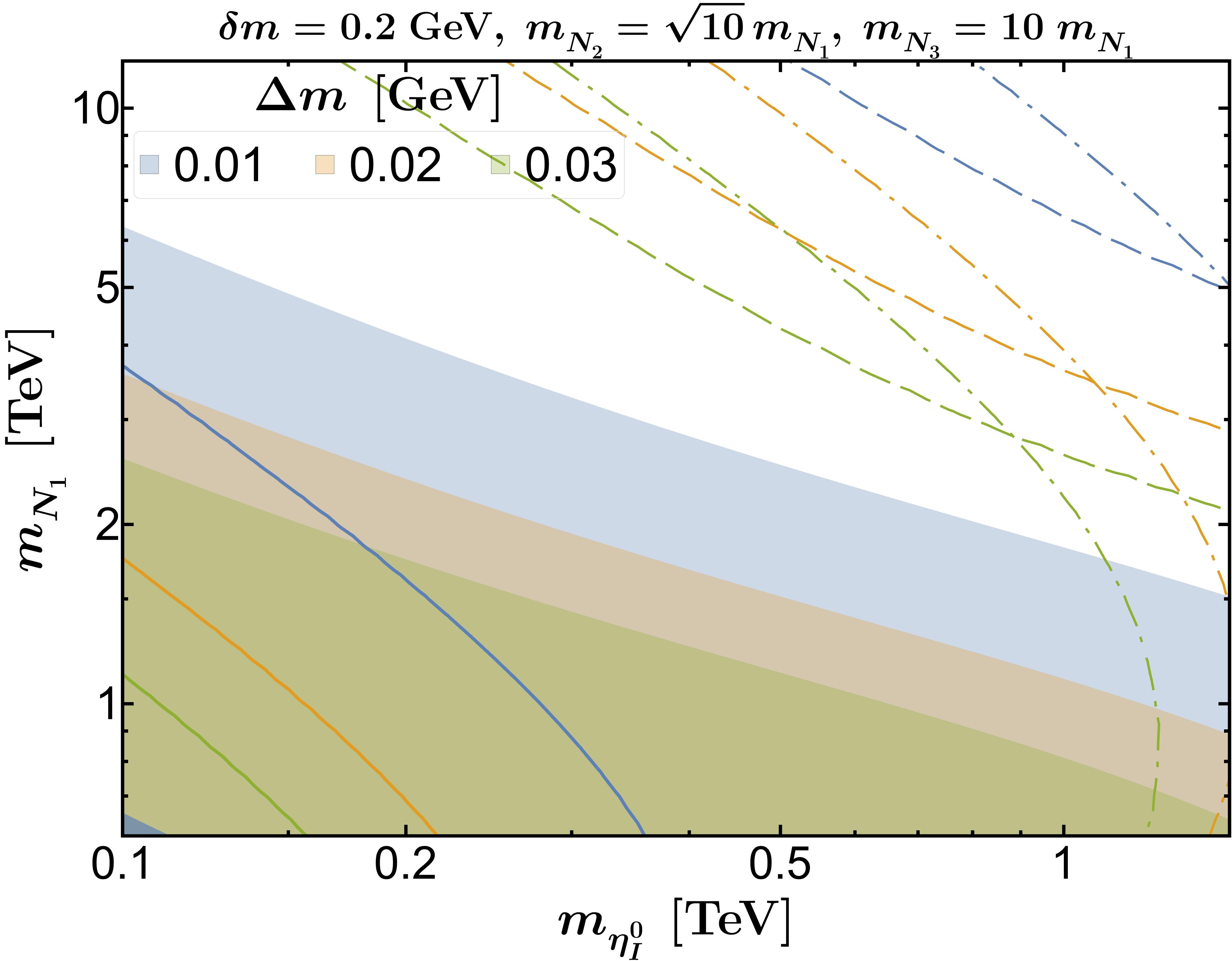}\quad
 \includegraphics[width=0.45\linewidth]{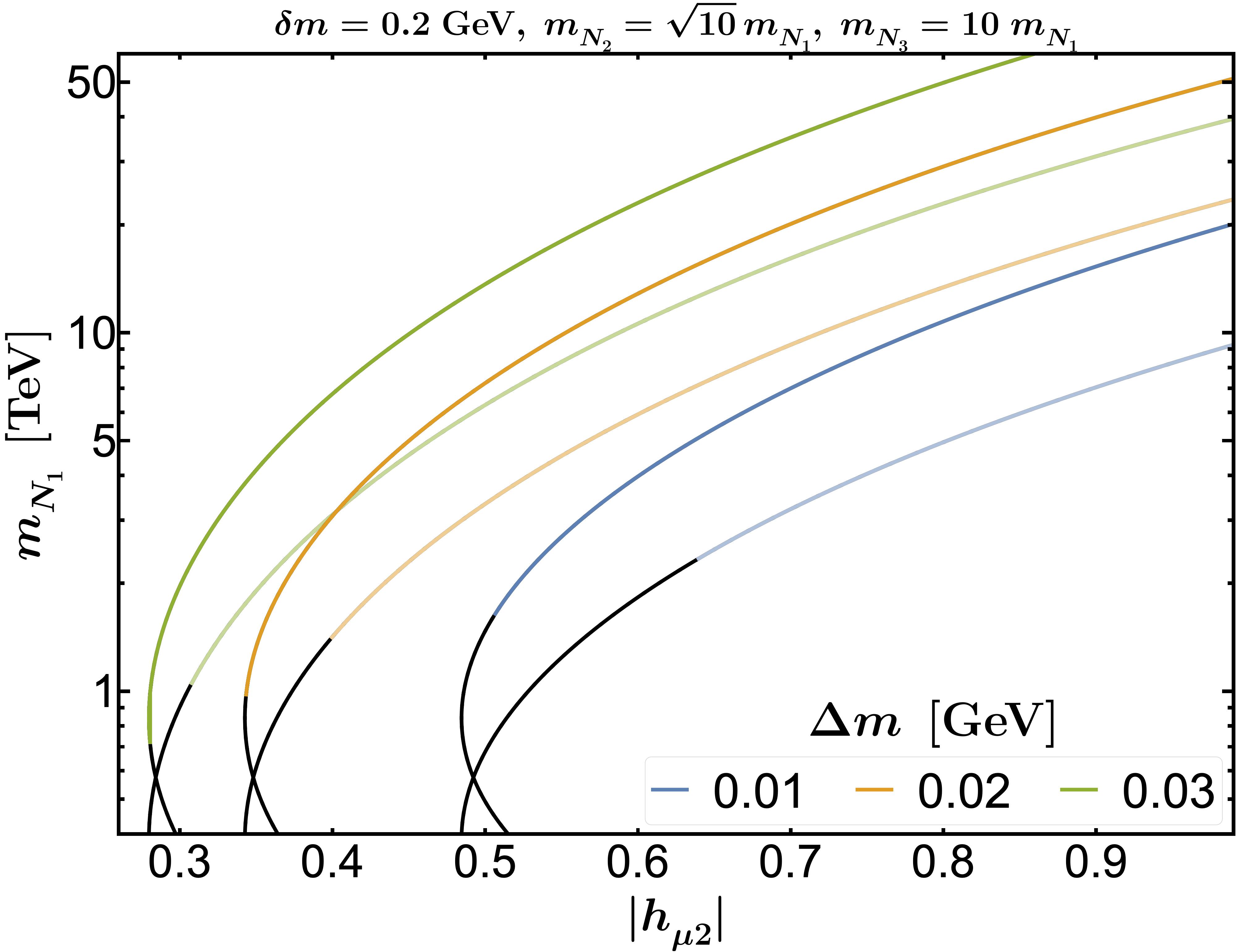}
 \caption{
Constraints from LFV observables and active neutrino mass limits in the VSM are depicted with large Yukawa coupling ($h_{\mu 2}$) and $m_{\nu^{}_{1}}=10^{-5}$ eV. \texttt{Left panel}: The shaded regions in $\metaI-\mnone$ plane are excluded by the current MEG II limit, with different colours corresponding to different values of $\Delta m$ mentioned in the plot legend. The dashed curves denote the projected Mu3e sensitivity, while the thick solid and dot-dashed curves represent the present SINDRUM-II bound and the projected COMET Phase-I sensitivity, respectively. \texttt{Right panel}: The coloured curves in $|h_{\mu2}|-\mnone$ plane corresponding to specific $\Delta m$ as mentioned in the legend, satisfy active neutrino mass constraints for $\metaI=1.26~\mathrm{TeV}$ (shown by dark colored curves) and $\metaI=580~\mathrm{GeV}$ (shown by light color curves). The black segments are excluded from the MEG II experimental limit.}
\label{fig:mueg_neutrino}
 \end{figure}

Let us summarise all the LFV and neutrino mass constraints here, as shown in \fig\ref{fig:mueg_neutrino}. In VSM, the active neutrino masses are primarily governed by the parameters $$\{h_{\alpha\beta},m_{N_k},\letaHpp,\metaR,\metaI\}.$$ One can rewrite the expression for the branching ratio of $\ell_{\alpha}\to\ell_{\beta}\gamma$ as,
\begin{align}
\mathcal{BR}(\ell_{\alpha}\to\ell_{\beta}\gamma)\simeq\dfrac{3(4\pi)^3\alpha_{\rm em}}{16 }\left(\frac{1}{G_F}\,\sum_{i=1}^3\dfrac{|\mathfrak{M}_{\alpha\beta}|}{\Delta m\,\metaI\,m_{N_{i}^{}}}\right)^2~\mathcal{BR}(\ell_{\alpha}\to\ell_{\beta}\nu_{\alpha}\overline{\nu_{\beta}})\,.
\label{eq:muegf}
\end{align}
The expression for $\mathfrak{M}_{\alpha\beta}$ can be found in \eq\eqref{eq:muegap}. From \eq\eqref{eq:muegf}, it follows that, for a fixed mass splitting $\Deltam$, the LFV branching ratios increase as $\metaI$ decreases. Consequently, to satisfy the current upper bound on $\mathcal{BR}(\mu\to e\gamma)$ requires larger $\mnone$ in the low $\metaI$ regime (see the left plot of \fig\ref{fig:mueg_neutrino}). On the other hand, for a fixed value of $\metaI$, increasing the mass splitting $\Delta m$ suppresses the LFV branching ratio, thereby allowing smaller $\mnone$ while remaining consistent with the existing experimental limits.
 

Right panel of \fig\ref{fig:mueg_neutrino}, displays the parameter space compatible with active neutrino masses for $\metaI=360$ GeV (dark shaded curves) and $\metaI=580$ (light shaded curves). The green, orange, and blue curves correspond to $\Deltam=\{0.03,~0.02,~0.01\}~\GeV$, respectively. Since both $\metaI$ and $\Delta m$ are fixed for each curve, the neutrino mass constraints uniquely determine the Yukawa coupling $h$ for a given value of $\mnone$ (see \eq\eqref{eq:CI}). Along each curve, the coloured segments denote parameter points that respect both the neutrino oscillation data and the $\mu\to e\gamma$ bound, whereas the black segments falls into the exclusion region of $\mathcal{BR}(\mu\to e\gamma)>1.5 \times 10^{-13}$. For fixed values of $\Delta m$ and $\metaI$, lowering $\mnone$ generally requires a reduction in the Yukawa couplings (see \eq\eqref{eq:Numassap}), in order to maintain the correct active neutrino mass spectrum.\footnote{This behavior holds for $\mnone\gtrsim\metaI$. For $\metaI>\mnone$, the dependence is reversed and Yukawa coupling increases as $\mnone$ decreases (see \eq\eqref{eq:Numassap}.} However, a smaller RHN mass simultaneously enhances the $\mu\to e\gamma$ branching ratio (see \eq\eqref{eq:muegf}). Therefore, below a certain threshold value of $\mnone$, the LFV rates exceed the experimental upper limits ($\mathcal{BR}(\mu\to e\gamma)>3.1 \times 10^{-13}$), rendering that region of parameter space phenomenologically unacceptable. This behavior is reflected by the black portions of the curves in the right panel of \fig\ref{fig:mueg_neutrino}. As illustrated in these figures, most of the parameter space of the model remains consistent with existing experimental constraints and lies within the prospective reach of future runs. VSM contributes also to Electric Dipole Moment (EDM), however our model parameter space that is susceptible for future collider probes lie far from the EDM experimental sensitivities, see Appendix \ref{sec:EDM} for details.
\subsection{Electroweak precision data}
Electroweak precision data \cite{Baak:2014ora}, in particular, the oblique parameters as computed in ref\,.~\cite{Barbieri:2006dq, Hambye:2009pw, Abada:2018zra}, constrains the DM mass splitting with the charged scalar by \cite{Barbieri:2006dq},
\begin{align}
|\deltam|\lesssim 140\,\GeV\,.
\end{align}
LEP precision data rule out parameter space with $m_{\eta_{R}^0}< 80$ GeV, $m_{\eta_{I}^0}< 100$ GeV and $\Deltam>8$ GeV\footnote{Note that the mass splitting $\Deltam$ cannot be arbitrarily small either: direct-detection constraints require $\Deltam \gtrsim \mathcal{O}(100\,\mathrm{keV})$ to suppress $Z$-mediated inelastic scattering \cite{Hambye:2009pw}, while IceCube solar neutrino data imply $\letaHpp\gtrsim 1.6\times10^{-5}(m_{\rm DM}/\mathrm{TeV})$, thereby setting a lower bound on $\Deltam$~\cite{deBoer:2021pon}.}  \cite{Lundstrom:2008ai}. 

\subsection{Collider constraints}
\label{sec:ppc}
Null observation at collider searches put constraints on the VSM model parameters. 
LEP and LHC bounds for slepton searches can be translated into a bound on charged inert scalar $\metap$. 
The current ATLAS lower bound translated on $\metap$ gives $\metap\gtrsim  270\,\GeV$ \cite{ATLAS:2014zve,Pierce:2007ut,Hashemi:2015swh, Kalinowski:2018ylg, ALEPH:2002ftu,Banerjee:2025qkb}. 
A recent study by the CMS Collaboration \cite{CMS:2026cjr,CMS:2026rhc} provides exclusion limits in the 
$(\metaI-\Deltam)$ plane, where the observed exclusion region reaches $\metaI = 108~\text{GeV}$ for 
$\Deltam = 78~\text{GeV}$, while for $\metaI = 70~\text{GeV}$, it covers the range $\Deltam = 40\text{-}90~\text{GeV}$.

Charged scalar Higgs searches at LEP~\cite{ALEPH:2002ftu} set a lower bound of $m_{H^\pm} > 72.5~\GeV$ in Type-I and $m_{H^\pm} > 80~\GeV$ in Type-II two-Higgs-doublet models (2HDM). These limits, however, apply only to the mass range below the top quark threshold. At the LHC, analogous searches have been performed both below~\cite{ATLAS:2014otc,CMS:2015lsf} and above~\cite{ATLAS:2018gfm} $m_t$. In the high-mass region, the bounds depend on the scalar mixing parameter $\tan\beta$; for example, for $\tan\beta = 10$, charged scalars lighter than $181~\GeV$ are excluded. These constraints rely on conventional 2HDM scenarios and assume prompt decays. 

In the Scotogenic model, the presence of a scalar dark matter candidate can yield a compressed spectrum, 
potentially eliminating prompt 
signatures. If the charged scalar becomes long-lived, it may instead produce disappearing tracks (DTs) observable at the LHC. This possibility has been examined in Ref.~\cite{Belyaev:2020wok}, where the ATLAS long-lived chargino search at $\sqrt{s} = 13~\text{TeV}$ with $36.1~\text{fb}^{-1}$~\cite{ATLAS:2017oal} was recast for several models, yielding exclusion contours in the $(m_{\eta^+}, \tau)$ plane for charged scalar pair production, where $\tau$ indicates decay life time.
\begin{figure}[htb!]
\centering
\includegraphics[width=0.475\linewidth]{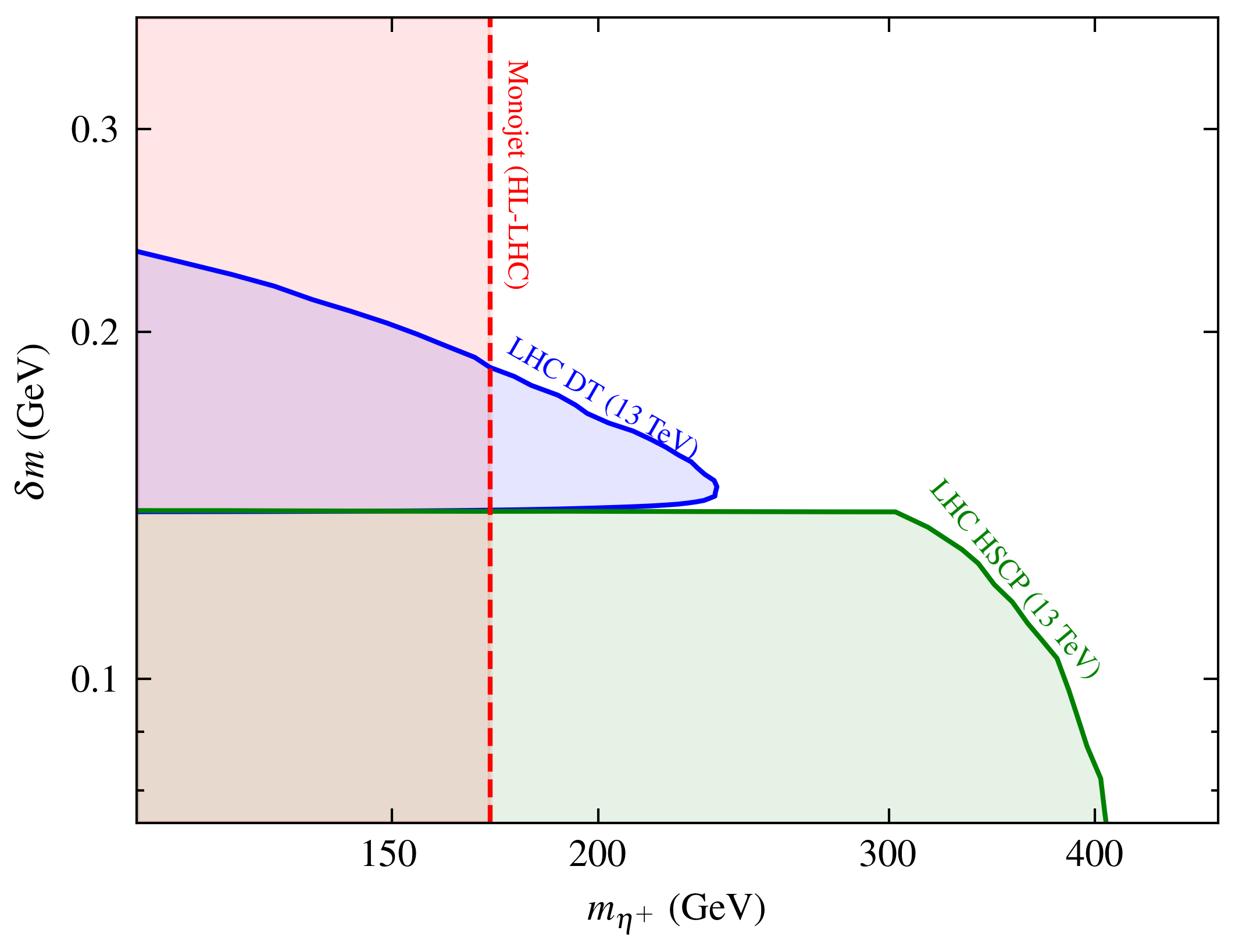}
\includegraphics[width=0.475\linewidth]{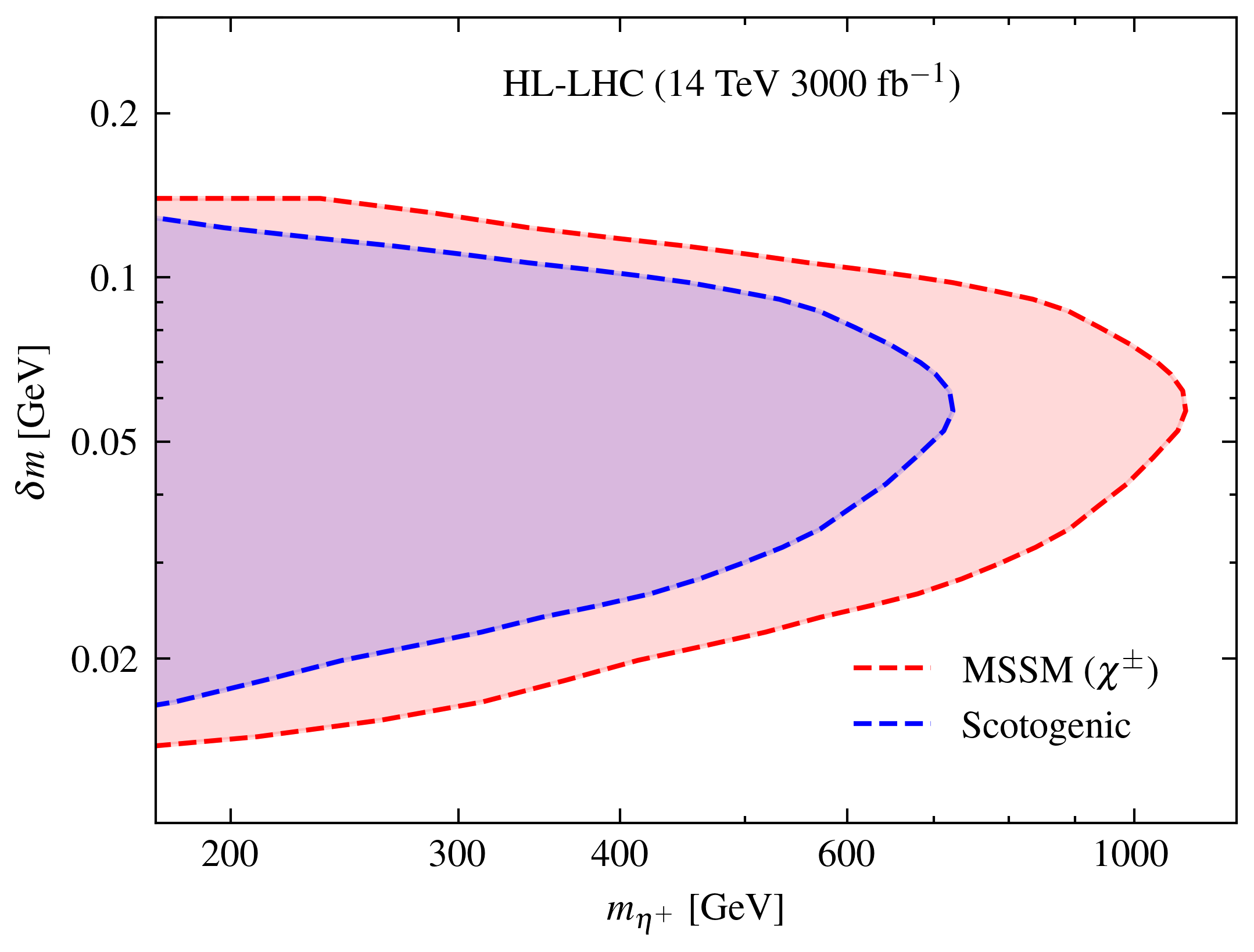}
\caption{\textit{Left}: 95\% C.L. exclusion limits in the $m_{\eta^{+}}-\delta m$ plane from LHC disappearing track (DT) shown by blue region, and heavy stable charged particle (HSCP) searches at $\sqrt{s}=13$ TeV by green region, together with the projected HL-LHC monojet + missing transverse energy sensitivity at $\sqrt{s}=14$ TeV with an integrated luminosity of 3 ab$^{-1}$ (\textit{red} dashed line). \textit{Right}: Projected 95\% C.L. on disappearing track exclusion in the $m_{\eta^{+}}-\delta m$ plane, obtained by translating the wino-chargino limits of Ref.~\cite{ATLAS:2018jjf} to the scotogenic scenario through lifetime matching and production cross-section rescaling at HL-LHC. The shaded regions indicate the excluded parameter space.}
\label{fig:lhc}
\end{figure}

In \fig\ref{fig:lhc}, we show the corresponding recast limit for the inert 2HDM, phenomenologically equivalent to the Scotogenic model in its LHC charged scalar sector, mapped onto the $(m_{\eta^+}, \delta m)$ plane. The maximum excluded mass reaches $\sim 237~\text{GeV}$. Similarly, at very small mass splittings, $\delta m \lesssim 0.15~\text{GeV}$, the charged scalar can become effectively stable at detector scales, giving rise to heavy stable charged particle (HSCP) signatures rather than disappearing tracks. Existing LHC HSCP searches at ATLAS with $\sqrt{s} = 13~\text{TeV}$ and luminosity $36.1~\text{fb}^{-1}$~\cite{ATLAS:2019gqq}, constrain this regime strongly, excluding charged scalar masses up to approximately $400~\text{GeV}$ for sufficiently long lifetimes. Charged scalar pair production can also yield a prompt mono jet signature from initial state 
radiation; current LHC limits for this channel are weaker than the LEP bound. The projected HL-LHC sensitivity from Ref.~\cite{Belyaev:2018ext} excludes masses up to $172~\text{GeV}$, also shown in \fig\ref{fig:lhc}.

\paragraph*{HL-LHC Disappearing Track Projection} The prospects for probing long-lived charged scalars at the HL-LHC can be estimated using the projected disappearing track sensitivity derived for wino-like charginos in Ref.~\cite{ATLAS:2018jjf}. Since the dominant decay mode of the charged scalar in the compressed regime, $\eta^\pm \to \eta^0_{I/R}\,\pi^\pm$, proceeds through the same weak current interaction as the wino ($\chi^{\pm}$) decay to neutralino ($\chi^{0}$): $\chi^\pm \to \chi^0\,\pi^\pm$, the charged scalar lifetime can be mapped directly onto the chargino lifetime through the corresponding mass splitting $\delta m$. To reinterpret the projected HL-LHC limits, we extract the 95\% C.L. exclusion contour in the $(m_{\chi^\pm},\tau)$ plane from Ref.~\cite{ATLAS:2018jjf}, for $pp \to \chi^{+}\chi^{-},\,\chi^{\pm}\chi^{0}$. For each point on the contour of the exclusion limit as in Ref.~\cite{ATLAS:2018jjf}, the corresponding wino production cross section is determined and matched to the inert-doublet/scotogenic charged scalar production cross section ($pp \to \eta^{+}\eta^{-},\,\eta^{\pm}\eta_{I/R}^{0}$), thereby obtaining an equivalent exclusion contour in the $(m_{\eta^\pm},\tau)$ plane. The resulting limit is subsequently expressed in the $(m_{\eta^\pm},\delta m)$ plane using the lifetime-mass-splitting relation.

The translated exclusion contour is shown in the right panel of \fig\ref{fig:lhc}. Owing to the smaller electroweak production cross section of scalar doublets compared to fermionic winos, the projected sensitivity is reduced after translation to the inert scalar/scotogenic scenario. Nevertheless, the HL-LHC is expected to significantly extend the present disappearing track reach, probing charged scalar masses up to maximum $700~\GeV$ for lifetimes corresponding to $\delta m \simeq (0.02-0.2)~\GeV$. The excluded region lies to the left of the projected contour, while larger masses remain beyond the anticipated HL-LHC sensitivity.
\section{Model phenomenology}
\label{sec:pheno}
The VSM frameowrk not only addresses the origin of non-zero neutrino masses and the DM relic density, but also provides possible explanations for other observed and unobserved phenomena in our Universe, such as lepton flavor violating decays and the baryon asymmetry, respectively. In the following subsection, we discuss the allowed parameter space that satisfies the exclusion limits from existing direct, indirect, and collider experiments, while simultaneously accounting for the aforementioned phenomena. In the next section, we additionally assess the detectability of the model parameters in future DM search experiments at future lepton colliders.

\subsection{Dark matter relic density and search prospect}
\label{sec:darkmatter}
\begin{figure}[htb!]
\centering
\includegraphics[width=0.475\linewidth]{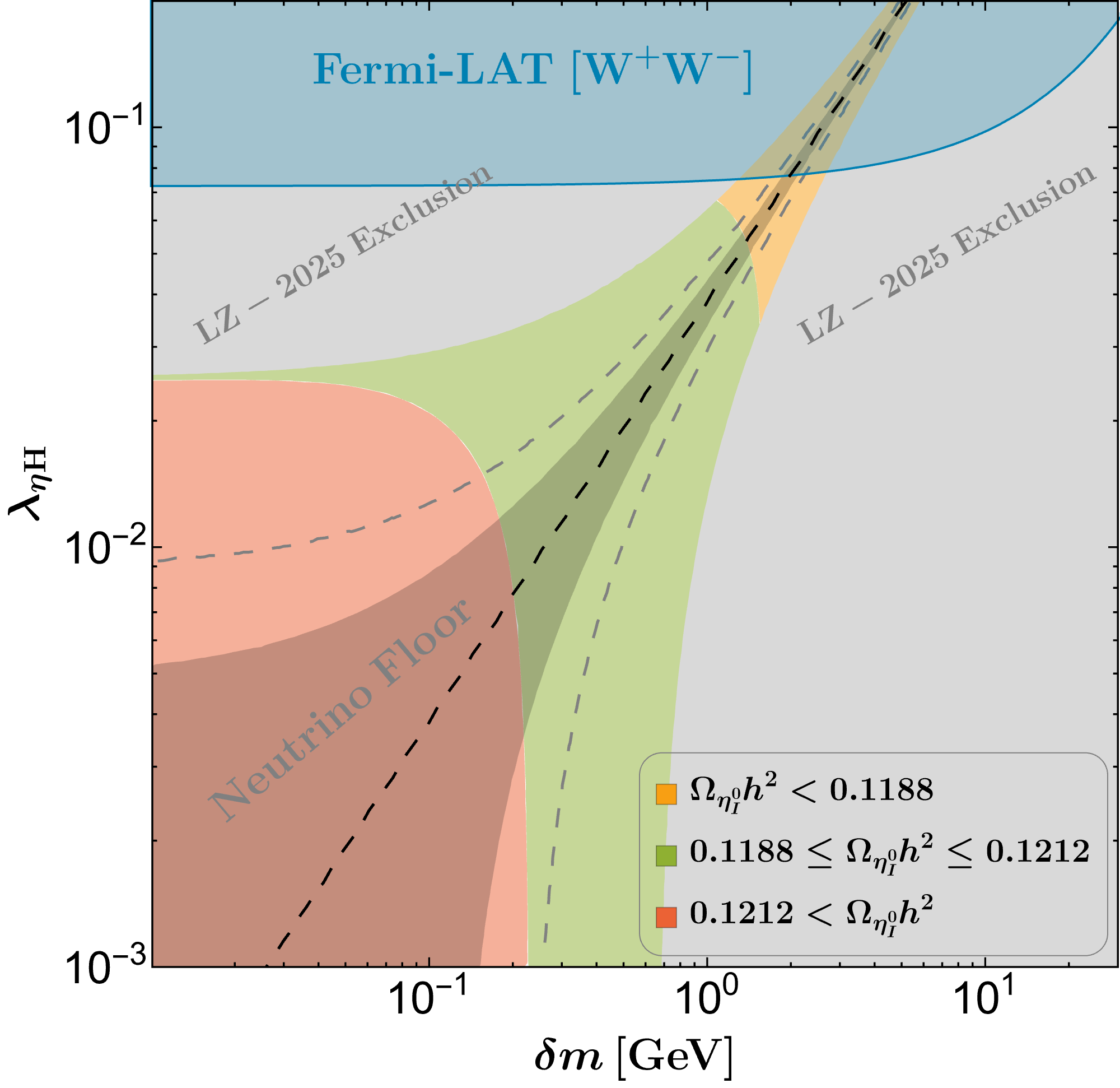}~~
\includegraphics[width=0.475\linewidth]{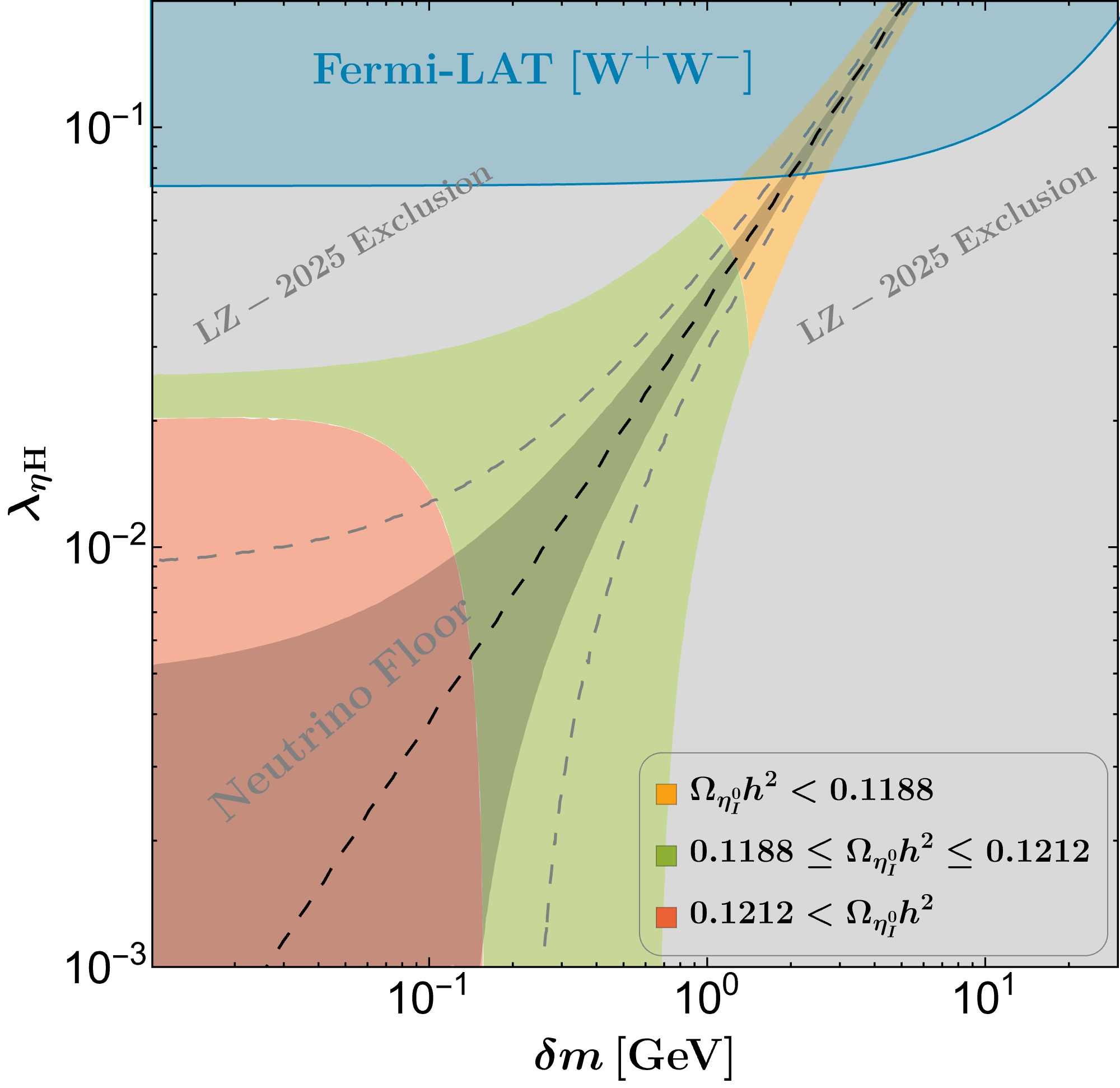}
\caption{The yellow, light green, and light red color shade represent the relic density regimes corresponding to under-abundance, correct abundance, and over-abundance, respectively. The light gray shaded region is excluded by the LZ-2025 constraint on the spin-independent DM–nucleon scattering cross section, while the gray band below the DD exclusion corresponds to the neutrino floor. The gray dashed lines denote the projected sensitivity of the XLZD Collaboration to the DM-nucleon scattering cross section. The black dashed line indicates the locus, $\letaHpp=\letaHp+\letaH$, along which the effective Higgs-DM coupling vanishes. The analysis is performed with $\metaI = 582~\GeV$, $\mnone= 4.0~\TeV,~\mntwo=3.0~\mnone,~\mnthree=10.0~\mnone$, $|h_{i\ell}|=0.1$, $\metap=\metaI+\deltam$ and $\metaR-\metaI=\Deltam = 0.01~(0.1)~\GeV$ for left (right) figure. Additionally, all points comply with the exclusion limits from the radiative muon decay process, $\mu^+ \to e^+ \gamma$. The deep gray-shaded region over the DD-allowed regime indicates the neutrino-floor. The indirect detection constraint from Fermi-LAT is shown by the light sky-blue shaded region.}
\label{fig:relic_DD}
\end{figure}

In this analysis, we assume that the CP-odd neutral component of the inert doublet serves as the real scalar DM candidate. Since the inert doublet carries electroweak charge, the DM particle annihilates into SM particles through both gauge and Higgs portal interactions. The relic density depends dominantly on the mass splittings $\Deltam$\footnote{ The lower limit on the splitting between the real and imaginary neutral components, $\Delta m \sim 100~\mathrm{keV}$, follows from constraints on $Z$-mediated inelastic DM-nucleon scattering, $\eta_I^0 \mathbb{N}\to \eta_R^0 \mathbb{N}$, which becomes kinematically forbidden for splittings larger than the typical halo DM kinetic energy \cite{Barbieri:2006dq, Hambye:2009pw,LopezHonorez:2006gr}.} and $\deltam$, as well as on the Higgs portal coupling $\letaH$. The parameters $\delta m$ and $\letaH$ also determine the spin-independent DM–nucleon scattering cross section, $\sigma^{\rm SI}_{\eta_I^0 N}$, and also control the annihilation into $W^+W^-$, which is constrained by indirect detection limits from Fermi-LAT. Owing to the presence of gauge-mediated annihilation channels, the parameter space with DM mass below $\sim 550~\GeV$ is mostly underabundant, except in the vicinity of the Higgs resonance, while masses above $\sim 650~\GeV$ lead to an overabundant relic density. Consequently, our analysis is restricted to this intermediate mass regime, and the correlations among the relevant parameters are shown in \fig\ref{fig:relic_DD}. We evaluate the present DM relic density by solving the Boltzmann equation (BEQ) using the micrOMEGAs package \cite{Alguero:2023zol}. The spin-independent DM–nucleon scattering cross section, $\sigma^{\rm SI}_{\eta_I^0 N}$, is also computed with micrOMEGAs, while the DM annihilation cross sections are calculated using CalcHEP \cite{Belyaev:2012qa}.

In the left (right) panel of \fig\ref{fig:relic_DD}, we show the over-, correct-, and under-abundant DM regions in the $\delta m$–$\lambda_{\eta H}$ plane, represented by red, green, and yellow shaded regions, respectively, for $\Delta m = 0.01~(0.1)~\GeV$. The DM relic density increases gradually with increasing $\delta m$, as the effective Higgs-portal coupling increases accordingly; however, this simultaneously enhances $\sigma^{\rm SI}_{\rm \eta_I^0 N}$, which is evident in both panels. Since $\Delta m$ is relevant only for the DM relic density, variations due to $\Delta m$ appear solely through changes in the relic density regime because of DM co-annihilation processes.
Although a significant portion of the relic-density allowed parameter space with $\delta m \gtrsim 1~\mathrm{GeV}$ is excluded by the recent LZ-2025 direct-detection results, as indicated by the deep-gray shaded region, the gray dashed lines denote the projected sensitivity of XLZD, and some regions lie below the neutrino floor \cite{Billard:2013qya, OHare:2021utq}. Nevertheless, the remaining parameter space between the LZ exclusion limit and the neutrino floor remains viable for our phenomenological analysis.
In addition, indirect-detection limits on DM annihilation into $W^+W^-$ from Fermi-LAT impose strong constraints on the parameter space allowed by both direct detection and relic density; however, these bounds could be relaxed for $\delta m \gtrsim 10~\GeV$ and entirely allowed for $\letaH\gtrsim 10^{-2}$.
Notably, in the neutrino-floor, if the DM direct-detection sensitivity is unable to distinguish between DM and neutrino signals, then the collider searches and indirect-detection prospects provide viable alternative avenues to probe this VSM framework. We also note that \fig\ref{fig:relic_DD} corresponds to a benchmark choice of DM mass $\metaI = 565~\GeV$; the results will be slightly modified for other mass choices of BSM particles within the aforementioned regime. Moreover, in the right panel of \fig\ref{fig:relic_DD}, we perform a similar exercise for $\Delta m = 0.1~\mathrm{GeV}$, which provides some relaxation in $\lambda_{\eta H}$ and $\delta m$ within the relic-allowed parameter space after imposing the DD constraints.
\begin{figure}[htb!]
\centering
\includegraphics[width=0.475\linewidth]{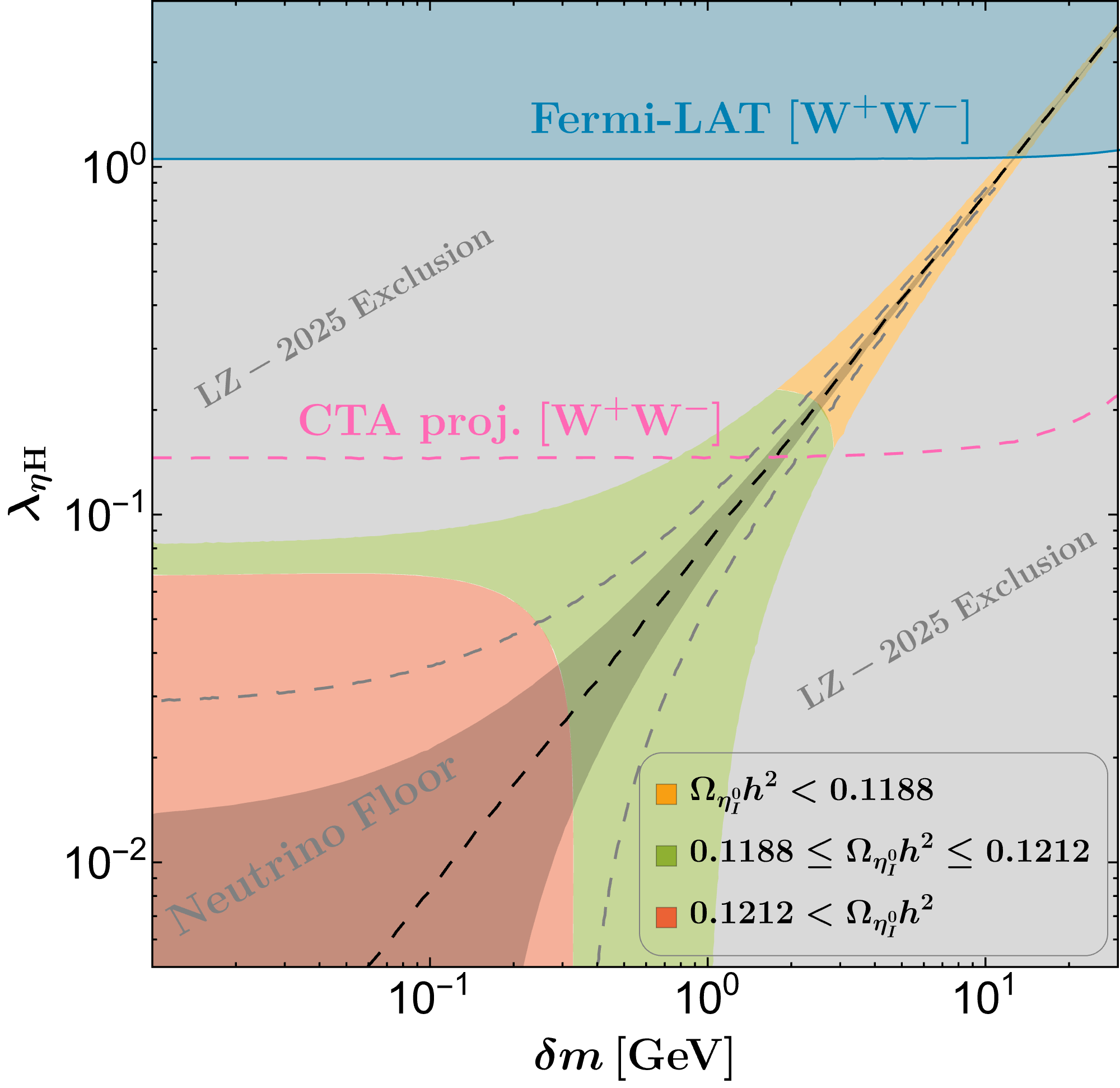}~~
\includegraphics[width=0.475\linewidth]{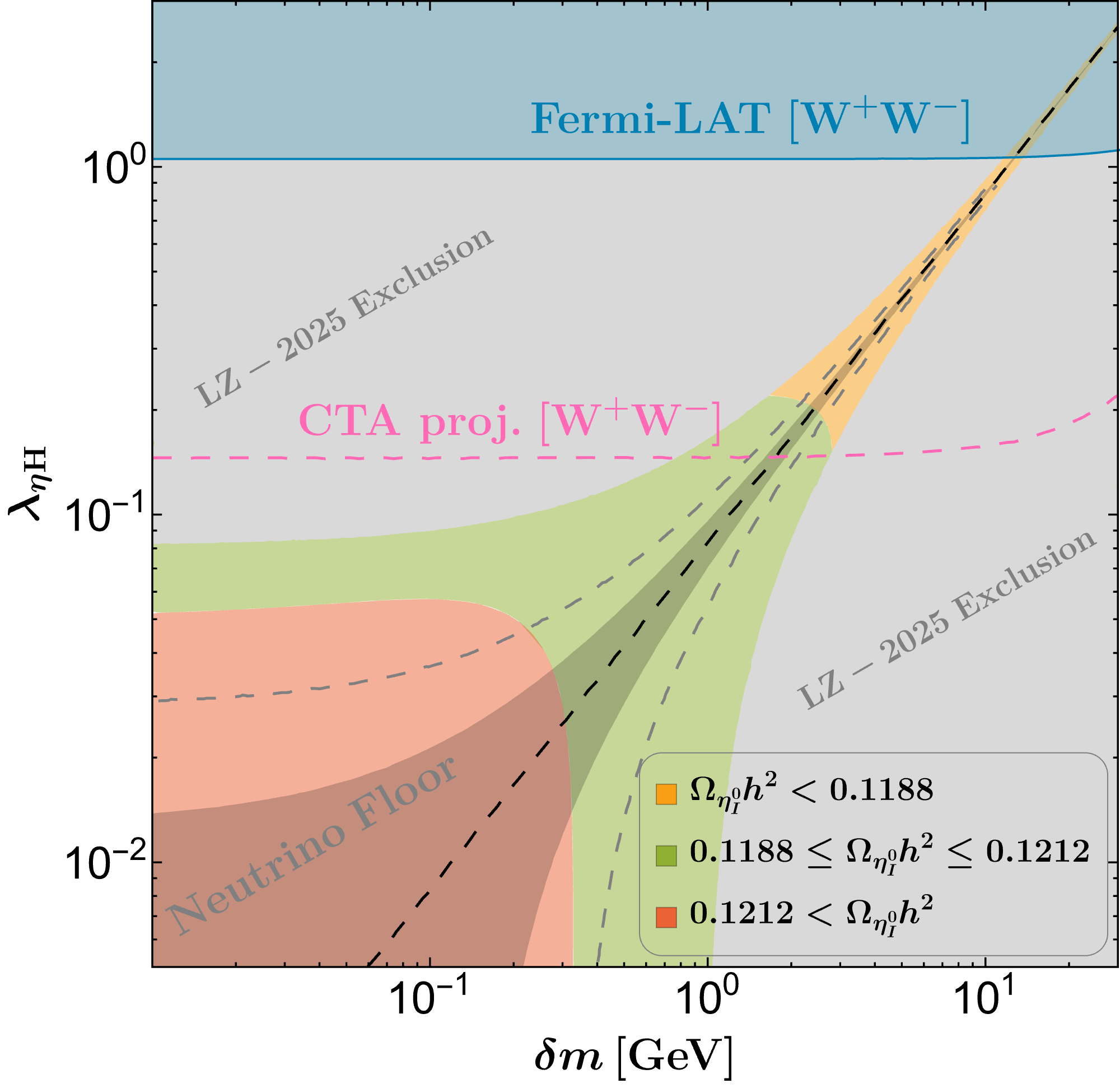}
\caption{Same as in \fig\ref{fig:relic_DD}, but with different parameters as $\metaI = 1.26~\TeV$, $\mntwo= 6.4~\TeV,~\mnone=\mntwo/\sqrt{10},~\mnthree=\sqrt{10}~\mntwo,~|h_{\ell 1}|=10^{-6},~|h_{e 2}|=|h_{e3}|=0.08,~|h_{\mu 2}|=|h_{\mu 3}|=0.33,~|h_{\tau 2}|=|h_{\tau 3}|=0.34$, $\metap=\metaI+\deltam$ and $\metaR-\metaI=\Deltam = 0.03~(0.3)~\GeV$ for left (right) figure. The indirect detection constraint on the DM annihilation to $W^+W^-$ from Fermi-LAT is shown by the light sky-blue shaded region, and the CTA projection \cite{CTAConsortium:2017dvg} is shown by pink dashed line.}
\label{fig:relic_DD2}
\end{figure}

The effects of Yukawa couplings can be incorporated by considering relatively light right-handed neutrino masses 
($\sim $few $\TeV$) and large Yukawa coupling values as shown in \fig\ref{fig:relic_DD2}. This is similar to \fig\ref{fig:relic_DD},
 the key difference arises in the DM mass, which must now be significantly larger ($\gg 565~\mathrm{GeV}$) to reproduce the observed relic density; otherwise, the DM remains under abundant. In this regime, the dominant DM number-changing process is the $t$-channel annihilation $\eta \eta^\dagger \to L \bar{L}$. All other discussions regarding DD and ID constraints remain unchanged. 
 
\subsection{Leptogenesis and matter anti-matter asymmetry}
\label{sec:leptogensis}
\begin{figure}[htb!]
\centering
\begin{adjustbox}{width=\textwidth}
\begin{tcolorbox}[colback=gray!5, colframe=black!10, boxrule=1pt, arc=4mm, boxsep=0pt, left=0pt, right=0pt, top=0pt, bottom=0pt, width=1\textwidth, halign=center]
\begin{tikzpicture}
\begin{feynman}
\vertex(a2){\(\color{black}{N_{1}}\)};
\vertex[right=2cm of a2] (b);
\vertex[above right=1.0cm and 1cm of b] (b1);
\vertex[below right=1.0cm and 1cm of b] (b2);
\diagram*{
(a2) -- [line width=0.25mm, plain, arrow size=1.2pt, style=bostonuniversityred, ultra thick] (b),
(b) -- [line width=0.25mm, charged scalar,arrow size=1.2pt, style=mediumtealblue, ultra thick] (b2),
(b) -- [line width=0.25mm, fermion,arrow size=1.2pt, style=mediumtealblue,ultra thick] (b1)};
\vertex[above right=1.0cm and 1cm of b] {\(\color{black}{L}\)};
\vertex[below right=1.0cm and 1cm of b] {\(\color{black}{\eta}\)};
\node at (b)[circle,fill,style=mediumtealblue,inner sep=1pt]{};
\end{feynman}
\end{tikzpicture}\hfill
\begin{tikzpicture}
\begin{feynman}
\vertex (a){\(\color{black}{N_1}\)};
\vertex[right=1.5cm of a] (a1);
\vertex[right=1cm of a1] (a2);
\vertex[right=1cm of a2] (b);
\vertex[above right=1.0cm and 1cm of b] (b1);
\vertex[below right=1.0cm and 1cm of b] (b2);
\diagram*{
(a)-- [line width=0.25mm, plain, arrow size=1.2pt, style=bostonuniversityred,ultra thick] (a1),
(a1) -- [line width=0.25mm, charged scalar, half left,arrow size=1.2pt, style=black,ultra thick, edge label={\(\color{black}{\eta}\)}] (a2),
(a2) -- [line width=0.25mm, fermion,half left, arrow size=1.2pt, style=black, ultra thick, edge label={\(\color{black}{L}\)}] (a1),
(a2) -- [line width=0.25mm, plain, arrow size=1.2pt, style=gray, ultra thick, edge label={\(\color{black}{N_{2,3}}\)}] (b),
(b) -- [line width=0.25mm, charged scalar,arrow size=1.2pt, style=mediumtealblue, ultra thick] (b2),
(b) -- [line width=0.25mm, fermion,arrow size=1.2pt, style=mediumtealblue,ultra thick] (b1)};
\vertex[above right=1.0cm and 1cm of b] {\(\color{black}{L}\)};
\vertex[below right=1.0cm and 1cm of b] {\(\color{black}{\eta}\)};
\node at (b)[circle,fill,style=mediumtealblue,inner sep=1pt]{};
\node at (a1)[circle,fill,style=black,inner sep=1pt]{};
\node at (a2)[circle,fill,style=black,inner sep=1pt]{};
\end{feynman}
\end{tikzpicture}\hfill
\begin{tikzpicture}
\begin{feynman}
\vertex(a){\(\color{black}{N_{1}}\)};
\vertex[right=1.5cm of a] (b);
\vertex[above right=0.75cm and 0.75cm of b] (b1);
\vertex[below right=0.75cm and 0.75cm of b] (b2);
\vertex[above right=0.25cm and 0.5cm of b1] (b11){\(\color{black}{L}\)};
\vertex[below right=0.25cm and 0.5cm of b2] (b22){\(\color{black}{\eta}\)};
\diagram*{
(a) -- [line width=0.25mm, plain, arrow size=1.2pt, style=bostonuniversityred, ultra thick] (b),
(b) -- [line width=0.25mm,edge label={\(\color{black}{\eta}\)}, charged scalar,arrow size=1.2pt, style=black, ultra thick] (b1),
(b) -- [line width=0.25mm, fermion,edge label'={\(\color{black}{L}\)},arrow size=1.2pt, style=black,ultra thick] (b2),
(b2) -- [line width=0.25mm, charged scalar,arrow size=1.2pt, style=mediumtealblue, ultra thick] (b22),
(b1) -- [line width=0.25mm, fermion,arrow size=1.2pt, style=mediumtealblue,ultra thick] (b11),
(b2) -- [line width=0.25mm, plain,arrow size=1.2pt,edge label'={\(\color{black}{N_{2,3}}\)}, style=black, ultra thick] (b1),
};
\node at (b)[circle,fill,style=black,inner sep=1pt]{};
\node at (b1)[circle,fill,style=black,inner sep=1pt]{};
\node at (b2)[circle,fill,style=black,inner sep=1pt]{};
\end{feynman}
\end{tikzpicture}
\end{tcolorbox}
\end{adjustbox}
\caption{Feynman diagrams related to the CP-asymmetry generation.}
\label{feyn:CP-asymmetry}
\end{figure}
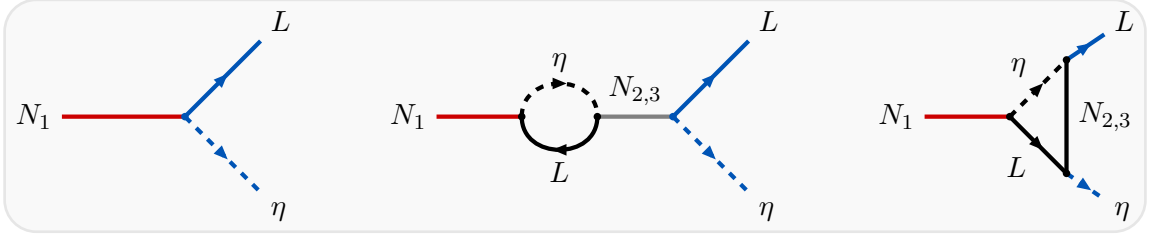
\noindent
 Despite the existence of several mechanisms for generating the baryon asymmetry of the Universe, leptogenesis is one of the most viable options. In this framework, a generated (B-L) asymmetry is partially converted into a baryon asymmetry when sphaleron processes decouple at around $130~\GeV$~\cite{DOnofrio:2014rug}. A nonzero CP asymmetry is generated through the interference between the tree-level and one-loop contributions to the decay ($N\to L\eta$)\footnote{Note $N\to L H$ is prohibited by $\mathcal{Z}_2$ symmetry.}, 
 as shown in \fig\ref{feyn:CP-asymmetry}.
In the VSM framework with three RHNs, successful leptogenesis can, in general, be realized at scales as low as a few tens of TeV~\cite{Hugle:2018qbw, Mahanta:2019gfe}. Furthermore, when spectator processes are properly taken into account and the inert doublet is assumed to be out of thermal equilibrium, leptogenesis can be achieved even at the $\mathcal{O}(\rm TeV)$ scale~\cite{Racker:2024fpn}. In contrast to that approach, in this work, we identify a previously unexplored choice of Casas-Ibarra (CI) parameters for which the 2nd and 3rd column ($N_2$ and $N_3$ coloumn, see \eq\eqref{eq:yukawa}) of  Yukawa coupling can be sizable $\mathcal{O}(0.1)$ while remaining within the perturbative regime~\footnote{Note that we keep $h_{1\alpha}\ll1$ so that we can have the sufficient amount of out-of equillibrium decay of $N_{1}$ to respect Shakarov's 3rd criteria~\cite{Sakharov:1967dj} to have successful leptogenesis.}. This increase will manifest as the increase in the CP asymmetry as $\varepsilon_{N_1}\sim\sum_{1\neq j}^{}{\rm Im}(h^\dagger h)^2_{1j}/(h^\dagger h)_{11}$, which in turn helps to lower the scale of the leptogenesis to TeV scale. In addition, this setup provides promising prospects for model detection in future experimental searches.

In our analysis, we use the same strategy as mentioned in \cite{Hugle:2018qbw} to get the maximum possible CP asymmetry. Here, we slightly modify the strategy; instead of optimizing $\lambda_{\eta H}^{\prime\prime}$, we optimize $\Delta m$, as this parameter plays a significant role in the DM phenomenology, while $\lambda_{\eta H}^{\prime\prime}$ can be retrieved from $\Delta m$ just by using \eq\eqref{eq:metaR2} and \eq\eqref{eq:metaI2}. The $\mathbb{R}$ matrix (see Appendix~\ref{sec:numass}) of the CI parametrization has three comlex angle $z_{12},z_{13}$ and $z_{23}$. The choice which produce the maximum CP asymmetry~\cite{Hugle:2018qbw} with $h_{\alpha2(3)}\sim\mathcal{O}(0.1)$, is mentioned below;
\begin{align}
    z_{12}=0,\quad z_{13_R^{}}=z_{13_R^{}}=\sqrt{\frac{m_{\nu^{}_{1}}}{2 m_{\nu^{}_{3}}}}\quad \text{and}\quad z_{23}=0+7i\,,
\end{align}
where $m_{\nu_{1}^{}}$ and $m_{\nu_{3}^{}}$ are respectively the lightest and heaviest active neutrino mass. Note that the choice of $z_{23_I^{}}=7$ is responsible for the large $h_{\alpha 2}$ and $h_{
\alpha 3}$. As an example, the structure of the Yukawa matrix for the choice of parameters $m_{\nu_{1}^{}}\sim 10^{-7}$ eV, $\Delta m=0.01$ GeV, $\metaI=580$ GeV, and $m_{\rm N_1^{}}^{}=1.4$ TeV, which generates the observed asymmetry, is as follows:
\begin{equation}
h=
\begin{pmatrix}
1.22\times10^{-6}+ 6.02\times10^{-8} i & 0.11 + 0.09 i & -0.13 + 0.15 i \\
2.66\times10^{-7} - 9.44\times10^{-7} i & 0.20 - 0.51 i & 0.71 + 0.28 i \\
1.74\times10^{-6}+ 1.00\times10^{-6} i & -0.16 - 0.55 i & 0.75 - 0.22 i
\end{pmatrix}\,.
\end{equation}

The comoving number density corresponding to the B-L asymmetry ($N_{B-L}$) is evaluated by solving the coupled BEQ and is given by,
\begin{eqnarray}\begin{split}
\dfrac{dN^{}_{N_{1}}}{dz}  = &-D_{N_{1}}\left(N_{N_{1}}-N_{N_{1},0}^{}\right),\\
\dfrac{dN_{B-L}^{}}{dz} = & -\varepsilon_{N_{1}}^{}D_{N_{1}}\left(N_{N_{1}}-N_{N_{1},0}^{}\right)-W_{\rm tot} N_{B-L}\,,
\label{eq:cbeq-lepto}
\end{split}
\end{eqnarray}
where $z=\mnone/T$, $N_{j,0}=N_j^{\eq}$, and the CP-asymmetry is dictated by $\varepsilon_{N_1}^{}$ (see eq:~\eqref{eq:cp asy}). Here $D_{N_{1}}^{}$ refers to decay and $W_{\rm tot}^{}$ indicates total wash-out contribution coming from inverse-decay plus $\Delta L=2$ scattering cross-section terms for $N_{1}$ as defined below:
\begin{eqnarray}
D_{N_{1}}^{}& = & K_{N_{1}}^{}z\dfrac{\kappa_{1}(z)}{\kappa_{2}(z)},\\
W_{\rm tot}^{}& = & W_{\rm ID} + W_{\Delta L=2},\\
W_{\rm ID}& = & \dfrac{1}{4}K_{N_{1}}^{}z^{3}\kappa_{1}(z),\\
W_{\Delta L=2}&=&\frac{\Gamma_{\Delta L=2}}{H z_1}= \frac{36\sqrt{5}M_P}{2\sqrt{\pi\,g_{\ast}}v^4}\frac{\mnone}{z^2}\frac{\overline{m_{\xi}}^2}{(\lambda_{\eta H}^{\prime\prime})^2}\,,
\label{eq:L2}
\end{eqnarray}
where $\overline{m_{\xi}}^2=\sum_{i,j}^{}\xi_{i}\xi_j{\rm Re}[(\mathbb{R}\mathbb{M}_\nu \mathbb{R^\dagger})_{ij}^{2}]$ with $\mathbb{M}_\nu$ as the diagonal active neutrino mass matrix (see Appendix~\ref{sec:numass}) and $K_{N_{1}}^{}=\Gamma_{N_{1}}^{}/H(z=1)$ is known as the decay parameter with the decay width $\Gamma_{N_{1}}^{}$ (see eq.~\eqref{eq:N1dw}), $H(z)$ defines Hubble parameter, and $\kappa_{i}$'s are the modified Bessel functions of second kind. In the following, we use baryon-to-photon ratio $\eta_{B}$ using $\eta_{B}\equiv 3/4\,\mathcal{C}_{\rm sp}\,g_{\ast s}/g_{\ast}\, N_{\rm B-L}^{\rm sp}$, where $g_{\ast s}=43/11$ and $g_{\ast }=116$ are respectively the effective degrees of freedom for energy density and entropy density. Also, $\mathcal{C}_{\rm sp}=8/23$~\cite{Harvey:1990qw} is the sphaleron conversion factor for two Higgs doublet models (2HDM) and $N_{B}^{\rm sp}$ is the $B-L$ asymmetry at the time of sphaleron decoupling.

\begin{figure}[htb!]
 \centering
 \includegraphics[width=1\linewidth]{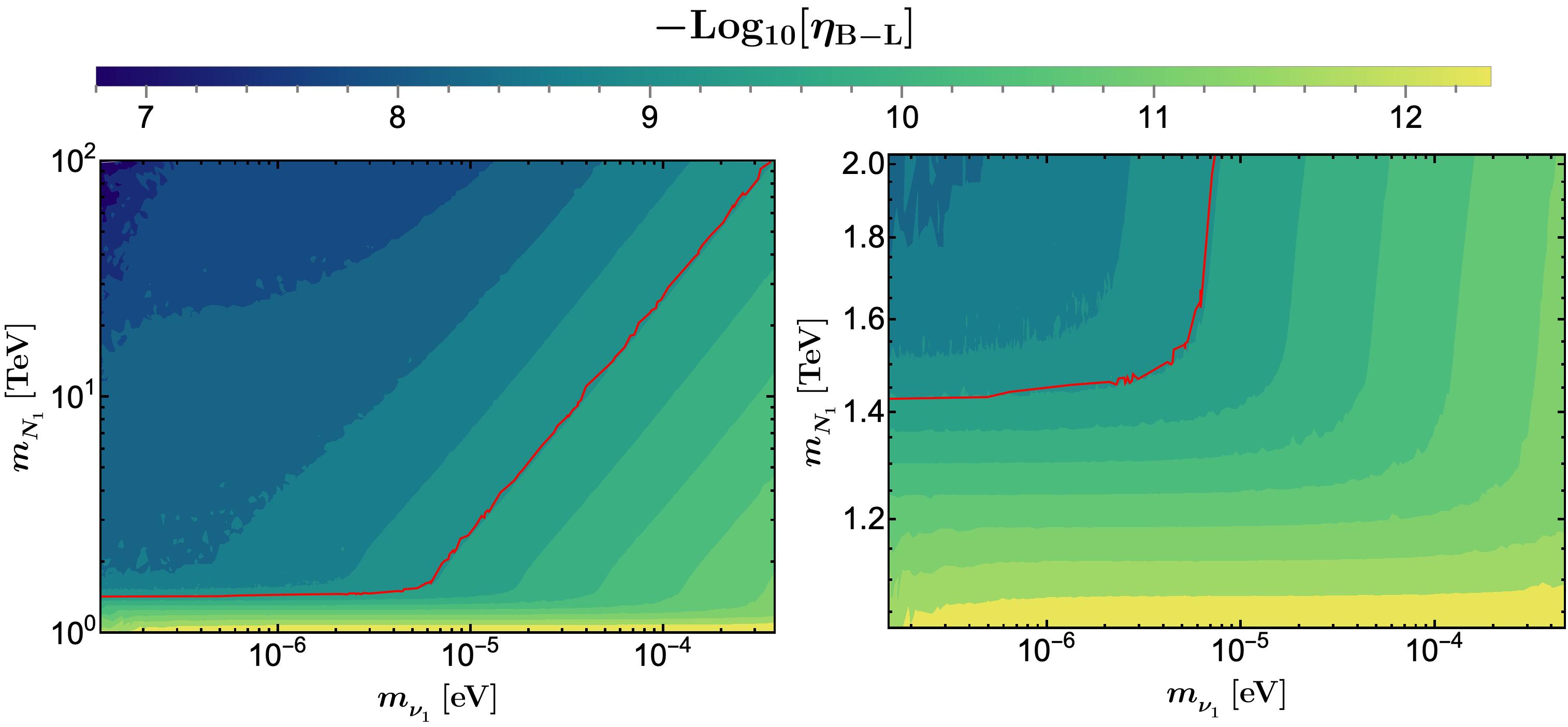}
\caption{In the above plots we depicts the maximized $\eta_{B}$ over the range of $\Delta m\in\{0.005,3\}$ in $m_{\nu_1^{}}$ vs $\mnone$ plane. We fixed the parameters at $\metaI=580$ GeV, $\lambda_{\eta H}=0.06$, $\delta m=0.2$ GeV. The red line represents the observed matter-antimatter asymmetry. The right-side plot is the magnified version of the left-side plot.}
 \label{fig:YBL}
 \end{figure}
The variation of $\eta_{B}$ in $m_{\nu_1^{}}-\mnone$ plane is shown in \fig\ref{fig:YBL} for a representative choice of model parameters, mentioned in the caption of the plot, with the observed value of the baryon asymmetry indicated by the red solid line. The $B-L$ asymmetry typically freezes in near $z\sim10$ , corresponding to a freeze-in temperature $T_{FI}\simeq\mnone/10$, whereas the sphaleron freeze-out temperature is $T_{sp}\sim131$ GeV~\cite{DOnofrio:2014rug}. This produced $B-L$ asymmetry must be partially converted into the baryon asymmetry before the sphaleron processes decouples from the thermal bath. Hence, we demand $T_{FI}\simeq\mnone/10\gtrsim T_{sp}$. This condition puts a lower bound on $\mnone$ quantified as $\mnone\gtrsim1.3$ TeV. Our numerical analysis provides a slightly stronger bound, $m_{N_1}^{f}\simeq1.4$ TeV (see \fig\ref{fig:YBL}). Here, $m_{\rm N_{1}^{}}^f$ is the lowest achievable smallest RHN mass that can successfully generate the observed $\eta_{B}^{}$ of the Universe. 
\begin{figure}[htb!]
 \centering
 \includegraphics[width=0.48\linewidth]{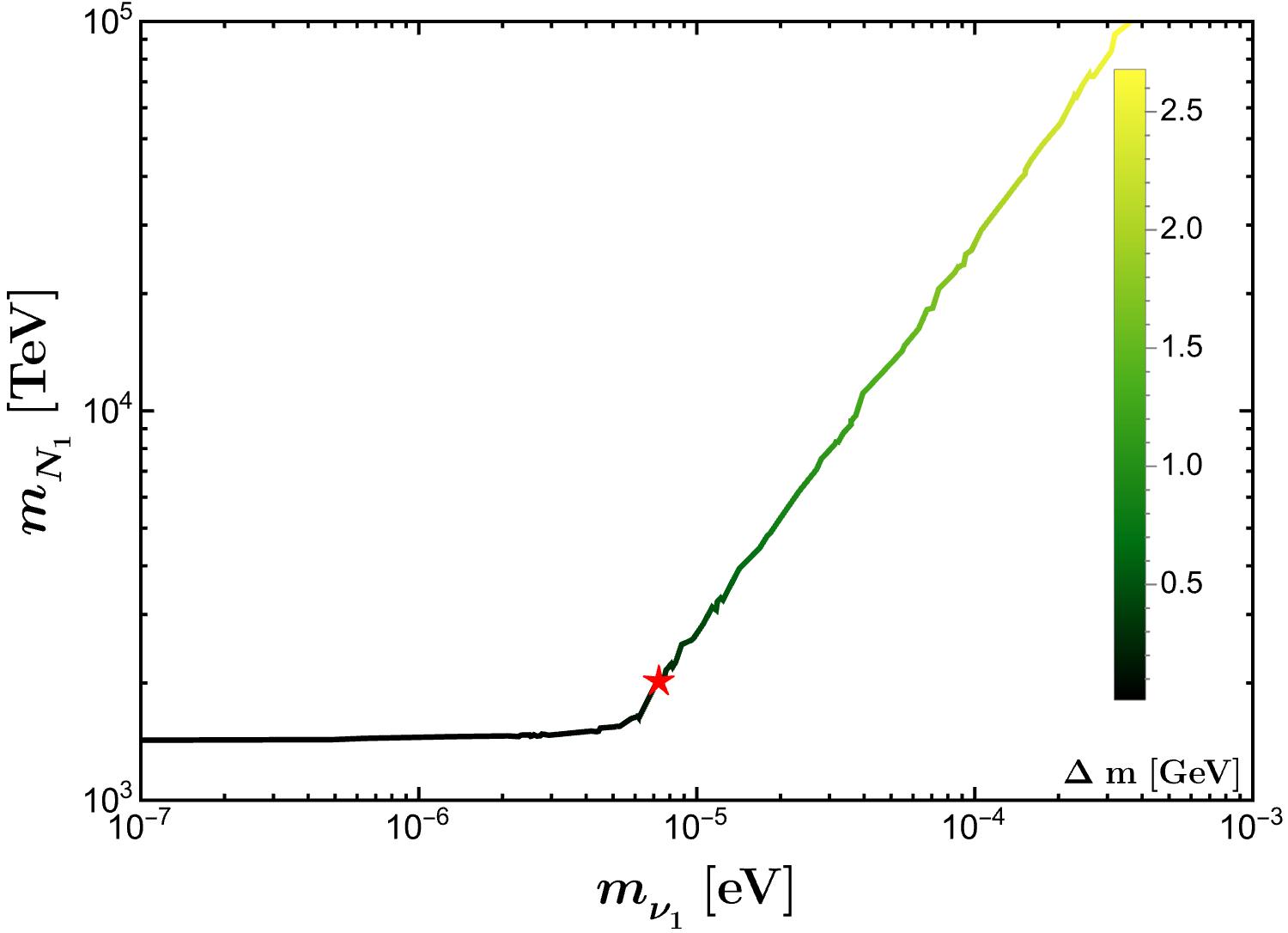} 
 \includegraphics[width=0.48\linewidth]{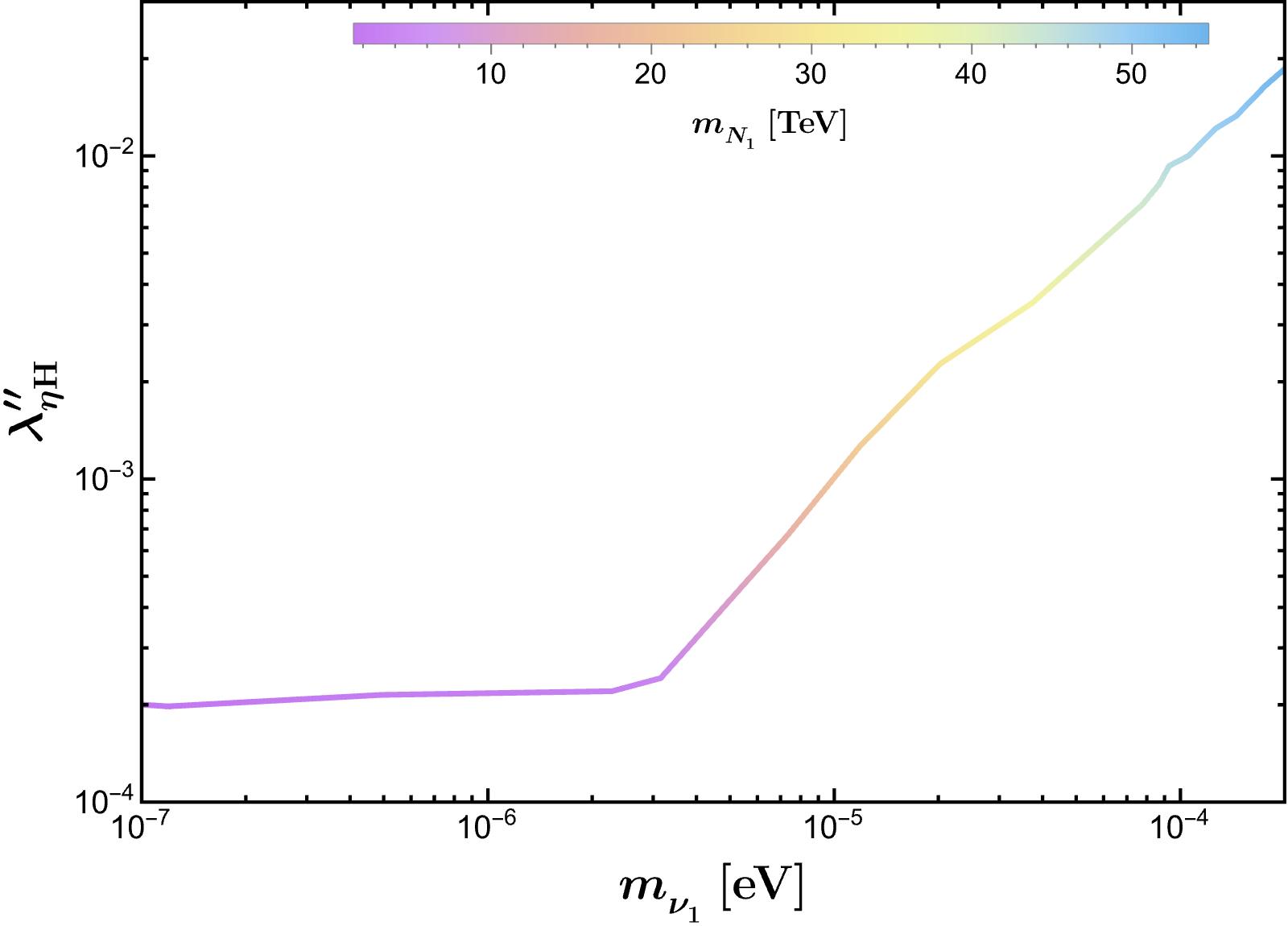}
\caption{
The left and right figures represent the baryon-asymmetry allowed regions in the $m_{\nu_1}-\mnone $ and $m_{\nu_1}-\lambda^{\prime\prime}_{\eta H}$ planes, respectively. The green–yellow color bar shows the variation of $\Delta m$, while the pastel color bar shows the variation of $\mnone$. The red star denotes the benchmark point that simultaneously reproduces the observed DM relic abundance and respects constraints from other DM detection experiments.}
 \label{fig:mnu1}
 \end{figure}
 
The successful reproduction of the observed baryon asymmetry establishes a characteristic correlation between $\mnone$ and the lightest active neutrino mass $m_{\nu_1^{}}$, as shown in the left panel of \fig\ref{fig:mnu1}. An increase in $m_{\nu_1^{}}$ enhances the Yukawa couplings $h_{\alpha 1}$ (see \eq\eqref{eq:Yuk1st}), thereby increasing $(h^\dagger h)_{11}$. At the same time, the colour gradient shows that a larger mass splitting $\Delta m$ is required to maintain the observed $\eta_B$, which in turn suppresses the couplings $h_{i\alpha}$ (to keep the active neutrino mass in correct ballpark, see \eq\eqref{eq:Numassap}). Since the CP asymmetry scales as $\varepsilon^{}_{N_1}\sim\sum_{1\neq j}^{}{\rm Im}(h^\dagger h)^2_{1j}/(h^\dagger h)_{11}$ with $(h^\dagger\,h)_{1j}$ approximately scaling as $(h^\dagger\,h)_{1j}\approx\left(\big(\mathtt{M}_{1j}\,\sqrt{m^{}_{N^{}_{1}}~m^{}_{N^{}_{j}}}\big)/(\Delta m\,\metaI)\right)$ (see \eq\eqref{eq:hdh1j}), the increase in $m_{\nu_1^{}}$, and the accompanying increase in $\Delta m$, leads to a suppression of $\varepsilon^{}_{N_1}$. A larger $\mnone$ is therefore needed to compensate for this suppression and reproduce the observed baryon asymmetry. The right panel of \fig\ref{fig:mnu1} shows the corresponding dependence of $\lambda_{\eta H}^{\prime\prime}$ on $m_{\nu_1^{}}$ for $\eta_B\simeq6.12\times10^{-10}$ \cite{Planck:2018vyg}, with the colour scale indicating the associated values of $\mnone$. The change in the slope around $m_{\nu_1^{}}\simeq10^{-5}$ eV reflects a change in the dominant washout dynamics, a behaviour also noted in Ref.~\cite{Hugle:2018qbw}. As $\lambda_{\eta H}^{\prime\prime}$ decreases, $\Delta L=2$ processes, such as $LL\to\eta^\ast\eta^\ast$ and $L\eta\to\bar{L}\eta^\ast$ , become increasingly dominant (see \eq\eqref{eq:L2}). For $\lambda_{\eta H}^{\prime\prime}\lesssim\mathcal{O}(10^{-4})$, the resulting washout becomes sufficiently strong that the observed baryon asymmetry can no longer be reproduced. Consequently, both the effect $\Delta L=2$ wash-out and sphaleron decoupling will provide the absolute lower bound on the smallest RHN mass. Also note here that although our leptogenesis analysis is carried out for $\metaI=580$ \GeV, the parameter space consistent with the observed baryon asymmetry remains largely unchanged as long as $\eta_{1}\ll1$ (see eq.~\eqref{eq:cp asy} and \eqref{eq:N1dw}).

To sum it up, a suitable choice of the Casas-Ibarra parameters allows sizable second- and third-generation Yukawa couplings enhancing the CP asymmetry, while suppressed first-generation Yukawa coupling helps avoiding excessive washout. Successful leptogenesis in VSM can consequently be achieved at scales as low as $\mathcal{O}(1)$ TeV, nearly an order of magnitude below the values typically found in the literature~\cite{Hugle:2018qbw} when spectator effects are neglected. Consistency with neutrino masses, LFV, DM phenomenology, and the observed baryon asymmetry via leptogenesis allows VSM model in such new incarnation to be probed at the future muon collider.

\section{Future muon colliders}
\label{sec:collider}
Future muon colliders provide a uniquely clean and controlled environment for probing models especially concerning dark matter or presence of lepton portals, offering several advantages over the LHC. Unlike $pp$ collisions at the LHC, muon collisions are free from overwhelming QCD backgrounds and parton distribution uncertainties, enabling precise measurements with reduced systematic effects. This absence of hadronic contamination significantly enhances the sensitivity to EW-scale NP, particularly scenarios where dark sector has coupling predominantly with the gauge bosons or leptons. Moreover, the clean final-state environment allows for detailed track-level analyses with excellent momentum resolution and minimal pile-up. Combined with next-generation detector technologies optimized for precision timing, vertexing, and tracking, future muon colliders can efficiently identify soft, displaced, or invisible signatures that are challenging to isolate at hadron machines. These features render muon colliders exceptionally powerful for probing compressed spectra, which is precisely the regime our analysis targets. In our analysis, we focus on charged scalar pair production mediated via both SM gauge and heavy neutrino. The presence of the heavy neutrino portal introduces interference effects which are absent in vanilla inert doublet model.

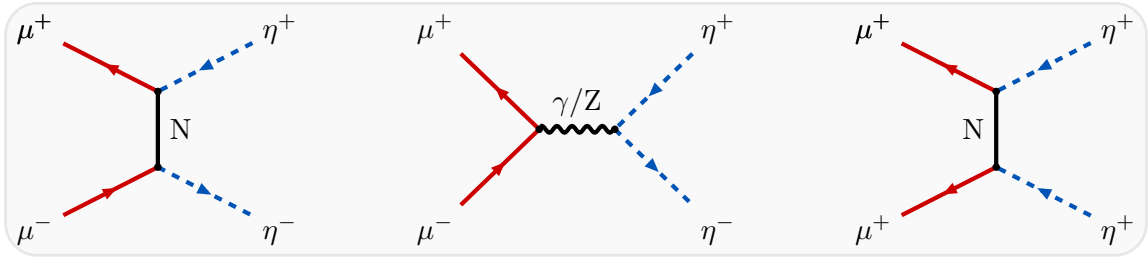
\begin{figure}[htb!]
\centering
\begin{adjustbox}{width=\textwidth}
\begin{tcolorbox}[colback=gray!5, colframe=black!10, boxrule=1pt, arc=4mm, boxsep=0pt, left=0pt, right=0pt, top=0pt, bottom=0pt, width=1\textwidth, halign=center]
\begin{tikzpicture}[baseline={(current bounding box.center)},style={scale=1.0, transform shape}]
\begin{feynman}
\vertex (a);
\vertex [above left=0.5 cm and 1.25cm of a] (a1){\(\rm \color{black}{\mu^+}\)};
\vertex [above right=0.5 cm and 1.25cm of a] (a2){\(\rm \color{black}{\eta^+}\)};
\vertex [below =1.0cm of a] (b);
\vertex [below left=0.5 cm and 1.25cm of b] (b1){\(\rm \color{black}{\mu^-}\)};
\vertex [below right=0.5 cm and 1.25cm of b] (b2){\(\rm \color{black}{\mathbb{\eta^-}}\)};
\diagram*{
(a2) -- [line width=0.25mm, charged scalar, arrow size=1.2pt, style=mediumtealblue,ultra thick] (a),
(a) -- [line width=0.25mm, fermion, arrow size=1.2pt, style=bostonuniversityred,ultra thick] (a1),
(b) -- [line width=0.25mm, charged scalar, arrow size=1.2pt, style=mediumtealblue,ultra thick] (b2),
(b1) -- [line width=0.25mm, fermion, arrow size=1.2pt, style=bostonuniversityred,ultra thick] (b),
(b) -- [line width=0.25mm, plain, arrow size=1.2pt, style=black,ultra thick, edge label'={\(\rm\color{black}{N}\)}] (a)};
\vertex [above left=0.5 cm and 1.25cm of a] (a1){\(\rm \color{black}{\mu^+}\)};
\node at (a)[circle,fill,style=black,inner sep=1pt]{};
\node at (b)[circle,fill,style=black,inner sep=1pt]{};
\end{feynman}
\end{tikzpicture}\hfill
\begin{tikzpicture}[baseline={(current bounding box.center)},style={scale=1.0, transform shape}]
\begin{feynman}
\vertex (a);
\vertex [above left=1.0 cm and 1.0cm of a] (a1){\(\rm \color{black}{\mu^+}\)};
\vertex [below left=1.0 cm and 1.0cm of a] (a2){\(\rm \color{black}{\mu^-}\)};
\vertex [right =1.0cm of a] (b);
\vertex [above right=1.0 cm and 1.0cm of b] (b1){\(\rm \color{black}{\eta^+}\)};
\vertex [below right=1.0 cm and 1.0cm of b] (b2){\(\rm \color{black}{\eta^-}\)};
\diagram*{
(a2) -- [line width=0.25mm, fermion, arrow size=1.2pt, style=bostonuniversityred,ultra thick] (a),
(a) -- [line width=0.25mm, fermion, arrow size=1.2pt, style=bostonuniversityred,edge label={\(\rm\color{black}{}\)}, ultra thick] (a1),
(b1) -- [line width=0.25mm, charged scalar, arrow size=1.2pt, style=mediumtealblue,ultra thick] (b),
(b) -- [line width=0.25mm, charged scalar, arrow size=1.2pt, style=mediumtealblue,ultra thick] (b2),
(b) -- [line width=0.25mm, boson, arrow size=1.2pt, style=black,ultra thick, edge label'={\(\rm\color{black}{\gamma/Z}\)}] (a)};
\node at (a)[circle,fill,style=black,inner sep=1pt]{};
\node at (b)[circle,fill,style=black,inner sep=1pt]{};
\end{feynman}
\end{tikzpicture}\hfill
\begin{tikzpicture}[baseline={(current bounding box.center)},style={scale=1.0, transform shape}]
\begin{feynman}
\vertex(a);
\vertex[above left=0.5 cm and 1.25cm of a] (a1){\(\rm \color{black}{\mu^+}\)};
\vertex[above right=0.5 cm and 1.25cm of a] (a2){\(\rm \color{black}{\eta^+}\)};
\vertex[below =1.0cm of a] (b);
\vertex[below left=0.5 cm and 1.25cm of b] (b1){\(\rm \color{black}{\mu^+}\)};
\vertex[below right=0.5 cm and 1.25cm of b] (b2){\(\rm \color{black}{\mathbb{\eta^+}}\)};
\diagram*{
(a) -- [line width=0.25mm, fermion, arrow size=1.2pt, style=bostonuniversityred,ultra thick] (a1),
(a2) -- [line width=0.25mm, charged scalar, arrow size=1.2pt, style=mediumtealblue,ultra thick] (a),
(b2) -- [line width=0.25mm, charged scalar, arrow size=1.2pt, style=mediumtealblue,ultra thick] (b),
(b) -- [line width=0.25mm, fermion, arrow size=1.2pt, style=bostonuniversityred,ultra thick] (b1),
(a) -- [line width=0.25mm, plain, arrow size=1.2pt, style=black,ultra thick, edge label'={\(\rm\color{black}{N}\)}] (b)};
\vertex [above left=0.5 cm and 1.25cm of a] (a1){\(\rm \color{black}{\mu^+}\)};
\node at (a)[circle,fill,style=black,inner sep=1pt]{};
\node at (b)[circle,fill,style=black,inner sep=1pt]{};
\end{feynman}
\end{tikzpicture}
\end{tcolorbox}
\end{adjustbox}
\caption{The Feynman diagrams relevant for charged scalar pair production at $\mu^+\mu^-$ (\textit{first} and \textit{second} diagrams) and $\mu^+\mu^+$ (\textit{third diagram}) colliders.}
\label{feyn:Ntau-N}
\end{figure}

Our study concerns four projected collider setups: Two opposite-sign muon colliders i.e. $\mu^{+}\mu^{-}$ colliders with $\sqrt{s} = 3$ TeV~\cite{MuonCollider:2022xlm} and $10$ TeV~\cite{Andreetto:2025mrd}, and two same-sign muon colliders i.e. $\mu^{+}\mu^{+}$ colliders with $\sqrt{s} = 6$ TeV~\cite{Hamada:2022mua} and $10$ TeV~\cite{Kitano:2025xaj}. The Feynman diagrams highlighting the Born production processes at future muon colliders are shown in \fig\ref{feyn:Ntau-N}. Although a same-sign muon collider with $\sqrt{s}=2$ TeV~\cite{Hamada:2022mua} has been proposed, pair production of $\eta^{+}$ is kinematically inaccessible for $m_{\eta^{+}} > 1$ TeV in such a setup.

\begin{figure}[htb!]
    \centering
    \includegraphics[width=0.4\linewidth]{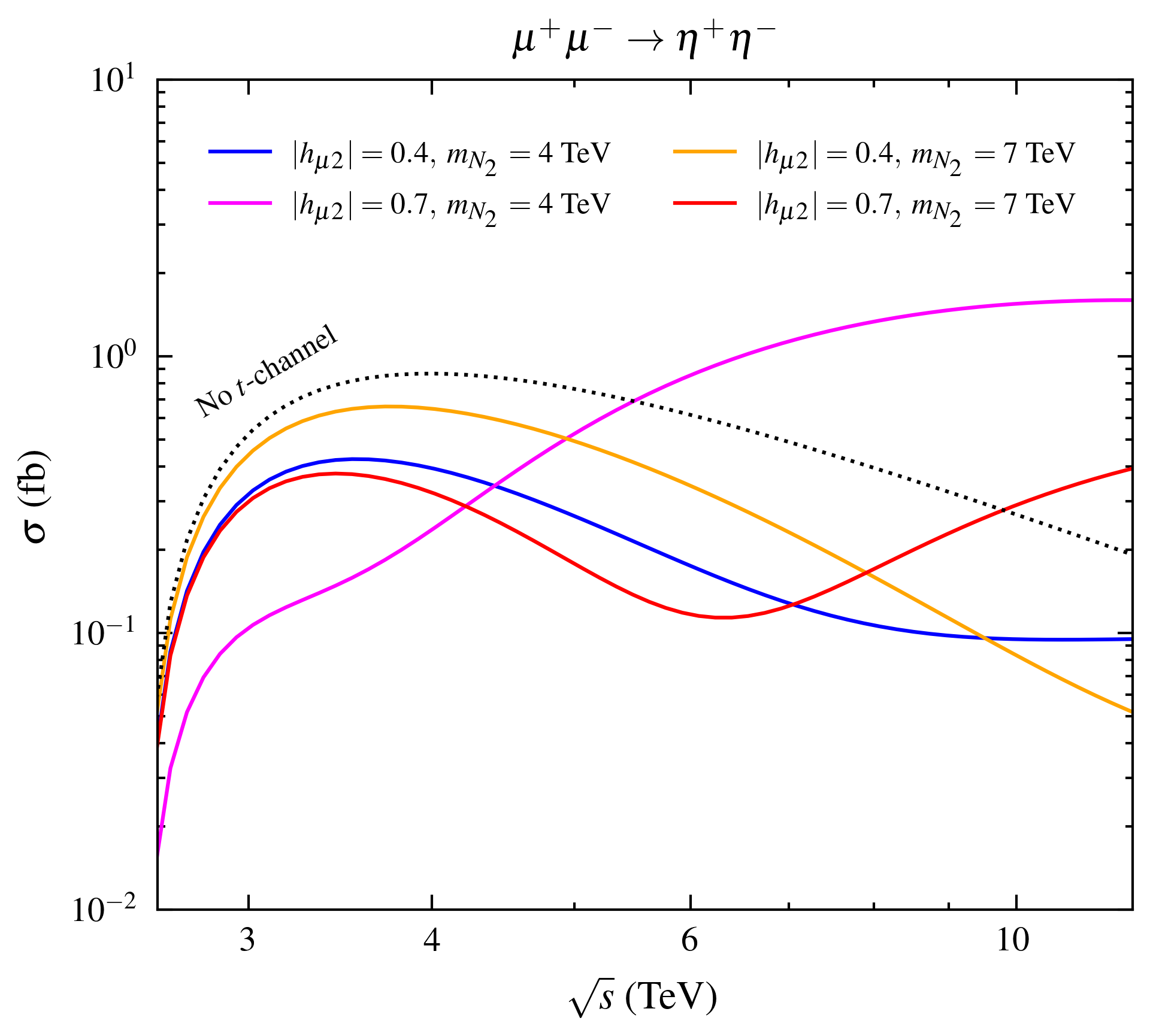}
    \includegraphics[width=0.4\linewidth]{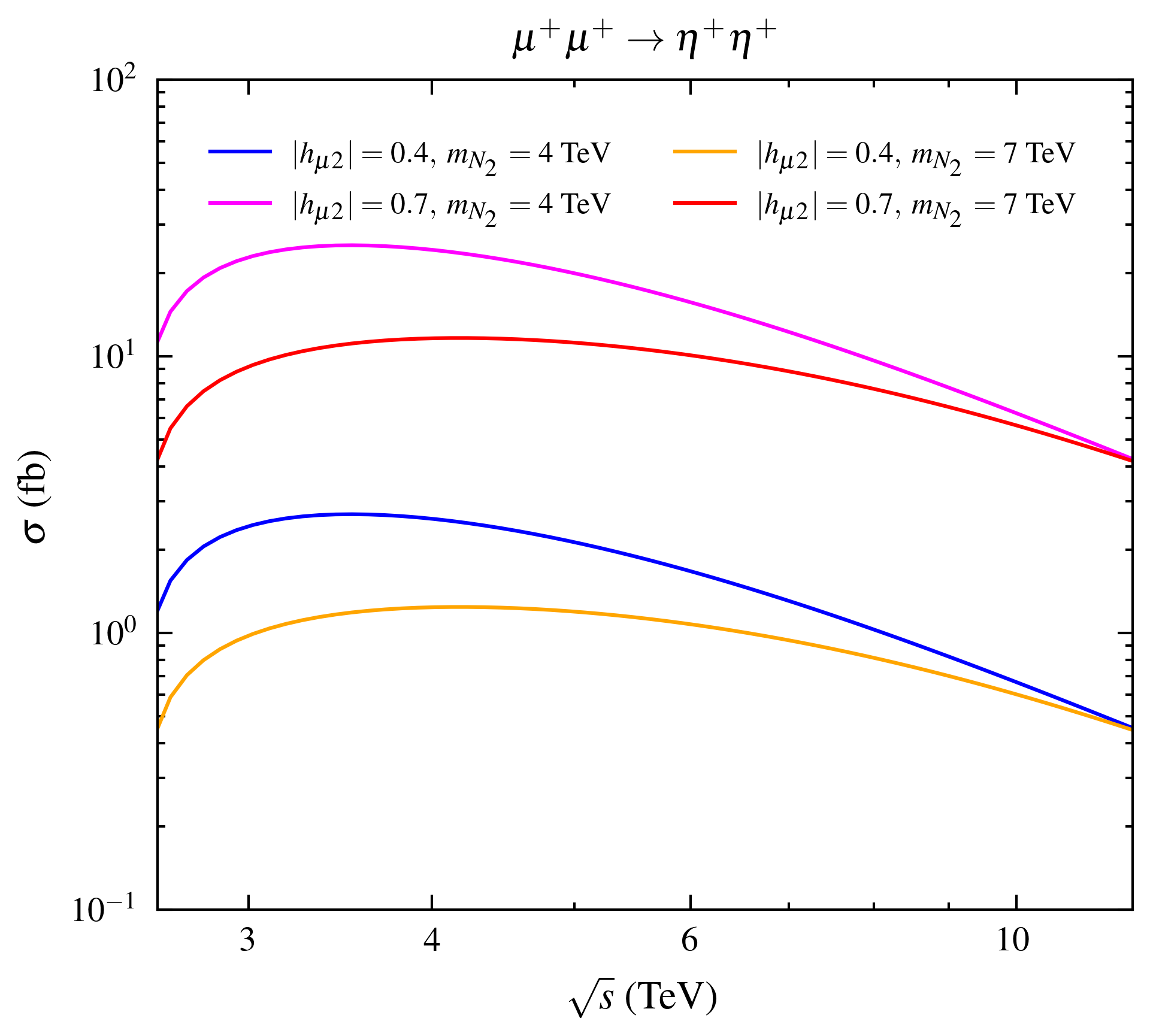}
    \includegraphics[width=0.4\linewidth]{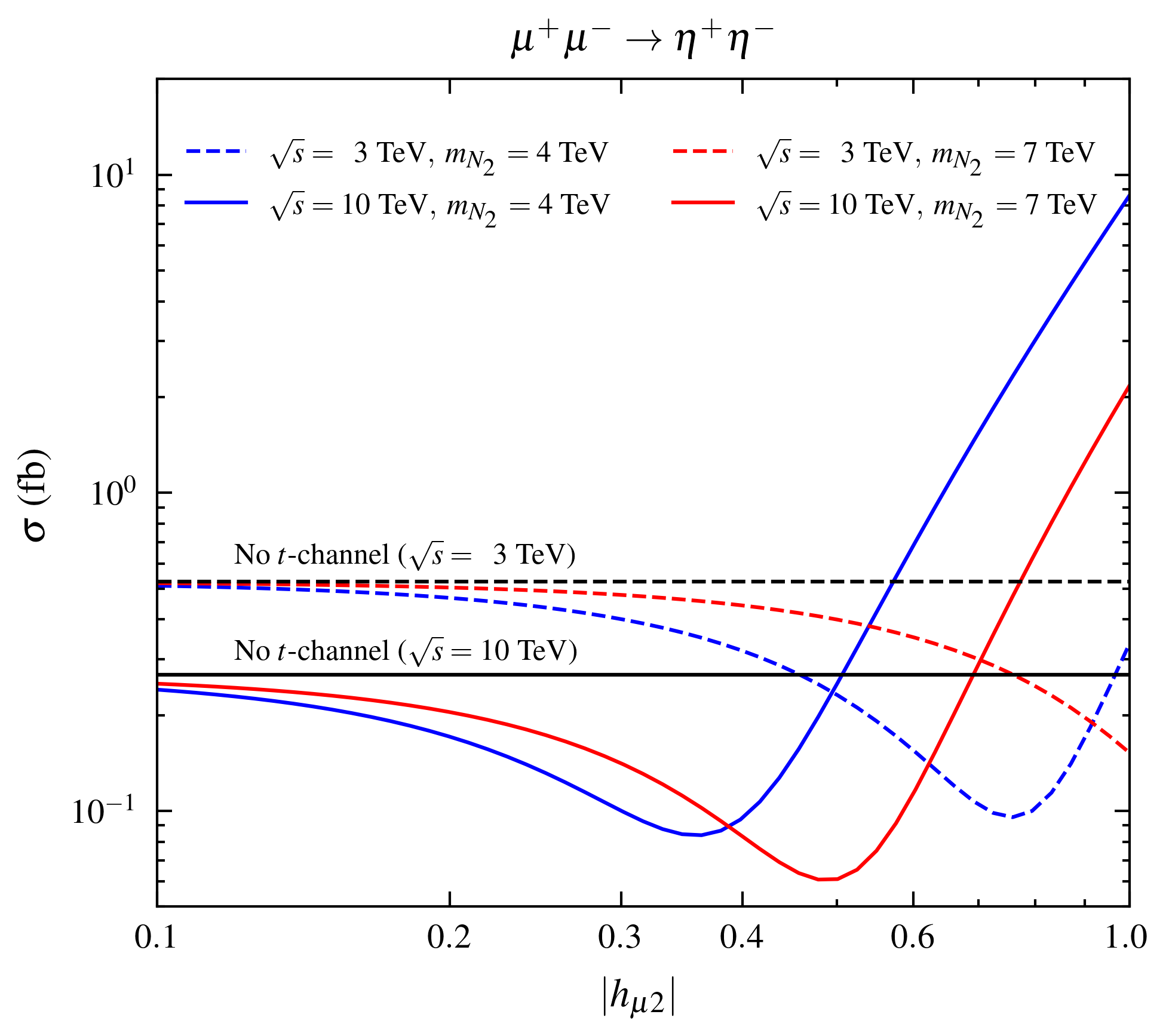}
    \includegraphics[width=0.4\linewidth]{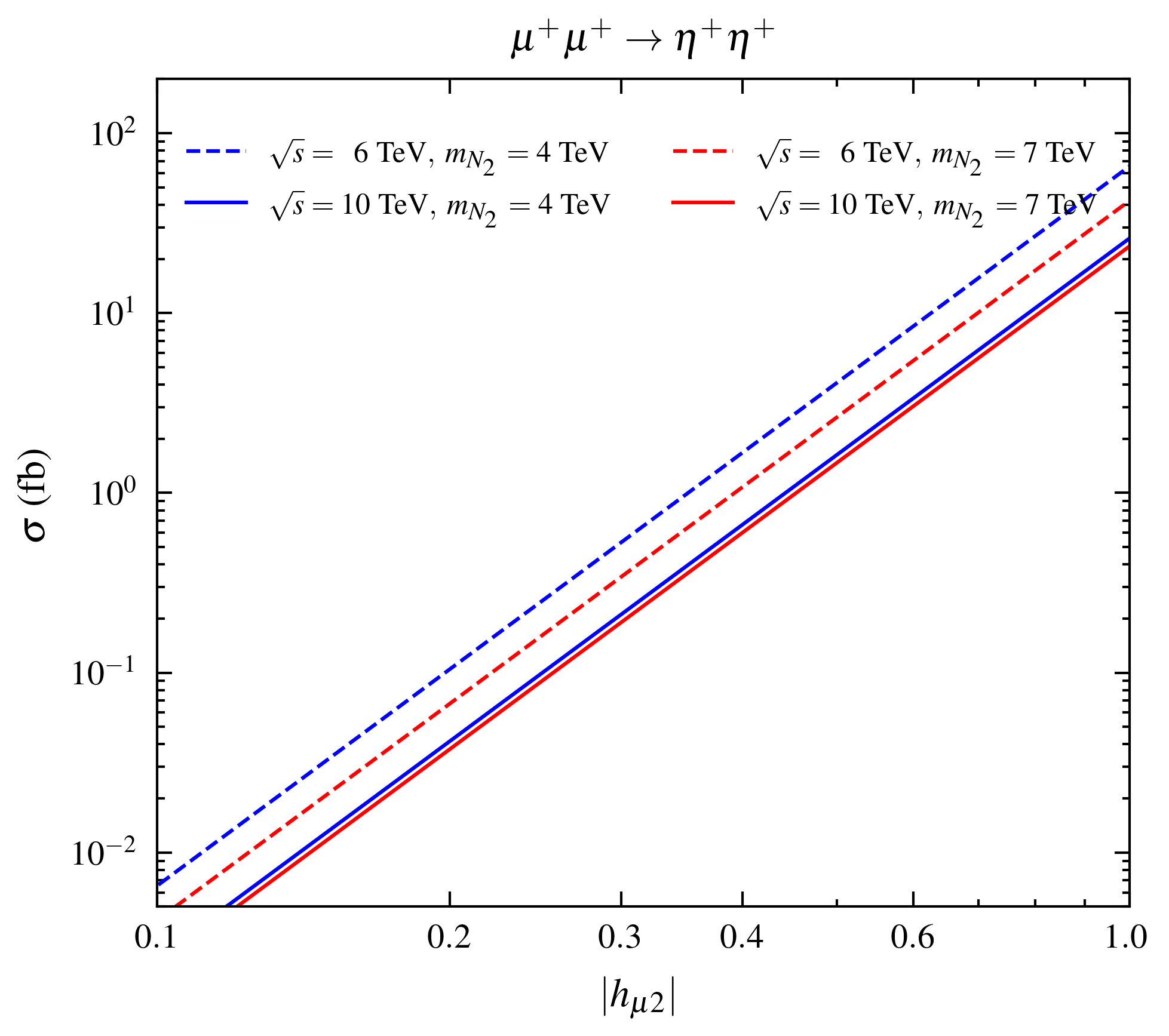}
    \includegraphics[width=0.4\linewidth]{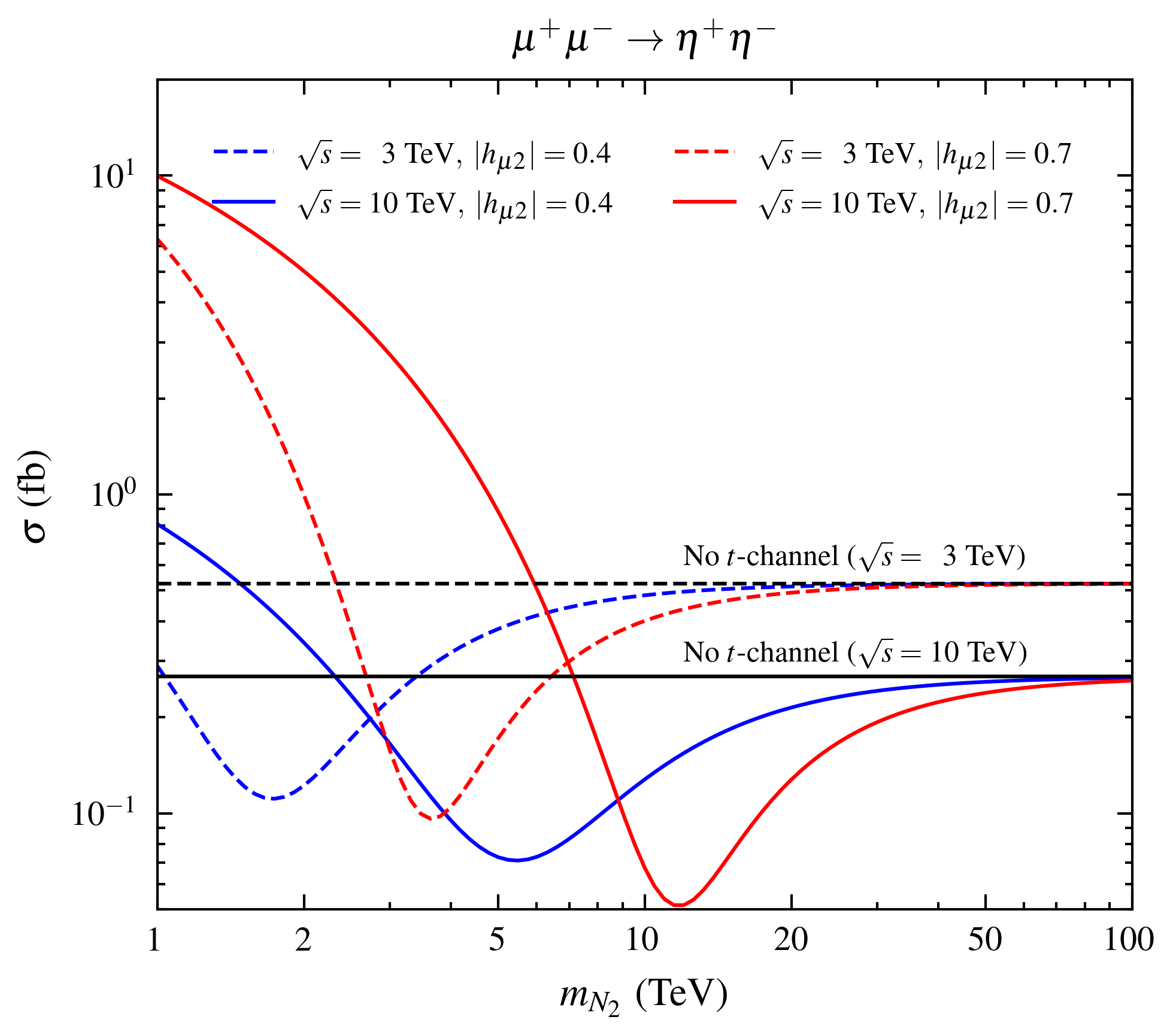}
    \includegraphics[width=0.4\linewidth]{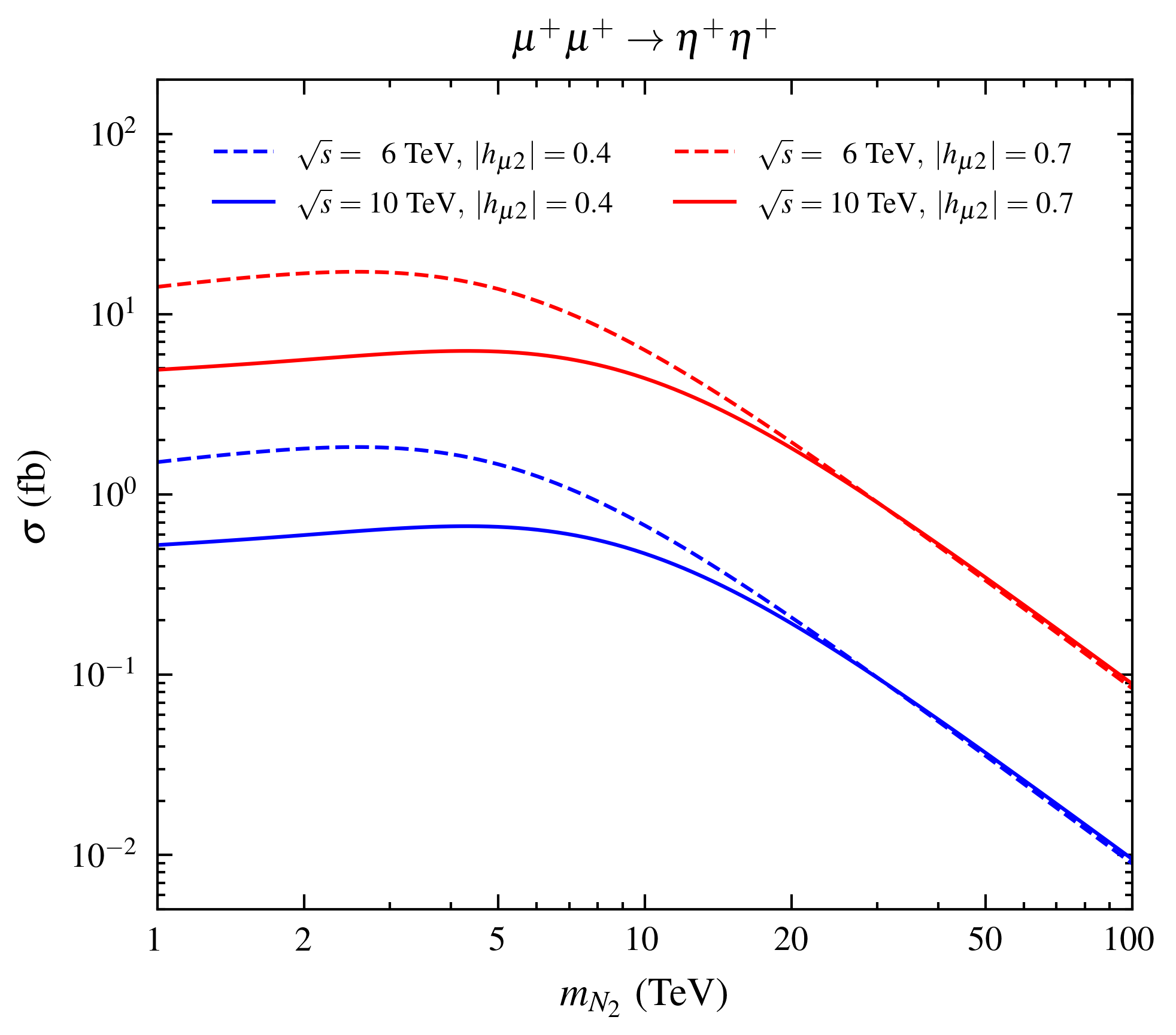}
    \caption{Cross section of $\mu^{+}\mu^{-} \to \eta^{+}\eta^{-}$ ($\mu^{+}\mu^{+} \to \eta^{+}\eta^{+}$) production at future opposite-sign (same-sign) muon colliders shown in \textit{left panel} (\textit{right panel}). Here, $m_{\eta^{+}}=1.26$ TeV. The different panels show variation of cross section $\sigma$ (fb) with $\sqrt{s}$ (TeV) (\textit{top panel}), $|h_{\mu 2}|$ (\textit{middle panel}) and $m_{N_{2}}$ (TeV) (\textit{bottom panel}).}
    \label{fig:mc}
\end{figure}

In \fig\ref{fig:mc}, we plot the production cross section of $\mu^{+}\mu^{-} \to \eta^{+}\eta^{-}$ ($\mu^{+}\mu^{+} \to \eta^{+}\eta^{+}$) production at future opposite-sign (same-sign) muon colliders. In case of opposite-sign muon colliders, the $s$- and $t$-channel diagrams interfere negatively, leading to a dip in the cross section, which doesn't occur in absence of the $t$-channel graph via RHN mediation. In case of same-sign colliders, the cross-section doen't drop, as only the $t$-channel process is present. In the top panel, we observe the variation of cross section ($\mu^{+}\mu^{-} \to \eta^{+}\eta^{-}$) with CM energy, $\sqrt{s}$. It is observed that the position of the dip shifts depending on the mass of the RHN, $m_{N_{2}}$ and the characteristic Yukawa coupling, $|h_{\mu 2}|$. As $m_{N_{2}}$ increases, the dip shifts to higher $\sqrt{s}$. Similarly as the Yukawa coupling increases, the dip shifts to lower $\sqrt{s}$. In the middle and bottom panels, we observe that with decrease of $|h_{\mu 2}|$ or increase of $m_{N_{2}}$, the t-channel process becomes subdominant, converging to the vanilla inert doublet model scenario. In case of same-sign muon colliders, the process $\mu^{+}\mu^{+} \to \eta^{+}\eta^{+}$ occurs in presence of Majorana RHN; hence observation of such process will act as signature of the Scotogenic model.

\begin{table}[htb!]
\centering
{\small
\begin{tabular}{c c c c c c c c}
\hline \hline 
\textbf{BPs} & $m_{\eta_{I}^{0}}\;(\text{TeV})$ & $\mntwo (\text{TeV})$  & $\Delta m$ (\text{GeV})&  $\delta m$ (\text{GeV})& $(|h_{2e}|,|h_{2\mu}|,|h_{2\tau}|)$ & $\lambda_{\eta H}^{\prime\prime}$ & $\lambda_{\eta H}$\\
\hline
\hline
\textbf{BP1} & 1.26 & $6.4$ & $0.03$ & $0.15$ & $(0.08,0.33,0.34)$ & $0.001$ & $0.075$\\
\textbf{BP2} & $1.26$ & $6.4$ & $0.03$ & $0.20$ & $(0.08,0.33,0.34)$ & $0.001$ & $0.060$\\
\textbf{BP3} & $1.26$ & $6.4$  & $0.03$ & $0.25$ & $(0.08,0.33,0.34)$ & $0.001$ & $0.045$\\
 \hline \hline
\end{tabular}}
\caption{The benchmark points \textbf{BP1-BP3} of the VSM model consistent with LFV, DM, Neutrino masses, and leptogenesis are used for collider analysis. The parameters which distinguish the benchmark points are, $\delta m$ and $\lambda_{\eta H}$. The other parameter kept fixed is $\mnone=\mntwo/\sqrt{10}$, and $\mnthree=\sqrt{10}\,\mntwo$.}
\label{tab:tcol}
\end{table}
For collider studies we consider three benchmark points listed in \cref{tab:tcol}, which address the dark matter relic density, neutrino masses, LFV, and the baryon asymmetry of the Universe. 

\subsection{Prompt search at muon colliders}
\label{sec:ppt}
\begin{figure}[htb!]
    \centering
    \includegraphics[width=0.325\linewidth]{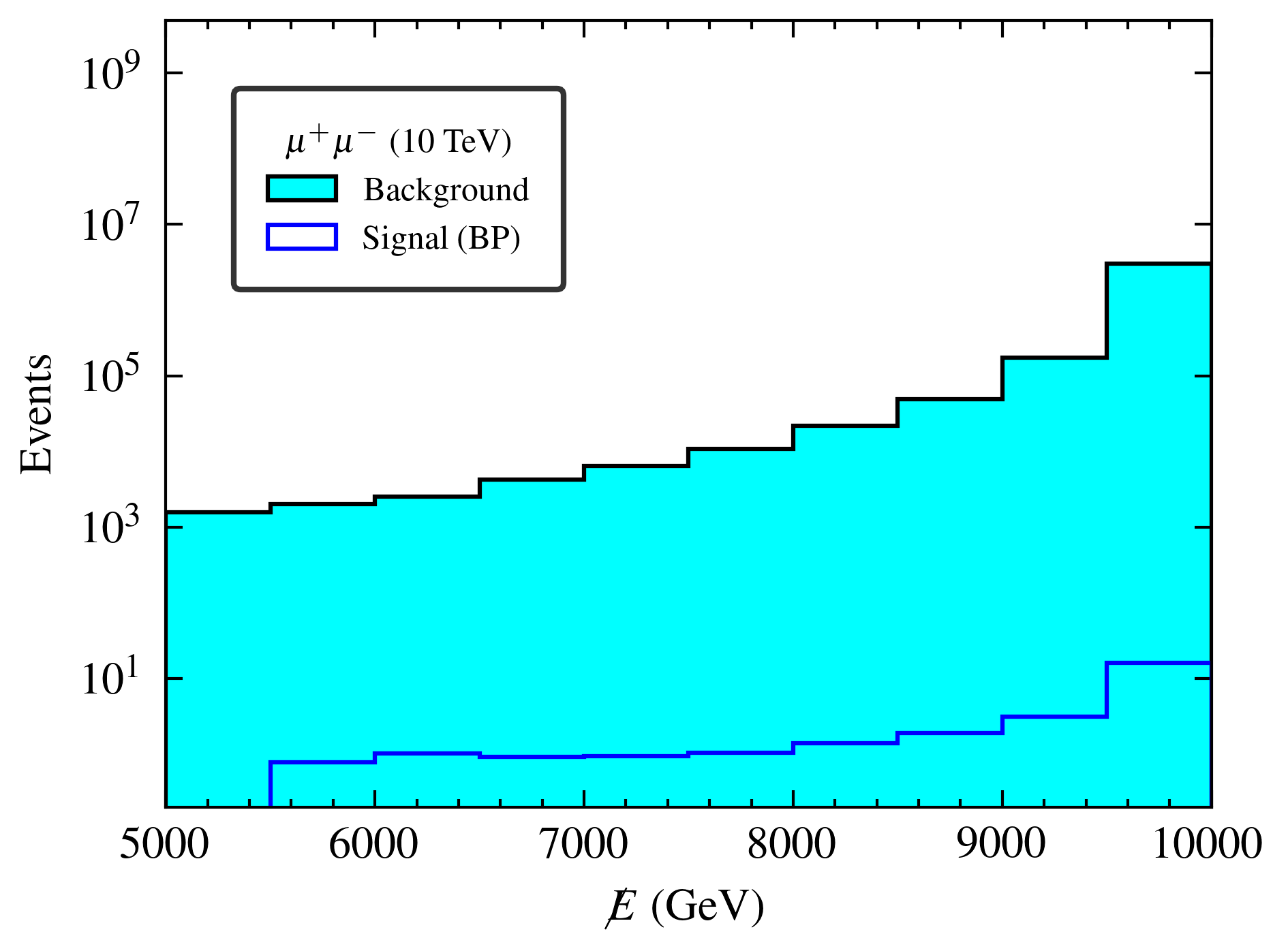}
    \includegraphics[width=0.325\linewidth]{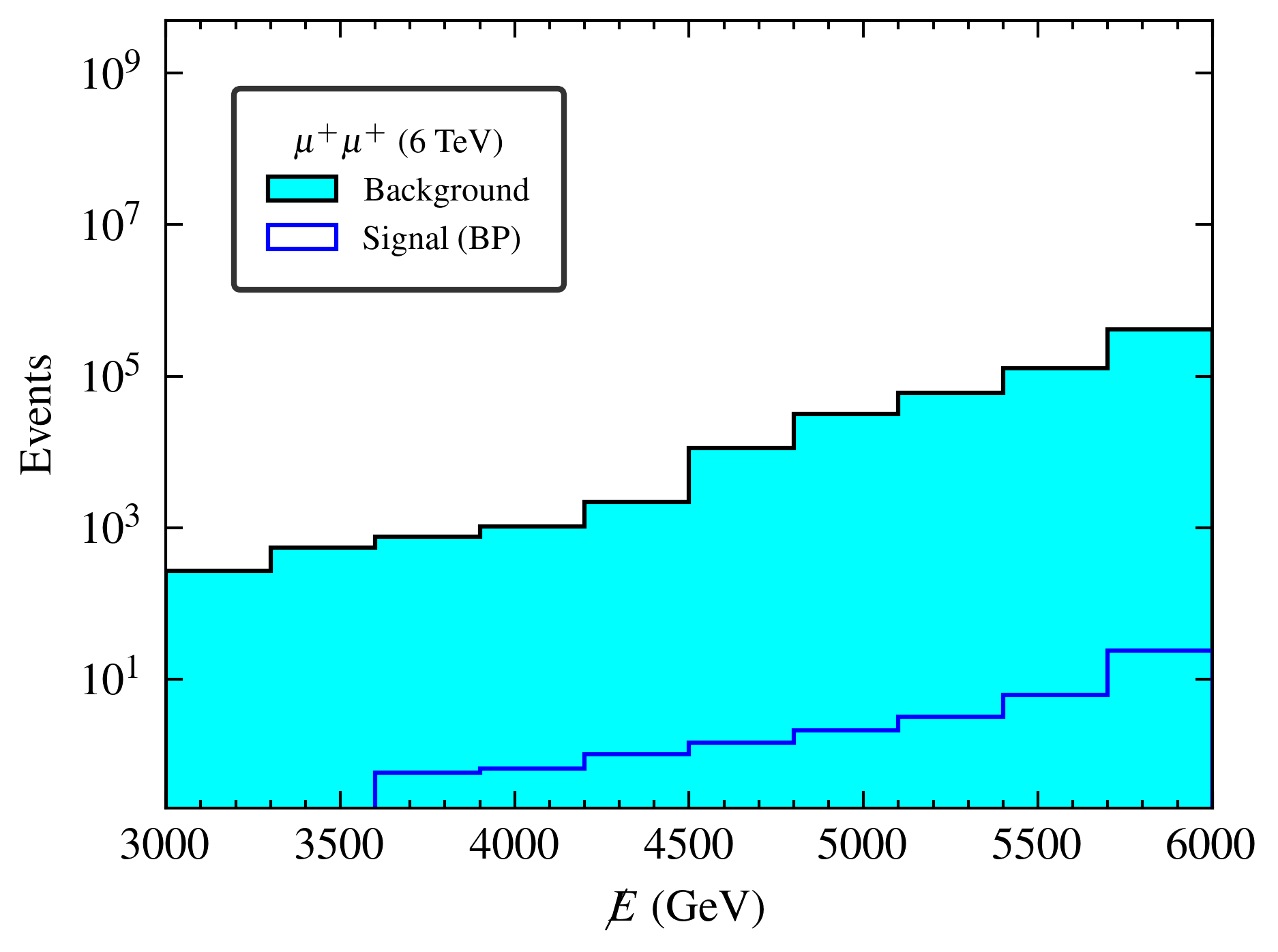}
    \includegraphics[width=0.325\linewidth]{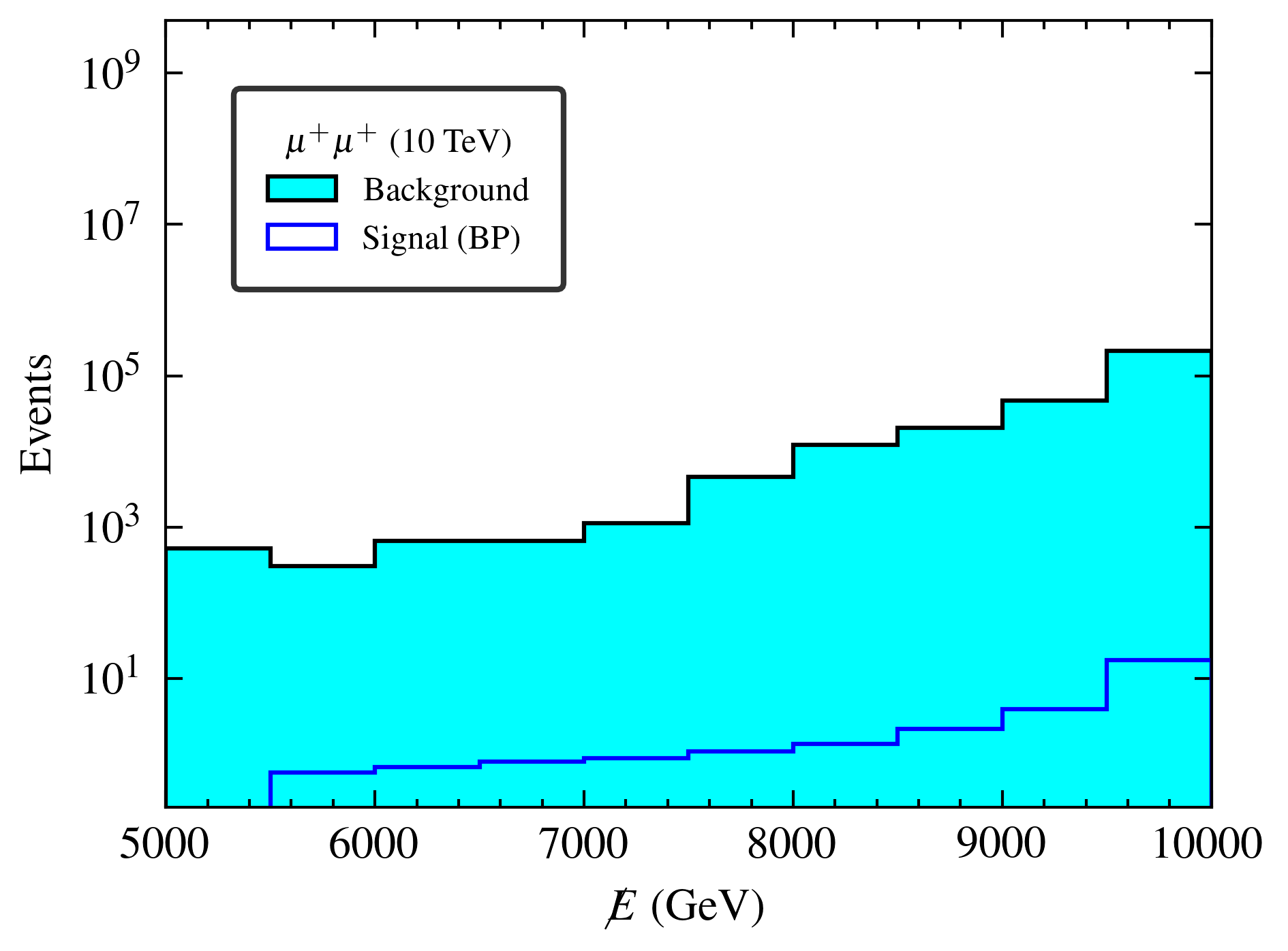}
    \caption{Missing energy distribution for mono-$\gamma$ signal for $\mu^{+}\mu^{\pm} \to \eta\,\eta\,\gamma$ production and dominant SM background at the future muon colliders. The CM energy for each cases have been mentioned in the figure inset.}
    \label{fig:mcx}
\end{figure}

In this section, we explore the prospects for \textbf{prompt searches} of heavy scalars. The Born-level processes for charged scalar production, shown in \fig\ref{feyn:Ntau-N}, involve particles that are long-lived and do not decay promptly. Moreover, their decays yield a compressed spectrum, making the visible decay products too soft to be efficiently detected by the calorimeters, effectively rendering them invisible, similar to neutrinos. Therefore, direct detection of these scalars at the prompt level is challenging. To overcome this, one must rely on the emission of an additional SM particle, whose recoil and the resulting missing energy can be used to infer the presence of the invisible heavy scalars. We investigate this possibility in the context of $\mu^{+}\mu^{-}$ ($\sqrt{s}=10$~TeV) and $\mu^{+}\mu^{+}$ ($\sqrt{s}=6$ and $10$~TeV) colliders in the following discussion. For the $\mu^{+}\mu^{-}$ collider with $\sqrt{s}=3$~TeV, however, the Scotogenic model parameters, the Yukawa coupling $h_{\mu 2}$ and the neutrino mass $m_{N_{2}}$ have negligible impact on the benchmark points, as can be seen from \fig\ref{fig:mc}.
\paragraph{Prompt signal ($\boldsymbol{\mu^{+}\mu^{-} \to \eta\, \eta\,\gamma}$):} Production processes contributing to the signal mono-$\gamma$ + missing energy are $\eta^{+}\eta^{-}\gamma$ and $\eta_{R}^{0}\,\eta_{I}^{0}\,\gamma$. The dominant SM background is $\nu \overline{\nu} \gamma$. For collider simulation, the model is implemented using \texttt{FeynRules}~\cite{Alloul:2013bka}. The generated model files are then interfaced with \texttt{MG5\_aMC}~\cite{Alwall:2011uj} for event generation. No generation-level cuts are imposed. The events are subsequently passed to \texttt{Pythia8}~\cite{Bierlich:2022pfr} for parton showering and hadronization, and to \texttt{Delphes3}~\cite{deFavereau:2013fsa} for detector simulation, where the default \texttt{Delphes3} card is used to model detector resolution, efficiencies, and smearing effects. We select events containing exactly one photon ($N_{\gamma} = 1$), with no additional identified electrons, muons, or jets ($N_{e} = N_{\mu} = N_{j} = 0$). Furthermore, to ensure the presence of invisible states carrying substantial energy, we impose the following missing energy selection cut:
\begin{equation}
    \slashed{E} > \dfrac{\sqrt{s}}{2}\,,\quad \text{ where,}\quad \slashed{E} = \sqrt{s} - \sum_{\rm visible} E\,.
\end{equation}
The missing energy distributions for the dominant SM background and the signal benchmarks are shown in the left panel of \fig\ref{fig:mcx}.

\paragraph{Prompt signal ($\boldsymbol{\mu^{+}\mu^{+} \to \eta^{+}\eta^{+}\gamma}$):} The dominant SM background is $\mu^{+} \mu^{+}\gamma$, where the muons are not identified. We follow the same collider simulation framework as in the $\mu^{+}\mu^{-}$ case, employing identical signal selection criteria and analysis cuts. The missing energy distributions for the dominant SM background and the signal benchmarks are shown in middle and right panel of \fig\ref{fig:mcx}.
We employ a likelihood-based framework to constrain the model parameters. In this scenario, the relevant parameters of the Scotogenic model are the Yukawa coupling $h_{\mu 2}$ and the neutrino mass $m_{N_{2}}$. The remaining parameters, namely the DM mass $m_{\eta^{0}_{I}}$ and the charged-neutral mass splitting $\delta m$, are fixed to $1.26~\text{TeV}$ and $0.25~\text{GeV}$, respectively, based on DM phenomenology discussed in Sec.~\ref{sec:darkmatter}.

\begin{figure}[htb!]
    \centering
    \includegraphics[width=0.475\linewidth]{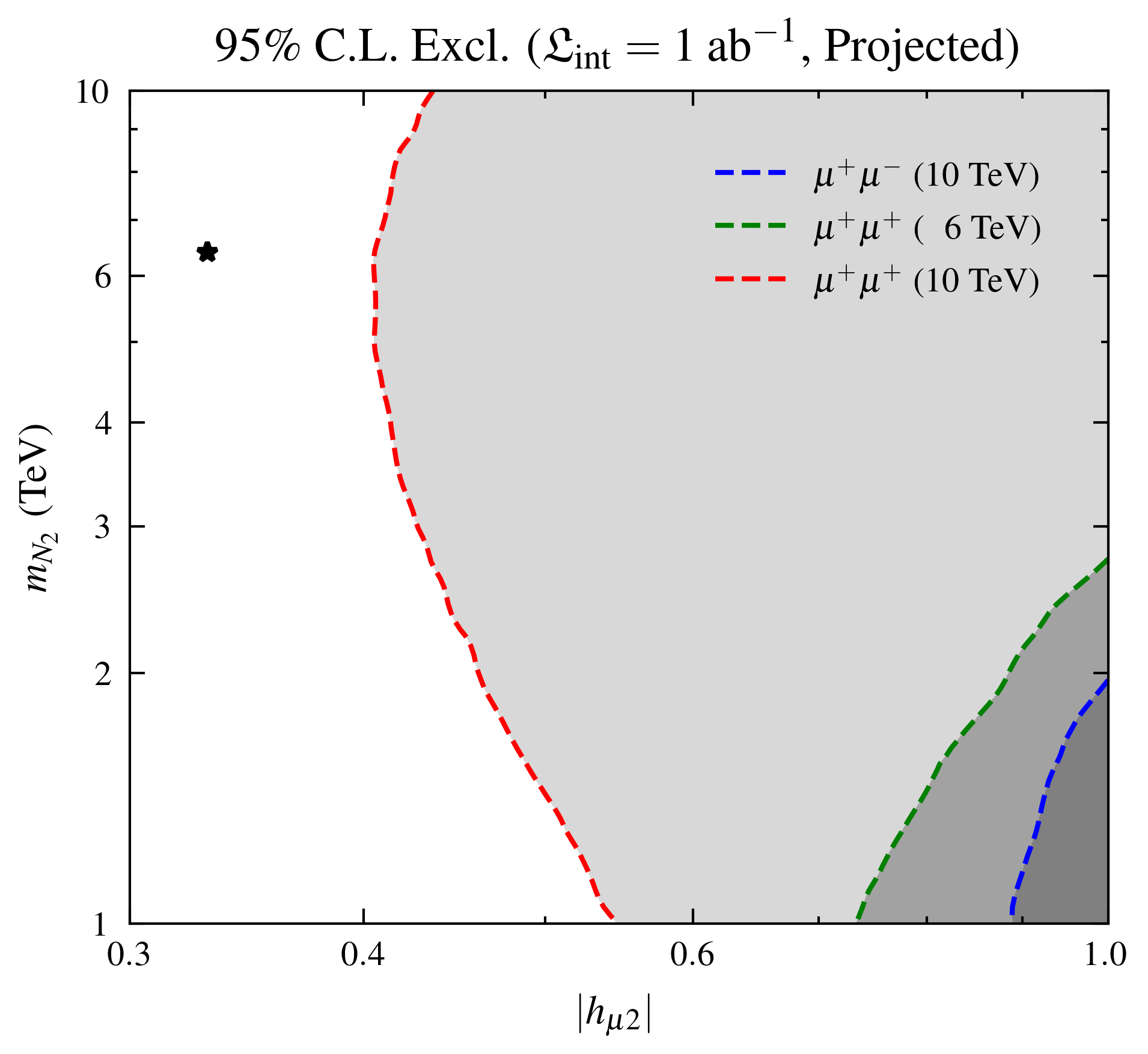}
    \includegraphics[width=0.475\linewidth]{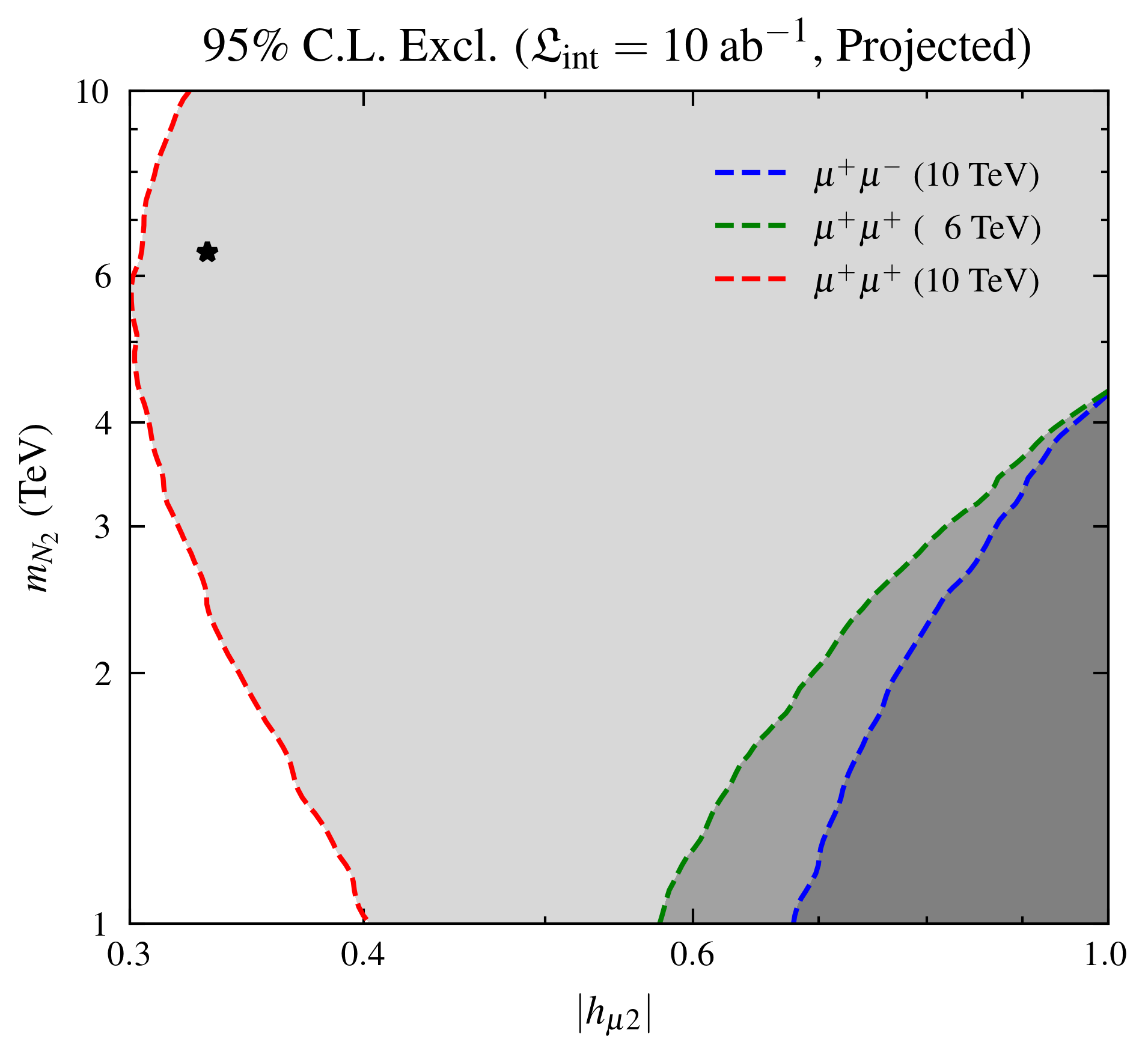}
    \caption{Projected 95\% C.L. exclusion limits on $|h_{\mu 2}|-m_{N_2}$ parameter space from mono-$\gamma$ signal for $\mu^{+}\mu^{\pm} \to \eta^{+}\eta^{\pm} \gamma\,/\,\eta_{I}^{0}\,\eta_{R}^{0} \gamma$ production at future muon colliders for integrated luminosities of 1 ab$^{-1}$ and 10 ab$^{-1}$, respectively. The star ($\star$) corresponds to the benchmark point, \textbf{BP}.}
    \label{fig:mcy}
\end{figure}

\paragraph{Projecting exclusion limits}
In this section, we present the projected exclusion limits on the $|h_{\mu 2}|-m_{N_2}$ using a binned likelihood analysis, by leveraging the shape information of the missing energy ($\slashed{E}$) distributions, thereby going beyond total rate measurements to extract stronger constraints. We outline the statistical framework used to quantify these projections below. The expected number of events in the ${\tt r}^{\rm th}$ bin of the $\slashed{E}$ distribution is given by:
\begin{equation}
    \mu_{\tt r} \left(|h_{\mu 2}|, m_{N_{2}}\right) = \mathfrak{L}_{\rm int} \times \left[\sigma^{s}_{\tt r} \left(|h_{\mu 2}|, m_{N_{2}}\right) + \sigma^{b}_{\tt r}\right]\,,
\end{equation}
where $\mathfrak{L}_{\rm int}$ is the integrated luminosity, $\sigma^{s}_{\tt r}$ is the cross section of the ${\tt r}^{\rm th}$ bin of the $\slashed{E}$ distribution, which depends on the model parameters. $\sigma^{b}_{\tt r}$ is the SM background cross section. The corresponding likelihood function \cite{ParticleDataGroup:2024cfk} is defined as a product of Poisson probabilities over all bins:
\begin{equation}
    \mathfrak{L} \left(|h_{\mu 2}|, m_{N_{2}}\right)  = \prod_{{\tt r}} \frac{\left[\mu_{\tt r} \left(|h_{\mu 2}|, m_{N_{2}}\right) \right]^{n_{\tt r}}e^{-\mu_{\tt r} \left(|h_{\mu 2}|, m_{N_{2}}\right)}}{n_{\tt r}!}\,,
\end{equation}
where, $n_{\tt r}$ is the observed number of events in the ${\tt r}^{\rm th}$ bin of the $\slashed{E}$ distribution. Since no observed data exist for future colliders, the observed event counts are assumed to equal to the SM yields. So, the observed events are taken as the background expectation:
\begin{equation}
    n_{\tt r} = \mathfrak{L}_{\rm int} \times \sigma^{b}_{\tt r}\,.
\end{equation}
The log-likelihood is then given by:
\begin{equation}
    \log{\mathfrak{L} \left(|h_{\mu 2}|, m_{N_{2}}\right)} = \sum_{{\tt r}} \big[n_{\tt r} \log{\mu_{\tt r} \left(|h_{\mu 2}|, m_{N_{2}}\right)} - \mu_{\tt r} \left(|h_{\mu 2}|, m_{N_{2}}\right) \big] + {\rm constant}.
\end{equation}
To quantify the sensitivity, we define the profile likelihood ratio:
\begin{equation} \label{eq:profile}
\begin{split}
    \lambda \left(|h_{\mu 2}|, m_{N_{2}}\right) &= \frac{\mathfrak{L} \left(|h_{\mu 2}|, m_{N_{2}}\right)}{\hat{\mathfrak{L}}_{b}} \,, \\
    \mathcal{Q} \left(|h_{\mu 2}|, m_{N_{2}}\right) &= -2 \log{\lambda \left(|h_{\mu 2}|, m_{N_{2}}\right)}\,,
\end{split}
\end{equation}
where, $\hat{\mathfrak{L}}_{b}$ denotes the likelihood for SM only scenario. The test statistic $\mathcal{Q}$ defined in \eq\eqref{eq:profile} is used to derive projected exclusion limits on the model parameters at a given confidence level. According to Wilks' theorem~\cite{Wilks:1938dza}, in the asymptotic limit, the distribution of $\mathcal{Q}$ approaches a chi-squared ($\chi^2$) distribution with degrees of freedom equal to the number of parameters being tested. Using the Wilks' theorem, we translate the value of $\mathcal{Q}$ into confidence intervals on the model parameters. For instance, the critical values of the test statistic $\mathcal{Q}$ corresponding to the 95\% C.L. is given by:
\begin{equation}
    \mathcal{Q} \leq \chi^2_{2,\,95\%} = 5.99\,,
\end{equation}
where $\chi^2_{f,\,p\%}$ denotes the $p^{\rm th}$ percentile of the chi-squared distribution with $f$ degrees of freedom. These thresholds define the regions of parameter space that are consistent with the background-only hypothesis at the corresponding confidence levels. We consider 10 bins of equal width of the $\slashed{E}$ distribution (of \fig\ref{fig:mcx}) for different collider setups and plot the projected 95\% C.L. exclusion limits on the $|h_{\mu 2}|-m_{N_2}$ plane in \fig\ref{fig:mcy}. The limits corresponds to integrated luminosities of 1 ab$^{-1}$ (\textit{left}) and 10 ab$^{-1}$ (\textit{right}), respectively.
We find that the $\mu^{+}\mu^{+}$ runs exhibit greater sensitivity to Scotogenic-specific parameters compared to the $\mu^{+}\mu^{-}$ runs. This can be understood from the absence of gauge-mediated diagrams in the $\mu^{+}\mu^{+}$ case. For the chosen benchmark point (BP), no sensitivity is achieved at an integrated luminosity of $1~\mathrm{ab}^{-1}$. However, increasing the luminosity to $10~\mathrm{ab}^{-1}$ brings the BP within the $95\%$ C.L. sensitivity reach for the $\mu^{+}\mu^{+}$ run at $\sqrt{s} = 10~\mathrm{TeV}$. Nevertheless, prompt searches remain largely insensitive to scenarios with smaller Yukawa couplings and heavier RHN masses.

\subsection{Long-lived search at muon colliders}
\label{sec:llp}
\begin{figure}[htb!]
    \centering
    \includegraphics[width=0.475\linewidth]{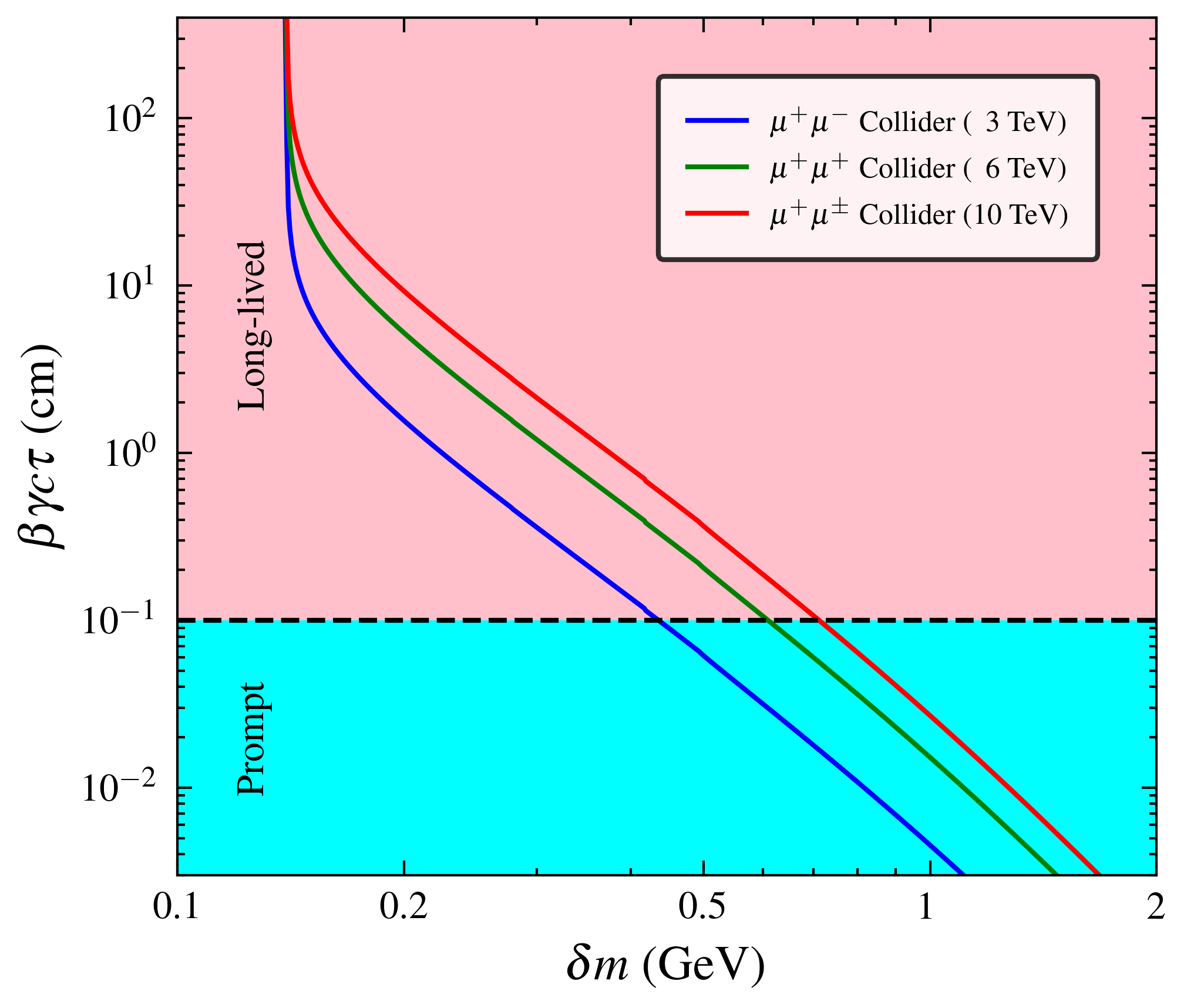}
    \caption{Proper length of $\eta^{\pm}$ decay vs.\ charged-neutral scalar mass splitting, $\delta m$ for different muon collider runs. The cyan and pink regions correspond to prompt decay ($\beta \gamma c \tau < 0.1$ cm) and long-lived search ($\beta \gamma c \tau > 0.1$ cm) regimes, respectively.}
    \label{fig:dt}
\end{figure}

The compressed mass spectrum of the scalar sector naturally gives rise to long-lived charged states. In particular, for small mass splittings $\delta m$, the visible decay products of $\eta^{\pm}$ are very soft, causing the charged track to effectively \textit{disappear} inside the detector. Such disappearing track signatures provide a powerful probe of this scenario at muon colliders, offering a complementary search strategy to prompt and missing energy-based analyses. The $\eta^{+}$ decay is dominated by the 2-body decay channel: $\eta^{\pm} \to \eta_{I/R}^{0} + \pi^{\pm}$. Another 2-body decay channel is $\eta^{\pm} \to \eta_{I/R}^{0} + K^{\pm}$. The decay width for the 2-body processes is:
\begin{equation}
\begin{split}
    \Gamma^{(\pi)}_{\eta^{+}} &= \left(\frac{G^{2}_{F}}{\pi}\right)\, |V_{ud}|^{2}\, f^{2}_{\pi}\, \left(\delta m\right)^{3}\,\sqrt{1-\left(\frac{m_{\pi}}{\delta m}\right)^{2}}\,, \\
    \Gamma^{(K)}_{\eta^{+}} &= \left(\frac{G^{2}_{F}}{\pi}\right)\, |V_{us}|^{2}\, f^{2}_{K}\, \left(\delta m\right)^{3}\,\sqrt{1-\left(\frac{m_{K}}{\delta m}\right)^{2}}\,, \\
\end{split}
\end{equation}
where $f_{\pi/K} = 0.130/0.156~\text{GeV}$. The $K$ mode is CKM-suppressed. When the mass splitting $\delta m < m_{\pi}$, only the 3-body mode $\eta^{\pm} \to \eta_{I/R}^{0} + \ell^{\pm} + \nu_{\ell}$ is kinematically allowed, which is highly suppressed and leads to a long-lived $\eta^{+}$. The decay width for the 3-body process is:
\begin{equation}
    \Gamma^{\ell}_{\eta^{+}} = \left(\frac{G^{2}_{F}}{15\,\pi^{3}}\right)\,\left(\delta m\right)^{5}\,F\left(\frac{m_{\ell}}{\delta m}\right)\,,
\end{equation}
where $F(r) = 1 - 8r^2 + 8r^6 - r^8 - 12r^4 \ln{r^2}$. Proper length of $\eta^{\pm}$ decay vs.\ charged-neutral scalar mass splitting, $\delta m$ for different muon collider runs, is shown in \fig\ref{fig:dt}.

\begin{figure}[htb!]
    \centering
    \includegraphics[width=0.475\linewidth]{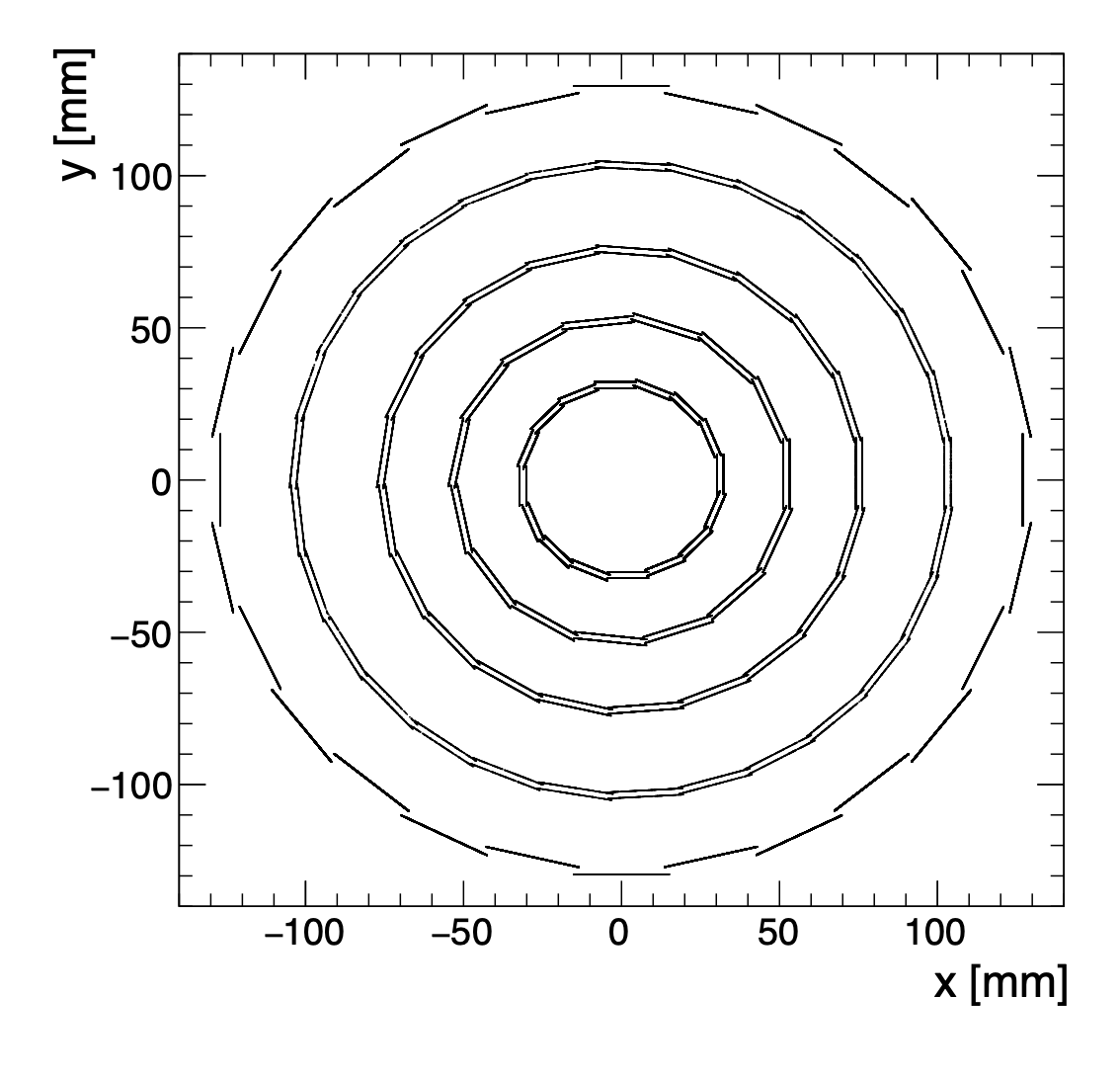}
    \includegraphics[width=0.475\linewidth]{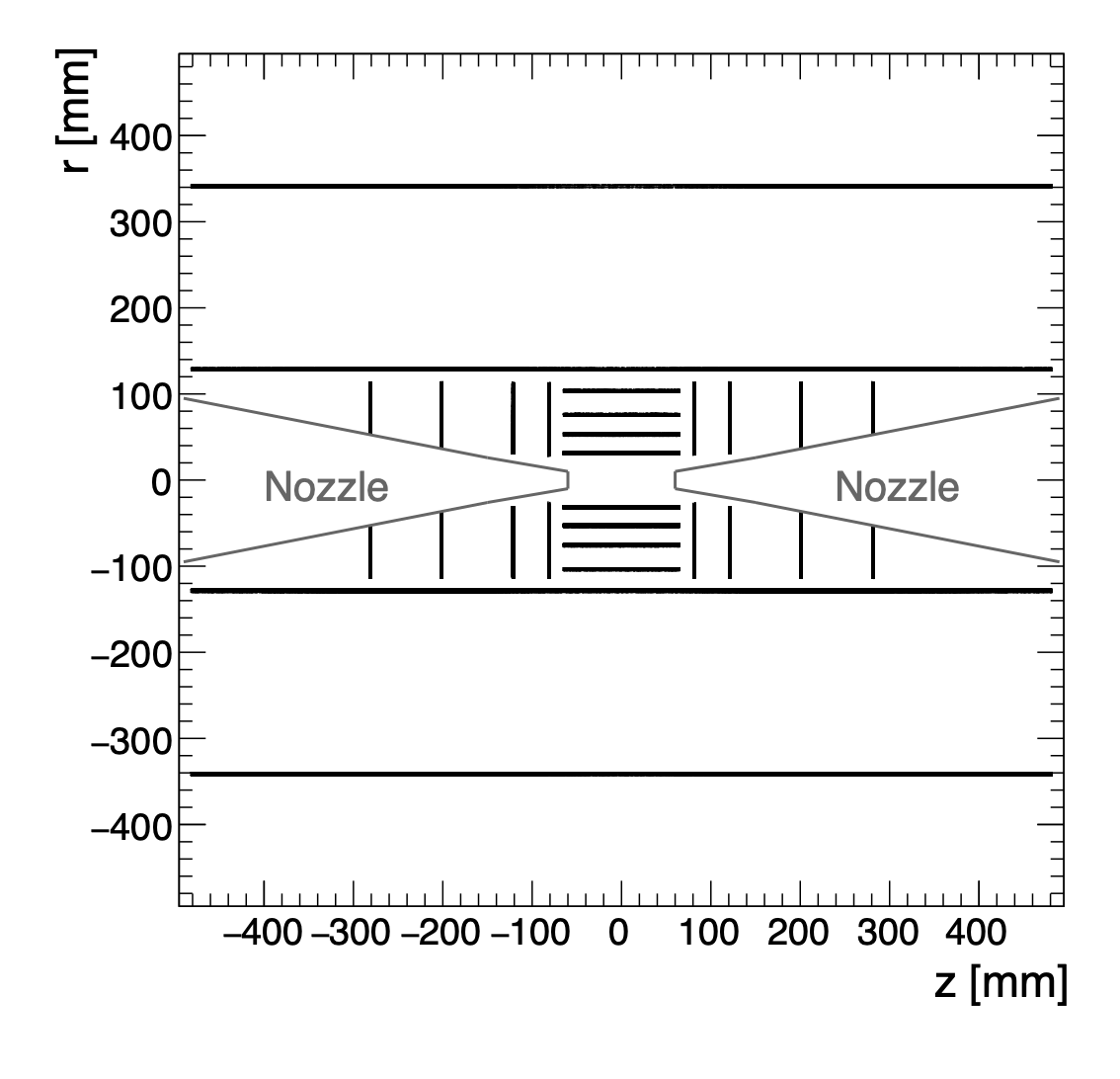}
    \caption{\textit{Muon collider trackers} ($x$-$y$ and $r$-$z$ sections): \textbf{VXD0-1} (30mm), \textbf{VXD2-3} (50mm), \textbf{VXD4-5} (75mm), \textbf{VXD6-7} (100mm), \textbf{IT0} (125mm) and \textbf{OT0} (340mm), adapted from~\cite{Capdevilla:2021fmj}.}
    \label{fig:detector}
\end{figure}

Fig.~\ref{fig:detector} illustrates the layout of the vertex detector and tracking system for a muon collider, adapted from~\cite{Capdevilla:2021fmj}, shown in both the transverse ($x$-$y$) and longitudinal ($r$-$z$) projections. The vertex detector (VXD) consists of multiple layers located at radii of 30~mm (\textbf{VXD0-1}), 50~mm (\textbf{VXD2-3}), 75~mm (\textbf{VXD4-5}), and 100~mm (\textbf{VXD6-7}). The tracking system further includes an inner tracker at 125~mm (\textbf{IT0}) and an outer tracker at 340~mm (\textbf{OT0}).

\begin{figure}
\centering
\includegraphics[width=0.325\linewidth]{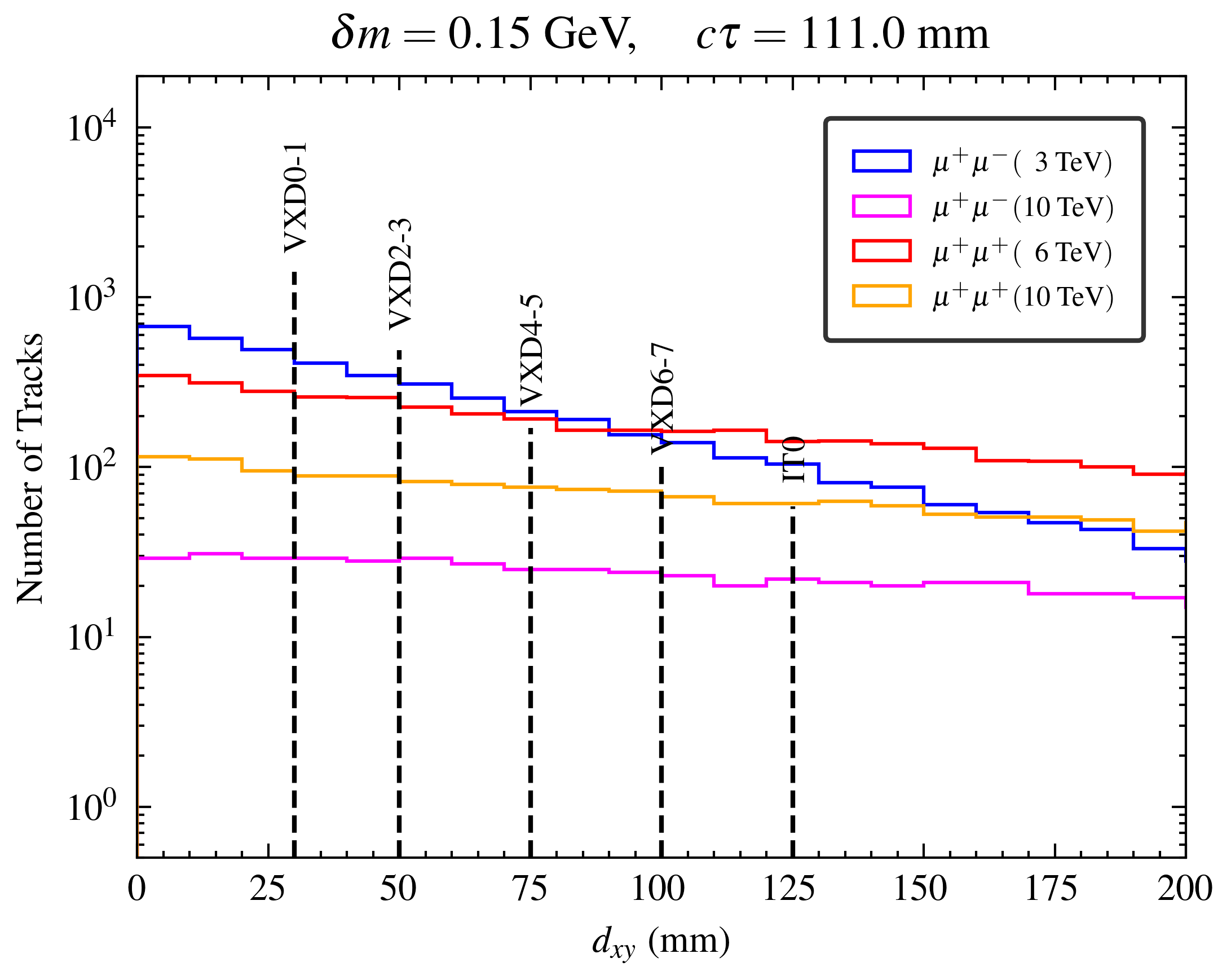}
\includegraphics[width=0.325\linewidth]{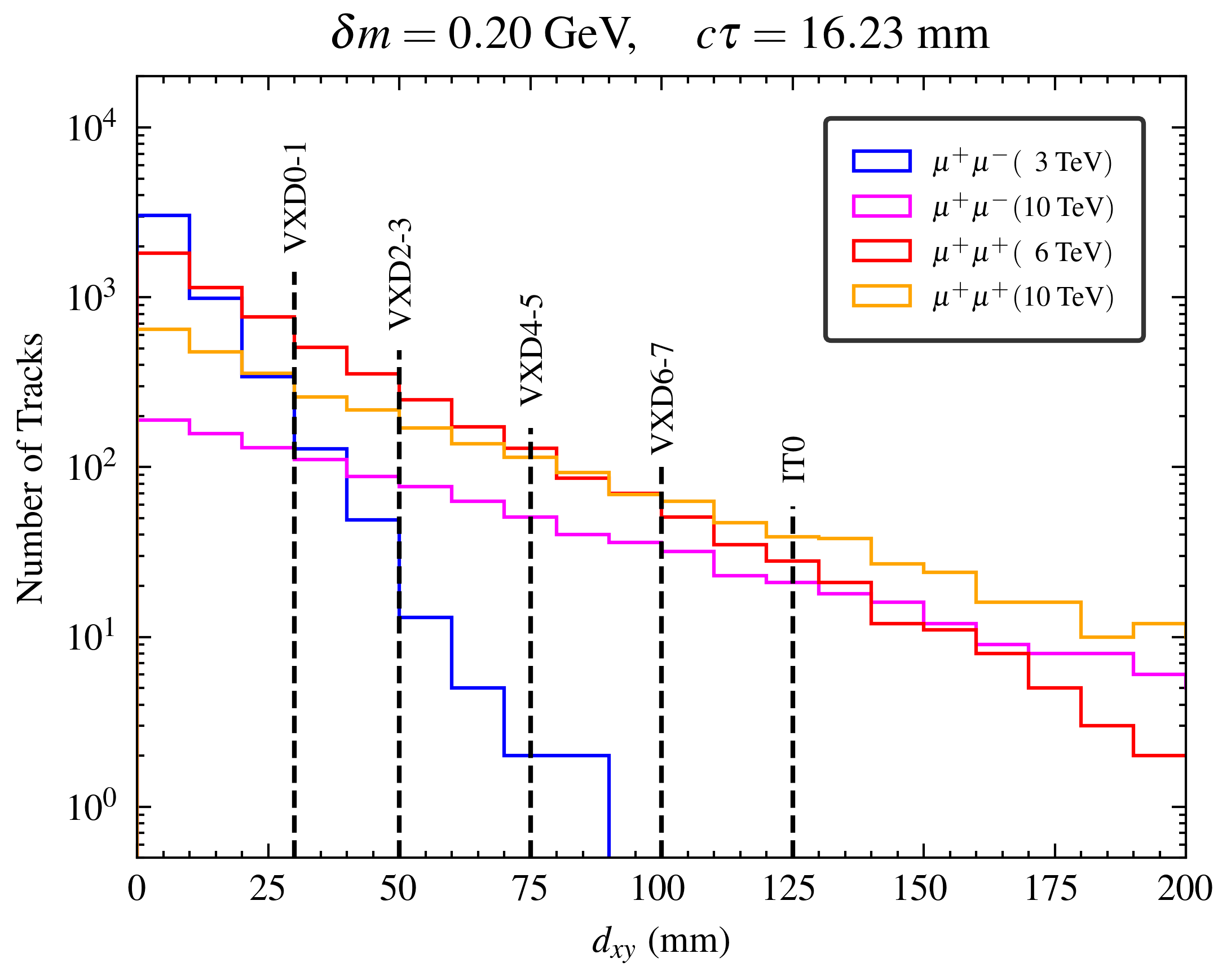}
\includegraphics[width=0.325\linewidth]{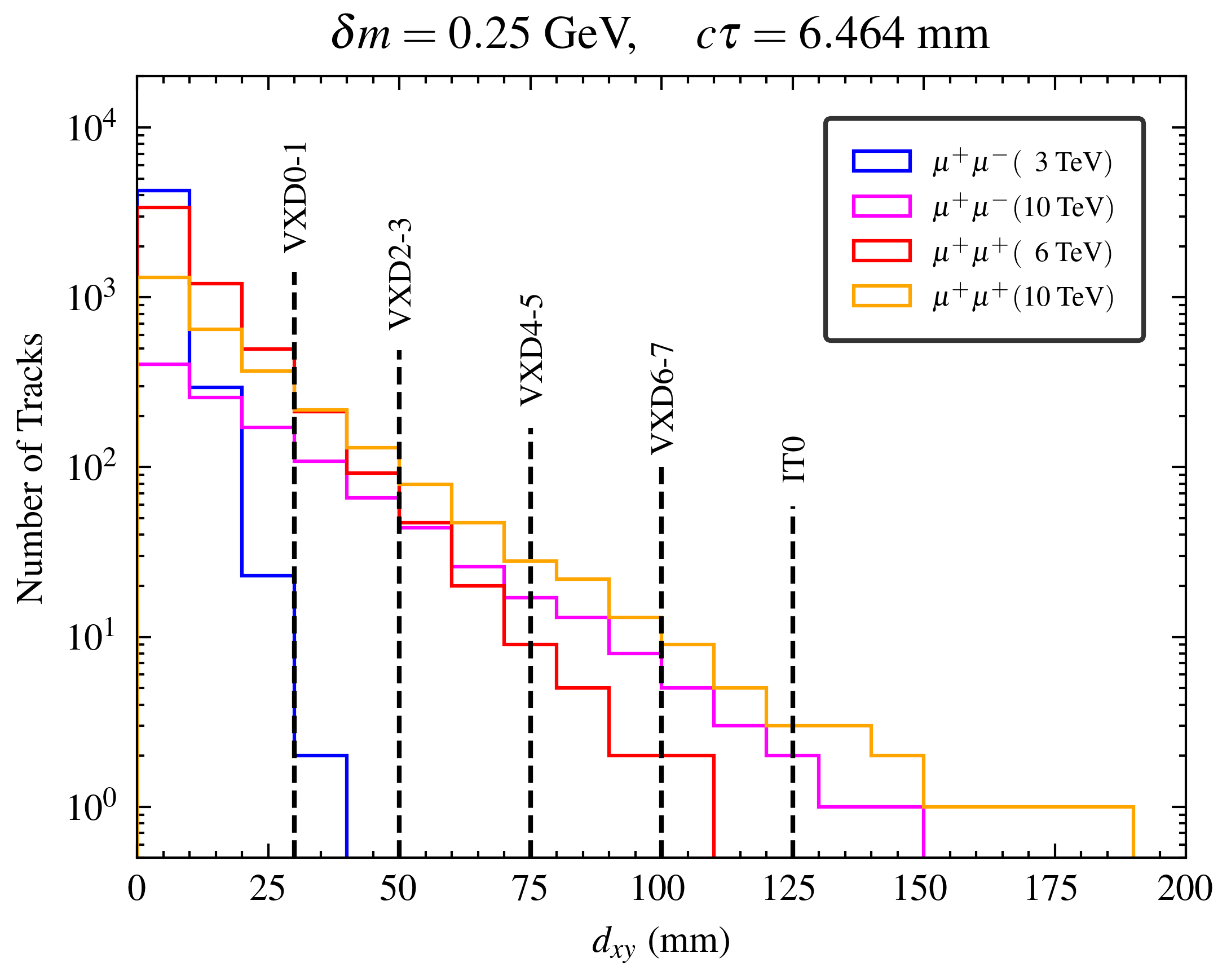}
\caption{Transverse track length ($d_{xy}$) distribution for different benchmarks: BP1 (\textit{left}), BP2 (\textit{middle}) and BP3 (\textit{right}) for different muon collider configurations with $10~\mathrm{ab}^{-1}$ integrated luminosity. No layer or angular efficiency is included.}
\label{fig:dxy-only}
\end{figure}

\begin{figure}
\centering
\includegraphics[width=0.475\linewidth]{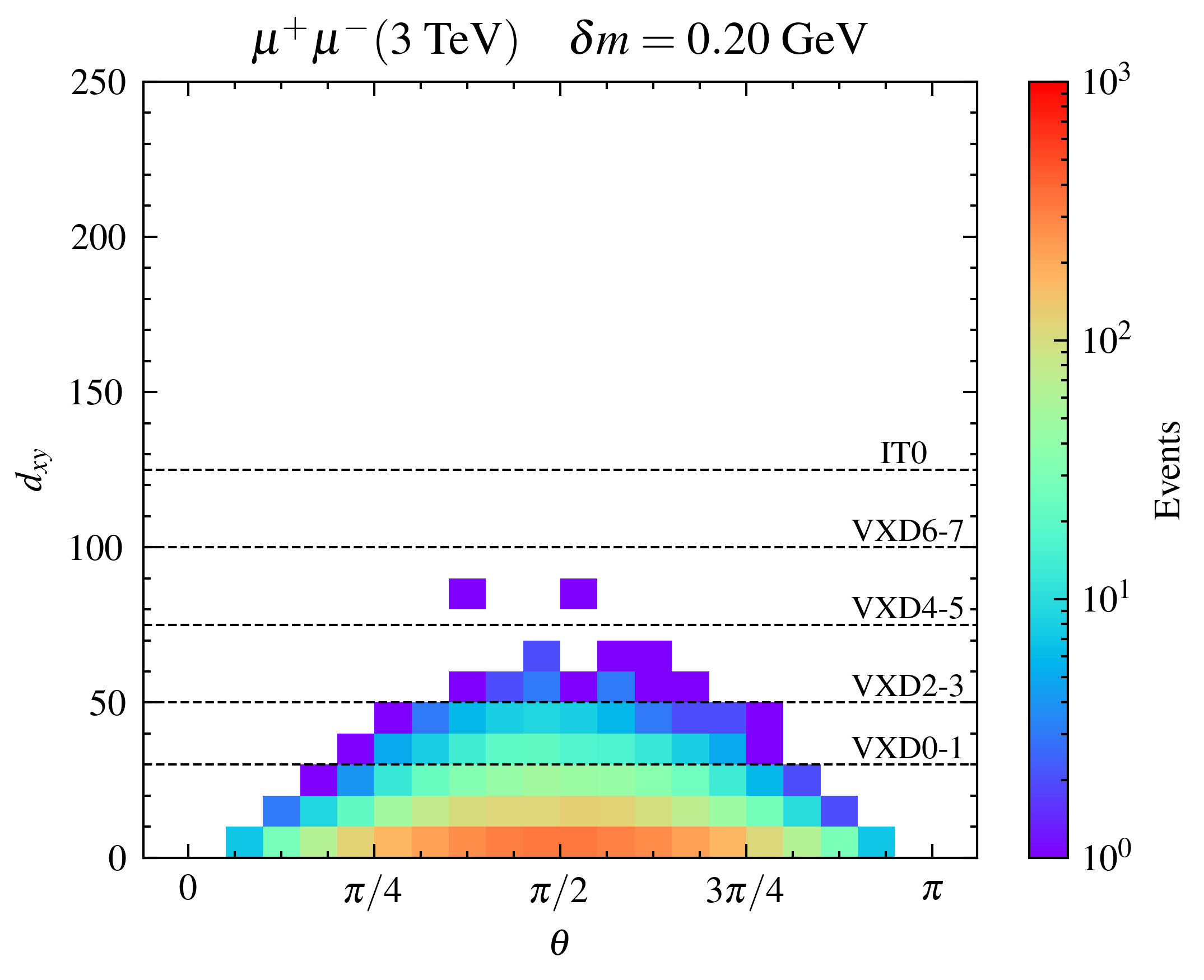}
\includegraphics[width=0.475\linewidth]{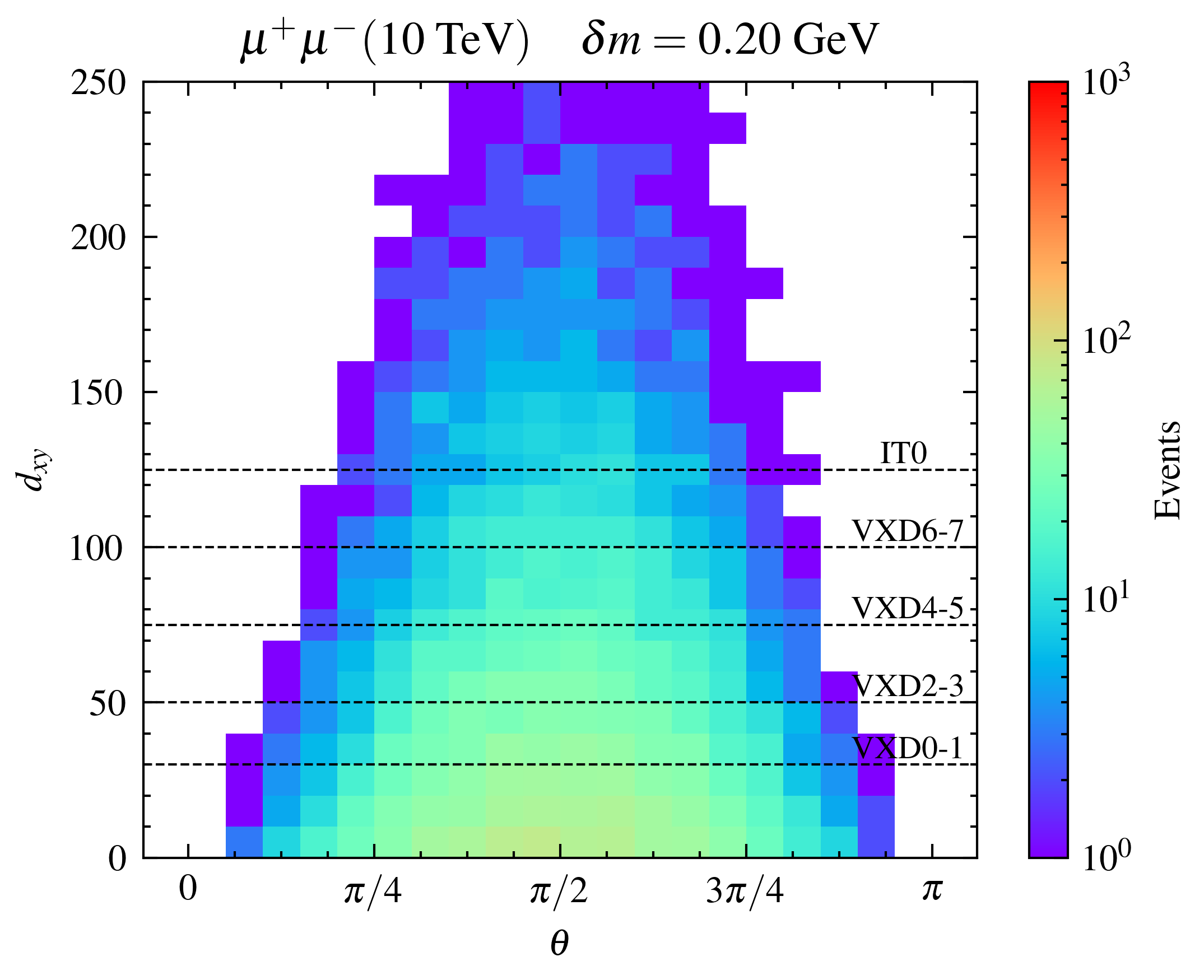} \
\includegraphics[width=0.475\linewidth]{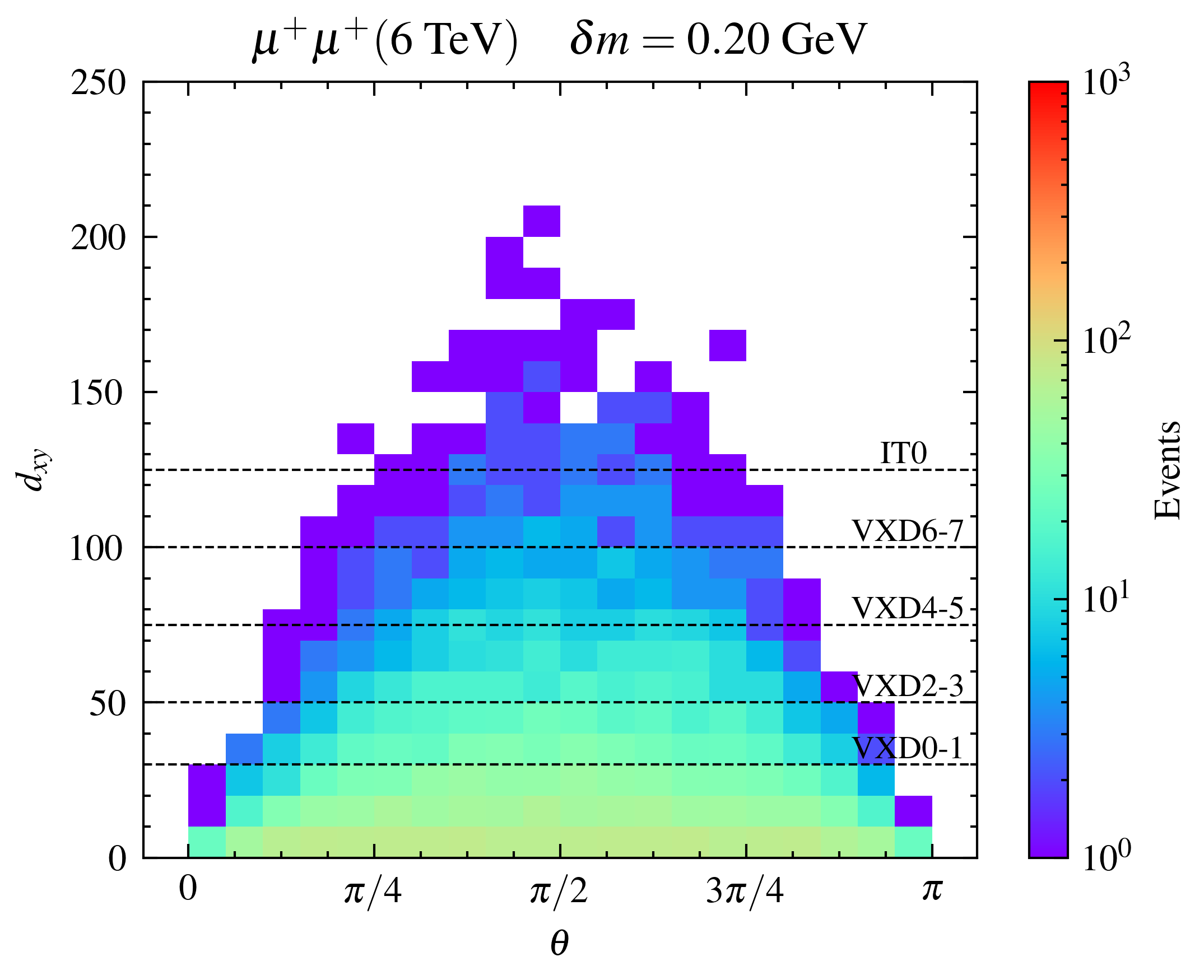}
\includegraphics[width=0.475\linewidth]{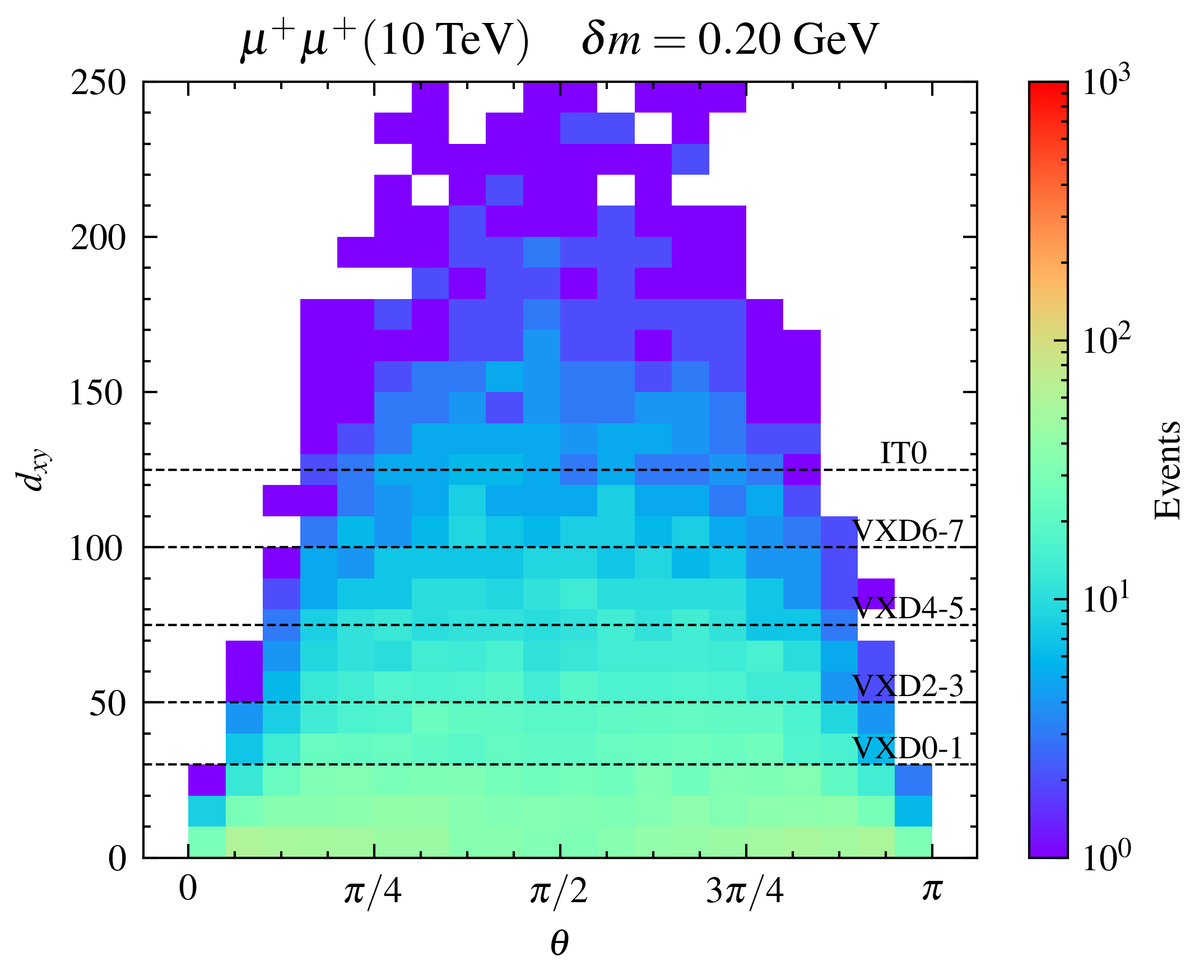}
\caption{Two-dimensional distributions of $\theta$ versus $d_{xy}$ for BP1 in different muon collider configurations with $10~\mathrm{ab}^{-1}$ integrated luminosity. No layer or angular efficiency is included.}
\label{fig:dxy-costheta}
\end{figure}

In the following paragraphs, we discuss a track-level analysis in the context of $\mu^{+}\mu^{-}$ and $\mu^{+}\mu^{+}$ colliders. Conventional track-based searches typically rely on additional objects, such as photons or jets, to provide triggers for detector readout. However, requiring such radiation suppresses the Born-level production cross section. Future muon colliders are expected to enable triggerless tracking capabilities for both stable and disappearing tracks. In this work, we adopt this triggerless approach, allowing us to retain the full event sample and thereby enhance the overall sensitivity of the analysis.
\paragraph{Disappearing track ($\boldsymbol{\mu^{+}\mu^{-} \to \eta^{+}\eta^{-}}$):}
The compressed mass spectrum of the scalar sector leads to long-lived charged particles whose decay products are typically too soft to be reconstructed. Consequently, the charged track abruptly terminates inside the tracker, giving rise to characteristic disappearing track (DT) signatures. Such signatures provide an important probe of compressed new physics scenarios at future muon colliders. To study this, we generate parton-level events for $\mu^{+}\mu^{-} \to \eta^{+}\eta^{-}$ using \texttt{MG5\_aMC}. Since the decay length of $\eta^{\pm}$ depends sensitively on the charged-neutral scalar mass splitting $\delta m$, the charged scalars are subsequently decayed by sampling from an exponential decay distribution governed by their proper lifetime. This procedure allows us to realistically model the spatial decay profile of the charged tracks inside the detector volume. We first examine the transverse decay length ($d_{xy}$) distributions for three benchmark points (BP1, BP2, BP3), corresponding to $\delta m = 0.15$, $0.20$, and $0.25$ GeV, respectively. The resulting distributions are shown in \fig\ref{fig:dxy-only}. As expected, smaller mass splittings correspond to longer-lived charged scalars, thereby enhancing the probability of observing tracks traversing multiple detector layers. In contrast, larger mass splittings lead to prompt decays, resulting in significantly shorter tracks.
To further understand the kinematic dependence, we present the two-dimensional distributions of $d_{xy}$ and $\cos\theta$ in \fig\ref{fig:dxy-costheta}. These plots illustrate the interplay between the boost of the charged scalar and its decay length in the laboratory frame. Tracks produced in the forward region generally possess larger boosts, leading to longer decay lengths and consequently larger probabilities of crossing multiple detector layers. This correlation plays an important role in determining the efficiency of disappearing track reconstruction.

To translate these kinematic distributions into observable event rates, we perform a track-level efficiency analysis. We assume a per-layer hit efficiency such that the probability of reconstructing a track with $N_{\rm hits}$ is given by
\begin{equation}
\epsilon(N_{\rm hits}) = (0.95)^{N_{\rm hits}},.
\end{equation}
Using this prescription, we compute the number of events containing at least one disappearing track crossing at least $N_{\rm hits}$ tracker layers. The resulting event yields for different mass splittings and integrated luminosities are summarized in Table~\ref{tab:layer1}.
\begin{table}[htb!]
\centering
\small
\begin{tabular}{|c|cc|cc|cc|}
\hline\hline
\multirow{2}*{$\boldsymbol{N_{\rm hits}}$}
& \multicolumn{2}{c|}{$\boldsymbol{\delta m = 0.15}$ \textbf{GeV}}
& \multicolumn{2}{c|}{$\boldsymbol{\delta m = 0.20}$ \textbf{GeV}}
& \multicolumn{2}{c|}{$\boldsymbol{\delta m = 0.25}$ \textbf{GeV}} \\
& $1$ ab$^{-1}$ & $10$ ab$^{-1}$
& $1$ ab$^{-1}$ & $10$ ab$^{-1}$
& $1$ ab$^{-1}$ & $10$ ab$^{-1}$ \\
\hline\hline
\multicolumn{7}{|c|}{$\boldsymbol{\mu^{+}\mu^{-}}$ \textbf{(3 TeV)}} \\ \hline\hline
$\geq 1$ & 159 & 1591 & 18 & 182 & 0 & 2 \\
$\geq 2$ & 126  & 1260  & 2 & 19  & 0  & 0  \\
$\geq 3$ & 89  & 888  & 0   & 3  & 0  & 0  \\
$\geq 4$ & 62  & 615  & 0   & 0  & 0  & 0  \\
$\geq 5$ & 42   & 421   & 0   & 0  & 0   & 0   \\
\hline\hline
\multicolumn{7}{|c|}{$\boldsymbol{\mu^{+}\mu^{-}}$ \textbf{(10 TeV)}} \\ \hline\hline
$\geq 1$ & 37 & 371 & 39 & 385 & 23 & 232 \\
$\geq 2$ & 36  & 365  & 29 & 289 & 10 & 100 \\
$\geq 3$ & 35  & 356  & 19 & 190 & 3 & 34 \\
$\geq 4$ & 34  & 345  & 12   & 125  & 1 & 11 \\
$\geq 5$ & 33  & 333  & 8   & 81 & 0  & 4 \\
\hline\hline
\end{tabular}
\caption{Number of events with at least $N_{\rm hits}$ tracker layer crossings for at least one track, for different mass splittings and luminosities at $\mu^{+}\mu^{-}$ colliders.}
\label{tab:layer1}
\end{table}

Several notable features emerge from these results. First of all, the event yield exhibits a strong dependence on the charged-neutral mass splitting $\delta m$. Smaller splittings lead to significantly longer lifetimes, thereby enhancing the probability of observing tracks traversing multiple tracker layers. This behavior is particularly evident for BP1 ($\delta m = 0.15$ GeV), where substantial event yields persist even for stringent hit requirements such as $N_{\rm hits} \geq 5$. Next, the collider CM energy plays an important role via boost effects. Although the production cross section at higher energies can be small, the large Lorentz boost enhances the decay length of the charged scalar substantially. Consequently, the $10$ TeV configuration retains a sizeable event yield for larger hit requirements, even for moderately larger mass splittings. We also observe that the DT signal rapidly diminishes for $\delta m = 0.25$ GeV, particularly at lower CM energy. In this regime, the charged scalar decays promptly, leading to very short tracks that fail to traverse through multiple layers of the tracker. This demonstrates the strong sensitivity of DT searches to compressed spectra. Finally, increasing the integrated luminosity from $1~\mathrm{ab}^{-1}$ to $10~\mathrm{ab}^{-1}$ significantly improves the reach of the analysis, especially in regions where the signal rate is otherwise statistically limited. Overall, disappearing track searches at muon colliders provide a powerful probe of highly compressed scalar sectors, with sensitivity driven primarily by the mass splitting, boost effects, and detector-level tracking capabilities.
\paragraph{Disappearing track ($\boldsymbol{\mu^{+}\mu^{+} \to \eta^{+}\eta^{+}}$):}
We perform a similar disappearing track analysis for the same-sign process $\mu^{+}\mu^{+} \to \eta^{+}\eta^{+}$. The event generation, decay treatment, and detector-level efficiency modeling follow the same procedure described for the $\mu^{+}\mu^{-}$ channel. The resulting event yields are summarized in Table~\ref{tab:layer2}.
\begin{table}[htb!]
\centering
\small
\begin{tabular}{|c|cc|cc|cc|}
\hline\hline
\multirow{2}*{$\boldsymbol{N_{\rm hits}}$}
& \multicolumn{2}{c|}{$\boldsymbol{\delta m = 0.15}$ \textbf{GeV}}
& \multicolumn{2}{c|}{$\boldsymbol{\delta m = 0.20}$ \textbf{GeV}}
& \multicolumn{2}{c|}{$\boldsymbol{\delta m = 0.25}$ \textbf{GeV}} \\
& $1$ ab$^{-1}$ & $10$ ab$^{-1}$
& $1$ ab$^{-1}$ & $10$ ab$^{-1}$
& $1$ ab$^{-1}$ & $10$ ab$^{-1}$ \\
\hline\hline
\multicolumn{7}{|c|}{$\boldsymbol{\mu^{+}\mu^{+}}$ \textbf{(6 TeV)}} \\ \hline\hline
$\geq 1$ & 189 & 1891 & 128 & 1280 & 34 & 345 \\
$\geq 2$ & 177  & 1770 & 68 & 680 & 8  & 76  \\
$\geq 3$ & 161  & 1609 & 30 & 302  & 1  & 12  \\
$\geq 4$ & 147  & 1468 & 14 & 137  & 0  & 2  \\
$\geq 5$ & 132  & 1320 & 6 & 61  & 0  & 0   \\
\hline\hline
\multicolumn{7}{|c|}{$\boldsymbol{\mu^{+}\mu^{+}}$ \textbf{(10 TeV)}} \\ \hline\hline
$\geq 1$ & 96 & 958 & 87 & 874 & 45 & 453 \\
$\geq 2$ & 93 & 928 & 61 & 613  & 18 & 179 \\
$\geq 3$ & 89 & 890  & 38  & 382  & 6 & 58 \\
$\geq 4$ & 84 & 845  & 24 & 238  & 2 & 20 \\
$\geq 5$ & 80 & 801  & 15  & 148  & 1 & 8 \\
\hline\hline
\end{tabular}
\caption{Number of events with at least $N_{\rm hits}$ tracker layer crossings for at least one track, for different mass splittings and luminosities at $\mu^{+}\mu^{+}$ colliders.}
\label{tab:layer2}
\end{table}

The same qualitative behavior observed in the $\mu^{+}\mu^{-}$ channel persists in the same-sign collider mode. We observe that smaller mass splittings produce longer-lived heavy charged scalars, resulting in substantially enhanced DT rates. In particular, the $\delta m = 0.15$ GeV benchmark yields large event numbers even for stringent tracker-layer requirements. Compared to the opposite-sign channel, the $\mu^{+}\mu^{+}$ collider run exhibits noticeably larger DT rates. This enhancement arises primarily from the absence of gauge-mediated contributions, leading to stronger sensitivity to the underlying NP interactions. Moreover, the larger boost at higher CM energy improves the probability of observing longer DTs. Overall, the same-sign muon collider mode provides a promising environment for probing compressed spectra through DT signatures, with significant event yields surviving even after requiring multiple tracker-layer crossings.
\paragraph{Energy deposition and timing of tracks:}
Although DTs provide one of the most distinctive signatures of compressed NP scenarios, their experimental identification at a muon collider presents unique challenges. SM processes rarely produce long-lived charged final states capable of generating highly energetic DTs, aside from a few cases involving heavy-flavor hadrons, secondary interactions, or even detector effects. Consequently, the observation of a DT would constitute strong evidence for BSM physics. Moreover, the comparatively clean collision environment and considerably reduced QCD activity at muon colliders are expected to offer superior tracking and vertex identification capabilities relative to hadron colliders, further enhancing the sensitivity to such LLP signatures. A major challenge, however, arises from beam-induced backgrounds (BIBs), which are expected to dominate detector occupancies in regions close to the interaction point. These backgrounds originate primarily from the decay of beam muons within the accelerator lattice. The decay products subsequently interact with the components of the detector and generate large numbers of secondary particles and tracks that populate the inner tracking layers. As a result, the hits produced by NP particles can become difficult to distinguish from BIB-induced activity, especially in the inner layers of the detector. Detailed investigations of BIB effects and mitigation strategies for muon collider detectors can be found in~\cite{Capdevilla:2021fmj}. An additional source of background arises from the shielding system itself. To suppress BIB contamination, the muon beams are tightly collimated and pass through a narrow tungsten nozzle surrounding the interaction region, as illustrated in \fig\ref{fig:detector}. Nevertheless, a fraction of beam particles and secondary radiation interact with the nozzle material, producing copious neutrons and other secondary particles that further contribute to detector occupancies. Consequently, purely hit-based track reconstruction may not always be sufficient to separate signal tracks from background-induced activity. In this context, complementary observables based on energy deposition and timing information have recently attracted considerable attention as powerful discriminants for heavy charged particles~\cite{MAIA:2025hzm,Littmann:2025ndn,Acanfora:2026aay}. The charged scalar $\eta^{\pm}$ considered in this work is significantly heavier than SM charged particles and can therefore exhibit distinct ionization and timing signatures, particularly when produced close to threshold. Since such particles travel more slowly than relativistic muons, they deposit energy differently in the tracking layers and arrive at detector elements with measurable time delays.
\begin{figure}[htb!]
\centering
\includegraphics[width=0.475\linewidth]{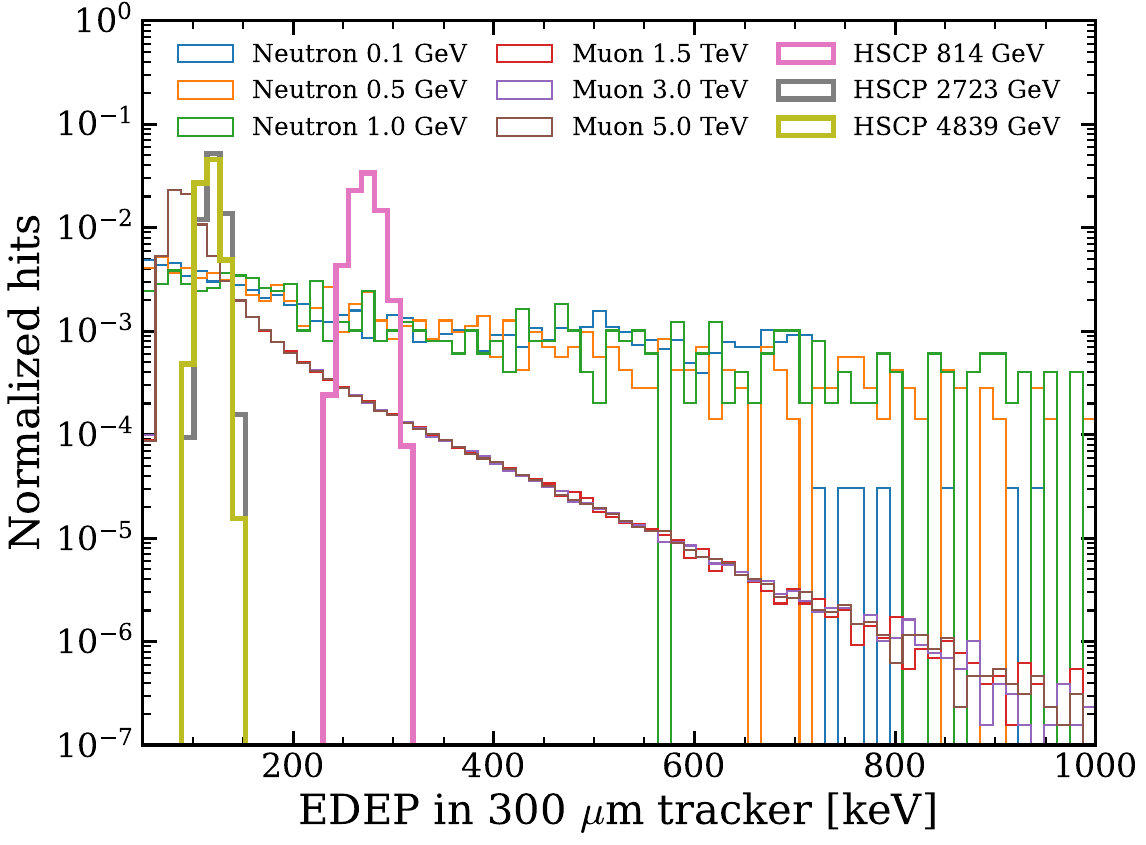}
\includegraphics[width=0.45\linewidth]{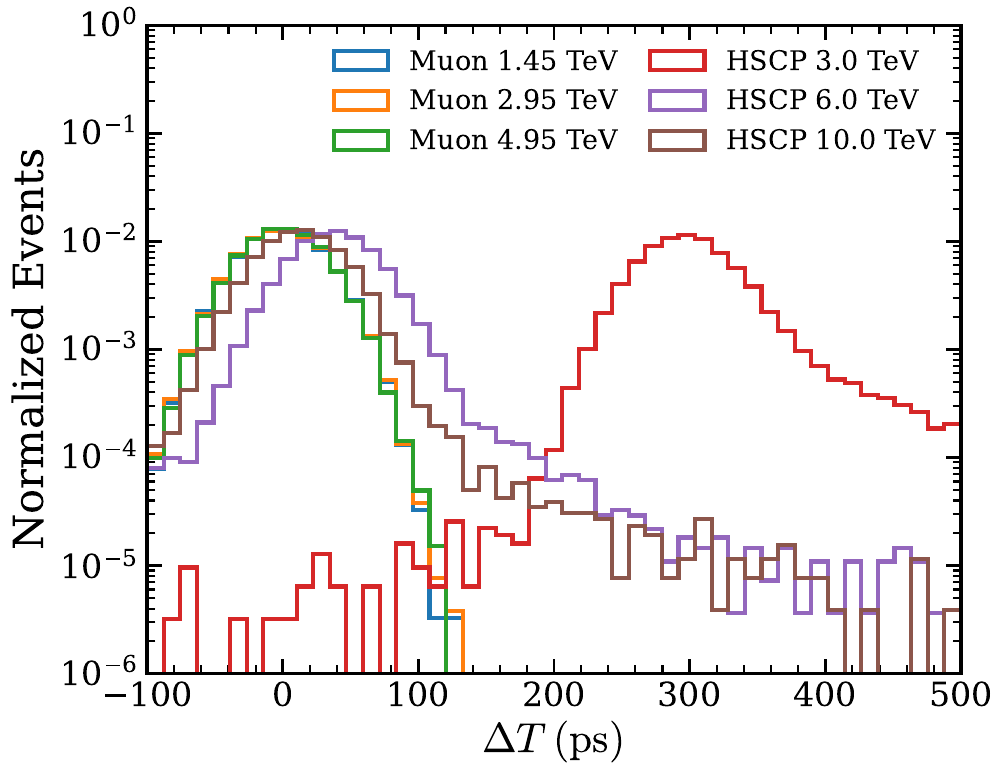}
\caption{Normalized hit counts for energy deposition (EDEP) on the fourth tracking layer by hits from heavy charged scalars (HSCPs), BIB muons, and neutrons (\textit{left}). Normalized event counts for the timing difference ($\Delta T$) of the fourth-layer hit between HSCPs and relativistic muons (\textit{right}).}
\label{fig:HSCP}
\end{figure}
The left panel of \fig\ref{fig:HSCP} shows the energy deposited in the fourth tracking layer by heavy charged scalars (denoted as HSCPs), BIB muons, and neutrons, implemented using GEANT4~\cite{GEANT4:2002zbu}. The muon-induced background exhibits a rapidly falling spectrum concentrated at low deposited energies, as expected for relativistic minimum-ionizing particles. In contrast, neutron-induced hits generate a broad and relatively featureless distribution that contributes to an approximately uniform background envelope. The HSCP signal displays a markedly different behavior, with a pronounced peak at larger deposited energies. Furthermore, this peak shifts toward higher values for lower collider energies, reflecting the reduced velocity of the produced charged scalars. Consequently, the separation between signal and background distributions becomes most pronounced for the $\sqrt{s} = 3$ TeV collider configuration, where the charged scalars are produced with comparatively small boosts and therefore exhibit enhanced ionization losses. Timing information provides an additional and largely independent discriminator. The right panel of \fig\ref{fig:HSCP} presents the distribution of the timing difference, $\Delta T$, between the arrival time of the HSCP at the fourth tracking layer and the arrival time of a relativistic particle traveling at the speed of light, implemented using Delphes3~\cite{deFavereau:2013fsa}. Explicitly, $\Delta T$ is defined as
\begin{equation}
\Delta T = T_{\rm HSCP} - T_{\rm light}\,.
\end{equation}
Since the heavy charged scalars travel with velocities significantly below the speed of light, they arrive later than relativistic muons, resulting in positive values of $\Delta T$. This effect becomes increasingly pronounced for lower-energy collider configurations, where the heavy scalars are produced closer to threshold and therefore possess smaller Lorentz boosts. As a result, the $\sqrt{s} = 3$ TeV setup again exhibits the strongest separation between signal and background populations. Taken together, the energy deposition and timing observables provide complementary handles for distinguishing heavy charged scalar tracks from beam-induced backgrounds. While a detailed detector-level analysis incorporating realistic background rates and reconstruction algorithms is beyond the scope of the present work, these results demonstrate that future muon collider detectors equipped with precise timing capabilities and high-granularity tracking systems could exploit both observables to significantly enhance the identification of disappearing-track and heavy charged particle signatures arising from compressed scalar sectors.
\section{Summary and Conclusion}
\label{sec:conclusion}
The minimal Scotogenic model with two generations of RHNs simultaneously explains observed DM relic density, active neutrino masses, and the baryon asymmetry of the Universe via leptogenesis. However, it typically requires right-handed neutrinos with masses $\sim\mathcal{O}(10^{10}\,\mathrm{GeV})$, leaving the new particles to remain inaccssible to upcoming sensitivities of direct, indirect, and collider searches. 

In this work, we consider an extension of the minimal Scotogenic framework with an additional RHN and investigate a scenario where the Yukawa couplings associated with the 2nd and 3rd generation RHNs can be relatively large, having the first-generation Yukawa coupling adequate enough to address successful leptogenesis. Thus the framework leads to observable signals at future muon colliders both due to the presence of large Yukawa couplings and low RHN mass, while simultaneously addressing lepton-flavor violation, DM, active neutrino masses, and leptogenesis. 

A larger second-generation Yukawa couplings also affects the DM freeze-out temperature if $\mntwo \sim \mathcal{O}(\TeV)$, through the dominant $t$-channel process $\eta\eta^\dagger\to \ell\bar{\ell}$ mediated by $N_2$. Consequently, the relic density satisfying region extends over the DM mass range $580~\mathrm{GeV}\lesssim\metaI\lesssim 1.3~\mathrm{TeV}$, while remaining consistent with the direct- and indirect-detection constraints.



Concerning leptogenesis, while a larger Yukawa coupling can generate larger asymmetry, it can also induce stronger washout effects via $\Delta L=2$ processes which we also take into account in our analysis. Notably in the hierarchical scenario ($\mnone<\mntwo< \mnthree$) such as ours, the asymmetry generated by heavier RHNs ($N_2$ and $N_3$) is washed out by the lightest RHN ($N_1$) as the heavier RHNs decouple early from the thermal bath, hence we only consider the effects of lightest $N_1$ in producing the lepton asymmetry. A scan in the $m_{N_1}$, $h_{i1}$, and $\Delta m$ plane reveals a minimal value of $m_{N_1}\sim 1.4~\TeV$ for successful leptogenesis with $\Delta m\sim 0.1$ or $\letaHpp\sim 2\times 10^{-4}$. 

On the collider front, prompt searches based on missing-energy signatures can probe scenarios with sizable Yukawa couplings and relatively light RHNs, however, requires large luminosity. On the other hand, the compressed scalar spectrum naturally predicts long-lived charged scalars, leading to disappearing track (DT) signatures that can serve as a distinctive feature of this model. Muon colliders, can achieve substantial sensitivity to DT through both track reconstruction and complementary observables such as energy deposition and timing information. The same-sign $\mu^{+}\mu^{+}$ collider, due to the absence of competing gauge-mediated contributions, can be even more effective in probing such signals. The combination of $\mu^{+}\mu^{-}$ and $\mu^{+}\mu^{+}$ runs, therefore, offers a unique opportunity to simultaneously probe the Scotogenic framework in the extended version. To the best of our knowledge, this work serves as one of the first of its kind to demonstrate such a possibility. 

%

\section*{Acknowledgments}
The authors would like to thank the IIT Hyderabad for its hospitality during WHEPP-2025, where part of this work was discussed. SB acknowledges the ANRF grant CRG/2023/000580.
\appendix
\newpage
\section{Neutrino mass and CP asymmetry}
\label{sec:numass}

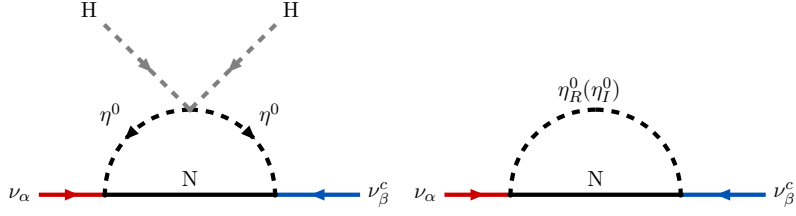
\begin{figure}[htb!]
\centering
\begin{tikzpicture}[baseline={(current bounding box.center)},style={scale=0.75, transform shape}]
\begin{feynman}
\vertex (a){\color{black}{$\nu_{\alpha}$}};
\vertex [right = 1.5cm of a] (b);
\vertex [right = 3cm of b] (c);
\vertex [right = 1.5cm of c] (d){\color{black}{$\nu_{\beta}^c$}};
\vertex [ above right = 1.5cm and 1.5cm of b] (e);
\vertex [ above left = 1.5cm and 1.5cm of e] (f){\color{black}{H}};
\vertex [ above right = 1.5cm and 1.5cm of e] (g){\color{black}{H}};
\diagram*{ 
(a) -- [line width=0.25mm, fermion, arrow size=1.2pt, style=bostonuniversityred, ultra thick] (b),
(b) -- [line width=0.25mm, plain,arrow size=1.2pt, style=black, ultra thick,edge label={$\color{black}{{\rm N} }$}] (c),
(d) -- [line width=0.25mm, fermion, arrow size=1.2pt, style=mediumtealblue, ultra thick] (c), 
(e) -- [line width=0.25mm, charged scalar, quarter right, style=black, ultra thick, arrow size=1.2pt,edge label'={$\color{black}{\eta^0 }$}] (b),
(e) -- [line width=0.25mm, charged scalar, quarter left, style=black, ultra thick, arrow size=1.2pt,edge label={$\color{black}{\eta^0 }$}] (c),
(f) -- [line width=0.25mm, charged scalar, style=gray, ultra thick, arrow size=1.2pt] (e),
(g) -- [line width=0.25mm, charged scalar,  style=gray, ultra thick, arrow size=1.2pt] (e)};
\node at (b)[circle,fill,style=black, inner sep=1pt]{};
\node at (c)[circle,fill,style=black, inner sep=1pt]{};
\node at (e)[circle,fill,style=gray, inner sep=1pt]{};
\end{feynman}
\end{tikzpicture}
\begin{tikzpicture}[baseline={(current bounding box.center)},style={scale=0.75, transform shape}]
\begin{feynman}
\vertex (a){\color{black}{$\nu_{\alpha}$}};
\vertex [right = 1.5cm of a] (b);
\vertex [right = 3cm of b] (c);
\vertex [right = 1.5cm of c] (d){\color{black}{$\nu_{\beta}^c$}};
\vertex [ above right = 1.5cm and 1.5cm of b] (e);
\vertex [ above right = 1.5cm and 0.7cm of b] (o){$\color{black}{\eta_{R}^0(\eta_{I}^0) }$};
\vertex [ above left = 1.5cm and 1.5cm of e] (f){\color{white}{H}};
\vertex [ above right = 1.5cm and 1.5cm of e] (g){\color{white}{H}};
\diagram*{
(a) -- [line width=0.25mm,fermion, arrow size=1.2pt, style=bostonuniversityred, ultra thick] (b),
(b) -- [line width=0.25mm, plain,style=black, ultra thick,edge label={$\color{black}{{\rm N} }$}] (c),
(d) -- [line width=0.25mm,fermion, arrow size=1.2pt, style=mediumtealblue, ultra thick] (c), 
(e) -- [line width=0.25mm,scalar, quarter right, style=black, ultra thick] (b),
(e) -- [line width=0.25mm,scalar, quarter left, style=black, ultra thick, arrow size=1.2pt] (c)};
\node at (b)[circle,fill,style=black, inner sep=1pt]{};
\node at (c)[circle,fill,style=black, inner sep=1pt]{};
\end{feynman}
\end{tikzpicture}
\caption{Radiative neutrino majorana mass generation where $i,j$ are generation indices. The left and right figures correspond to before and after EWSB, respectively.}
\label{fig:NuMass}
\end{figure}
The neutrino mass generated from the diagrams shown in fig\,.~\ref{fig:NuMass} is given by
\begin{eqnarray}
\left(  \mathcal{M}  \right)_{\alpha \beta} & = & \sum_{i=1}^{3} \dfrac{h^{*}_{ \alpha i} h^{*}_{\beta i} m^{}_{N^{}_{i}}}{32 \pi^{2}}  \left[  L(m_{\eta_{R}^0}^{2} )^{}_{i}  -L(m_{\eta_{I}^0}^{2})^{}_{i}\right],
\label{eq:Numass}
\end{eqnarray}
\noindent where the function $L$ has the following form 
\begin{eqnarray}
L(m^{2})^{}_{i} & = & \dfrac{m^{2}}{m^{2}-m_{N^{}_{i}}^{2}} \ln \left( \dfrac{m^{2}}{m_{N^{}_{i}}^{2}} \right)\,.
\end{eqnarray}
\begin{align}
\nonumber\mathbb{M}_{\alpha\beta}&=\sum_{k}h_{\alpha k}^*(\Lambda_{k})^{-1} h^*_{\beta k}
\\\nonumber&=h_{\alpha 1}^*(\Lambda_{1})^{-1} h^*_{\beta 1}+h_{\alpha 2}^*(\Lambda_{2})^{-1} h^*_{\beta 2}+h_{\alpha 3}^*(\Lambda_{3})^{-1} h^*_{\beta 3}
\\&=
\begin{pmatrix}
h^*_{\alpha 1} & h^*_{\alpha 2} & h^*_{\alpha 3}
\end{pmatrix} 
\begin{pmatrix}
1/\Lambda_1 & 0  & 0\\
0 & 1/\Lambda_2  & 0\\
0 & 0 & 1/\Lambda_3  \\
\end{pmatrix}
\begin{pmatrix}
h^*_{\beta 1} \\ h^*_{\beta 2} \\ h^*_{\beta 3}
\end{pmatrix} \,;
\end{align}

\begin{align}
\mathbb{M}=\begin{pmatrix}
h^*_{e1} & h^*_{e2} & h^*_{e3}\\
h^*_{\mu1} & h^*_{\mu2} & h^*_{\mu3}\\
h^*_{\tau1} & h^*_{\tau2} & h^*_{\tau 3}
\end{pmatrix} 
\begin{pmatrix}
1/\Lambda_1 & 0  & 0\\
0 & 1/\Lambda_2  & 0\\
0 & 0 & 1/\Lambda_3  \\
\end{pmatrix}
\begin{pmatrix}
h^*_{e1} & h^*_{\mu1}& h^*_{\tau1}\\ h^*_{e2}& h^*_{\mu2}& h^*_{\tau2}\\  h^*_{e3}& h^*_{\mu3}& h^*_{\tau3}\\
\end{pmatrix} \equiv h^* \Lambda^{-1} (h^*)^T=h^* \Lambda^{-1} h^{\dagger}\,,
\end{align}
where 
\begin{eqnarray}
h= 
\begin{pmatrix}
h_{e1} & h_{e2} & h_{e3} \\
h_{\mu1} & h_{\mu2} & h_{\mu3} \\
h_{\tau1} & h_{\tau2} & h_{\tau3} 
\end{pmatrix} ~{\rm and} ~
\Lambda^{-1}=
\begin{pmatrix}
1/\Lambda_1 & 0  & 0\\
0 & 1/\Lambda_2  & 0\\
0 & 0 & 1/\Lambda_3  \\
\end{pmatrix}\,,
\end{eqnarray}
\noindent and 

\begin{eqnarray}\label{eq:nLam}
\Lambda_{i} & = & \dfrac{4\pi^{2}}{m_{\eta_{R}}^{2}-m_{\eta_{I}}^{2}}  \xi_{i}\,m^{}_{N^{}_{i}} = \dfrac{4\pi^{2}}{\lambda^{\prime\prime}_{\eta H}}  \xi_{i} \dfrac{m^{}_{N^{}_{i}}}{v^2}.
\label{eq:yukawa}
\end{eqnarray}

\noindent The loop functions $\xi_{i}$ are given by

\begin{eqnarray}\label{eq:nzeta}
\xi_{i} & = & \bigg( \dfrac{1}{8} \dfrac{m_{N^{}_{i}}^{2}}{m_{\eta_{R}}^{2}-m_{\eta_{I}}^{2}} \left[L(m_{\eta_{R}}^{2})^{}_{i} -L(m_{\eta_{I}}^{2})^{}_{i}\right]  \bigg)^{-1}\,.
\end{eqnarray}
Now, the light neutrino mass is diagonalized using the usual PMNS matrix $\rm U $, with Majorana and Dirac phases, which is determined from neutrino oscillation data and
\begin{align}
\rm \mathbb{M}_{\nu}=diag(m^{}_{\nu^{}_{1}},m^{}_{\nu^{}_{2}},m^{}_{\nu^{}_{1}})=U^{\dagger} \mathbb{M} U^*=U^{\dagger}h^* \sqrt{\Lambda^{-1}}\sqrt{\Lambda^{-1}} h^{\dagger}  U^*\,,
\label{eq:numass}
\end{align}
$$\rm \mathbb{M}_{\sqrt{\nu}}\mathbb{M}_{\sqrt{\nu}}=U^{\dagger}h^* \sqrt{\Lambda^{-1}}\sqrt{\Lambda^{-1}} h^{\dagger}  U^*\,,$$
$$\rm \mathbb{I}=\mathbb{M}_{\sqrt{\nu^{-1}}}U^{\dagger}h^* \sqrt{\Lambda^{-1}}\sqrt{\Lambda^{-1}} h^{\dagger}  U^*\mathbb{M}_{\sqrt{\nu^{-1}}}=\left[\sqrt{\Lambda^{-1}} h^{\dagger}  U^*\mathbb{M}_{\sqrt{\nu^{-1}}}\right]^T\left[\sqrt{\Lambda^{-1}} h^{\dagger}  U^*\mathbb{M}_{\sqrt{\nu^{-1}}}\right]\equiv \mathbb{R}^T\mathbb{R}\,.$$
where $\mathbb{R}$ is any $3 \times 3$ orthogonal matrix. Then, the Yuakawa coupling matrix satisfying the neutrino data can be written as,

\begin{equation}
h=U^*\mathbb{M}_{\sqrt{\nu}}\mathbb{R}^{\dagger}\sqrt{\Lambda}
\label{eq:CI}
\end{equation}

where \cite{Antusch:2011nz},
\begin{eqnarray}
\mathbb{M}_{\sqrt{\nu}}= 
\begin{pmatrix}
\sqrt{m^{}_{\nu^{}_{1}}} & 0 & 0 \\
0 & \sqrt{m^{}_{\nu^{}_{2}}} & 0 \\
0 & 0 & \sqrt{m^{}_{\nu^{}_{3}}} \\
\end{pmatrix} ~\,,~
\sqrt{\Lambda}=
\begin{pmatrix}
\sqrt{\Lambda_{1}} & 0  & 0 \\
0 & \sqrt{\Lambda_{2}} & 0 \\
0 & 0 & \sqrt{\Lambda_{3}}
\end{pmatrix}\,,\\\,~
U=\begin{pmatrix}
c_{12}c_{13} & s_{12}c_{13}  & s_{13}e^{-i\delta} \\
-s_{12}c_{23}-c_{12}s_{23}s_{13}e^{i\delta} & c_{12}c_{23}-s_{12}s_{23}s_{13}e^{i\delta} & s_{23}c_{13} \\
s_{12}s_{23}-c_{12}c_{23}s_{13}e^{i\delta} & -c_{12}s_{23}-s_{12}c_{23}s_{13}e^{i\delta} & c_{23}c_{13} \\
\end{pmatrix}\,,~{\rm and }\,\\
\mathbb{R}=\begin{pmatrix}
\tilde{c}_{12}\tilde{c}_{13} & \tilde{s}_{12}\tilde{c}_{13}  & \tilde{s}_{13} \\
-\tilde{s}_{12}\tilde{c}_{23}-\tilde{c}_{12}\tilde{s}_{23}\tilde{s}_{13} & \tilde{c}_{12}\tilde{c}_{23}-\tilde{s}_{12}\tilde{s}_{23}\tilde{s}_{13} & \tilde{s}_{23}\tilde{c}_{13} \\
\tilde{s}_{12}\tilde{s}_{23}-\tilde{c}_{12}\tilde{c}_{23}\tilde{s}_{13} & -\tilde{c}_{12}\tilde{s}_{23}-\tilde{s}_{12}\tilde{c}_{23}\tilde{s}_{13} & \tilde{c}_{23}\tilde{c}_{13} \\
\end{pmatrix}\,
\end{eqnarray}
here, we use the shorthand notations $c_{ij}(s_{ij})=\text{cos}~\theta_{ij}(\text{sin}~\theta_{ij})$ ($\theta_{ij}$ are the mixing angles with values given in \eq\eqref{eq:neutrino_mass1} and \eq\eqref{eq:neutrino_mass2}) and $\tilde{c}_{ij}(\tilde{s}_{ij})=\text{cos}~z_{ij}(\text{sin}~z_{ij})$ (where $z_{ij}$ are the complex angles). Using the above matrices, one can find the first column of the Yukawa matrix as
\begin{equation}
    h_{\alpha 1}\simeq \left[\tilde{c}_{13}\sqrt{m^{}_{\nu^{}_{1}}}\,U^\ast_{11}\sqrt{\Lambda_1}\,,\tilde{c}_{13}\sqrt{m^{}_{\nu^{}_{1}}}\,U^\ast_{21}\sqrt{\Lambda_1}\,, \tilde{c}_{13}\sqrt{m^{}_{\nu^{}_{1}}}\,U^\ast_{31}
\sqrt{\Lambda_1}\right]\,,
\label{eq:Yuk1st}
\end{equation}
where, $\alpha$ is the flavor indices ranging from $1$ to $3$. It is clear from the above expression that with small $\theta_{ij}$ and fixed $z_{ij},$ and $\Lambda_{i}$, $h_{1\alpha}$ increases with increasing $m^{}_{\nu^{}_{1}}$. 
The CP asymmetry generated via the decay of $N_k$ generation of RHN (see \fig\ref{feyn:CP-asymmetry}), for the Scotogenic model, can be quantified as
\begin{eqnarray}
\varepsilon_{N_{i}}=\frac{1}{8\pi(h^\dagger h)_{ii}}\sum_{j\neq i}{\rm Im}\left[(h^\dagger h)^2_{ij}\right]\left[f\left(\eta_i,r_{ji}\right)-\frac{\sqrt{r_{ji}}}{r_{ji}-1}\left(1-\eta_{i}\right)\right]\,,
\end{eqnarray}
where the function $f(\eta_i,r_{ji})$ is the loop function generated from the interference between the tree level and the vertex correction one loop level decay of $N_{i}$ and quantified as
\begin{eqnarray}
    f(\eta_i,r_{ji})\equiv \sqrt{r_{ji}}\left(1+\frac{1+r_{ji}-2\eta_i}{(1-\eta_{i})^2}{\rm ln}\left[\frac{r_{ji}-\eta_{i}^2}{1+r_{ji}-2\eta_i}\right]\right)\,,
    \label{eq:cp asy}
\end{eqnarray}
with $r_{ji}=m_{N_{j}^{}}^{2}/m_{N_{i}^{}}^{2}$ and $\eta_{i}=\mu_{\eta}^2/m_{N_{i}^{}}^{2}\equiv\left[(\metaI+\delta m)^2-\lambda_{\eta H }v^2\right]/m_{N_{i}^{}}^{2}$. The lepton asymmetry is produced via the RHN ($N_{1}$) decay and the decay width is given by 
\begin{equation}\label{eq:N1dw}
\Gamma_{N_{1}}^{}=\frac{\mnone}{8\pi}(h^{\dagger}h)_{11}(1-\eta_{1})^2\,.
\end{equation}

Following \eq\eqref{eq:Numass}, the active neutrino mass matrix admits a particularly simple approximate form in the compressed-spectrum limit $\letaHpp\ll1$, for which $\metaR\approx\metaI=m_{\eta^{0}_{}}$. In this limit, we obtain~\cite{Toma:2013zsa} 
\begin{equation}
    \left(  \mathcal{M}  \right)_{\alpha \beta}\approx \sum_{i=1}^{3} \dfrac{h^{*}_{ \alpha i} h^{*}_{\beta i} \Delta m\,m_{\eta^{0}_{}}}{(4 \pi)^{2}~m^{}_{N^{}_{i}}}  \left[  \frac{m^{2}_{N^{}_{i}}}{m_{\eta^{0}_{}}^2-m^{2}_{N^{}_{i}}}  + \frac{m^{4}_{N^{}_{i}}}{\left(m_{\eta^{0}_{}}^2-m^{2}_{N^{}_{i}}\right)^2}\,\ln\left(\frac{m^{2}_{N^{}_{i}}}{m_{\eta^{0}_{}}^2}\right)\right]\,.
    \label{eq:Numassap}
\end{equation}
In this special scenario, one can define 
\begin{equation}
    \tilde{\Lambda}_i=\frac{(4 \pi)^{2}~m^{}_{N^{}_{i}}}{\Delta m\,m_{\eta^{0}_{}}}\left[  \frac{m^{2}_{N^{}_{i}}}{m_{\eta^{0}_{}}^2-m^{2}_{N^{}_{i}}}  + \frac{m^{4}_{N^{}_{i}}}{\left(m_{\eta^{0}_{}}^2-m^{2}_{N^{}_{i}}\right)^2}\,\ln\left(\frac{m^{2}_{N^{}_{i}}}{m_{\eta^{0}_{}}^2}\right)\right]^{-1}\,,
    \label{eq:tildel}
\end{equation}
Substituting \eq\eqref{eq:tildel} into \eq\eqref{eq:CI}, we obtain 
\begin{equation}
h^*_{\beta i}h_{\alpha i}=\tilde{\Lambda}_i\mathfrak{M}_{\alpha\beta}\propto    \frac{m^{}_{N^{}_{i}}}{\Delta m\,m_{\eta^{0}_{}}}\mathfrak{M}_{\alpha\beta}\,,
\label{eq:muegap}
\end{equation}
where, $\mathfrak{M}_{\alpha\beta}=\sum_{k,l=1}^{3}\sqrt{m_{\nu^{}_k}m_{\nu^{}_l}}U^{\ast}_{\alpha k}U_{\beta l}\mathbb{R}^\ast_{ik}\mathbb{R}_{il}$.
Furthermore, employing the unitarity of the PMNS matrix ($U^\dagger U=\mathbb{I}$) and performing the corresponding contractions, the off-diagonal combination of $h^\dagger h$ can be written as
\begin{equation}
    (h^\dagger h)_{1j}=\sqrt{\tilde{\Lambda}_1\tilde{\Lambda}_j}\,\mathtt{M}_{1j}\propto\frac{\sqrt{m^{}_{N^{}_{1}}~m^{}_{N^{}_{j}}}}{\Delta m\,m_{\eta^{0}_{}}}\mathtt{M}_{1j}\,,
    \label{eq:hdh1j}
\end{equation}
where, $\mathtt{M}_{1j}=\sum_{k=1}^3m_{\nu_{k}^{}}\mathbb{R}_{1k}\mathbb{R}_{jk}^{\ast}$.
\section{Expressions for loop factors and other quantities}
\label{sec:loop}
In this section, we present various analytical expressions for loop functions and other quantities used in the calculation of LFV processes. The expression for dipole from factor in \eq\eqref{eq:mueg} is given by,
\begin{align}
F_D=\sum_{i=1}^3\dfrac{h^*_{\beta i}h_{\alpha i}}{2(4\pi)^2}\dfrac{1}{m_{\eta^+}^2}G(x_{i})\simeq\sum_{i=1}^3\dfrac{\mathfrak{M}_{\alpha\beta}}{2\Delta m\,\metaI\,m_{N_{i}^{}}}\quad\text{for}\quad G(x_{i})\approx x_i^{-1}\,,
\label{eq:FDapp}
\end{align}
where $x_{k}^{}=m_{N_k^{}}^2/m_{\eta^+}^2$. Now we will provide the detailed expressions for various quantities used in \eq\eqref{eq:mu3e} as,
\begin{gather}
F_{RR}=\dfrac{F g_{R}^{\ell}}{g_2^2\sin^2\theta_W m_Z^2}\,,~
F_{RL}=\dfrac{F g_{L}^{\ell}}{g_2^2\sin^2\theta_W m_Z^2}\,,~
A_{ND}=\sum_{i=1}^3\dfrac{h^*_{\beta i}h_{\alpha i}}{6(4\pi)^2}\dfrac{1}{m_{\eta^+}^2}G_2(x_i)\,,~\text{with}\nonumber
\end{gather}
\begin{gather}
g_{L}^{\ell}=\dfrac{g_2}{\cos\theta_W}\left(\dfrac{1}{2}-\sin^2\theta_W\right),~g_{R}^{\ell}=-\dfrac{g_2}{\cos\theta_W}\sin^2\theta_W\,,~F=\sum_{i=1}^3\dfrac{h^*_{\beta i}h_{\alpha i}}{2(4\pi)^2}\dfrac{m_\alpha m_\beta}{m_{\eta^+}^2}\dfrac{g_2}{\cos\theta_W}G(x_i)\,,\nonumber
\end{gather}
where $g_2$ is the $SU(2)_L$ gauge coupling and $\theta_W$ is the weak mixing angle. The coefficient B in \eq\eqref{eq:mu3e} is expressed as
\begin{gather}
 B=\dfrac{1}{e(4\pi)^2m_{\eta^+}^2}\sum_{i,j=1}^3\left[\dfrac{1}{2}D_1(x_i,x_j)h_{i\alpha}h_{i\beta}^\ast h_{j\beta}h_{j\beta}^\ast+\sqrt{x_ix_j}D_2(x_i,x_j)h_{i\alpha}h_{i\beta}h_{j\beta}^\ast h_{j\beta}^\ast\right]\,.
\end{gather}
All the loop functions used in above repressions are given below
\begin{align}
   G(x)&=\dfrac{1-6x+3x^2+2x^3-6x^2~\log~x}{6(1-x)^4},  \\
   G_2(x)&=\dfrac{2-9x+18x^2-11x^3+6x^3\log x}{6(1-x)^4}\,,\\
D_1(x,y)&=-\dfrac{1}{(1-x)(1-y)}-\dfrac{x^2\log x}{(1-x)^2(x-y)}-\dfrac{y^2\log y}{(1-y)^2(y-x)}\,,\\
D_2(x,y)&=-\dfrac{1}{(1-x)(1-y)}-\dfrac{x\log x}{(1-x)^2(x-y)}-\dfrac{y\log y}{(1-y)^2(y-x)}\,.
\end{align}
It is worth noting that the loop functions remain finite and do not develop singularities. In the special limits $x,y \to 1$, they reduce to the following simple forms:
\begin{align}
G(1)=\frac{1}{12},\,~G_2(1)=\frac{1}{4}\,,~D_1(1,1) &= -\frac{1}{3}\,,~D_2(1,1)=\frac{1}{6}.
\label{eq:loop_limit1}
\end{align}
For equal arguments ($y \to x$), the functions $D_1$ and $D_2$ simplify to
\begin{align}
D_1(x,x)=-\frac{1-x^2+2x\log x}{(1-x)^3}\,,~D_2(x,x)=\frac{-2+2x-(1+x)\log x}{(1-x)^3}.
\label{eq:D2xx}
\end{align}
When one of the arguments is unity, one obtains
\begin{align}
D_1(x,1)=D_1(1,x)
&=
-\frac{1-4x+3x^2-2x^2\log x}
{2(1-x)^3}\,,\\
D_2(x,1)=D_2(1,x)
&=\frac{1-x^2+2x\log x}
{2(1-x)^3}.
\label{eq:D2x1}
\end{align}
The effective iso-scalar and iso-vector couplings in the $\mu\to e$ conversion process (see \eq\eqref{eq:mute}), denoted by
$g^{(0)}_{xk}$ and $g^{(1)}_{xk}$ respectively, are defined as
\begin{align}
g^{(0)}_{xk}&=\frac{1}{2}\sum_{q=u,d,s}\left[
\widetilde {g_{xk}^{} }(q)\,\mathbb{G}_{K}^{(q,p)}+
\widetilde{g_{xk}^{}}(q)\,\mathbb{G}_{K}^{(q,n)}
\right],\nonumber\\
g^{(1)}_{xk}&=\frac{1}{2}\sum_{q=u,d,s}\left[
\widetilde{g_{xk}}(q)\,\mathbb{G}_{K}^{(q,p)}-\widetilde{g_{xk}}(q)\,\mathbb{G}_{K}^{(q,n)}
\right],
\label{eq:g0g1}
\end{align}
where $x=L,R$ and $k=S,V$. The numerical values of the nucleon form factors
$\mathbb{G}_K$ are taken from the Ref.~\cite{Kuno:1999jp},\cite{Kosmas:2001mv}. Since the contribution from the Higgs-mediated processes is strongly suppressed due to the presence of the small Yukawa coupling, they can be safely neglected. Under this approximation, the relevant couplings reduce to
\begin{align}
\widetilde{g_{LV}}(q)&=\widetilde{g^{\gamma}_{LV}}(q)+\widetilde{g^{Z}_{LV}}(q),\nonumber\\
\widetilde{g_{RV}}(q)&=
\widetilde{g_{LV}}(q)\Big|_{L\leftrightarrow R},\nonumber\\
\widetilde{g_{LS}}(q)&\simeq 0,\qquad
\widetilde{g_{RS}}(q)\simeq 0\,.\nonumber
\label{eq:effective_couplings}
\end{align}
The photon and $Z$ mediated contributions arise from the one-loop diagrams shown in \fig\ref{feyn:lfv}. Retaining only the dominant terms, the corresponding effective couplings are given by 
\begin{align}
\widetilde{g^{\gamma}_{LV}}(q)&=
\frac{\sqrt{2}}{G_F}\,
e^2 Q_q
\left(A_{ND}-F_D\right),\\
\widetilde{g^{Z}_{RV}}(q)&=
-\frac{g_L^{q}+g_R^{q}}{\sqrt{2}G_F}\,
\frac{F}{m_Z^2}\,.
\label{eq:z_coupling}
\end{align}
Here $Q_q$ denotes the electric charge of the quark $q$, while the tree-level
couplings of the $Z$ boson to quarks are
\begin{align}
g_L^{q}
&=
\frac{g_2}{\cos\theta_W}
\left(Q_q\sin^2\theta_W
-T_3^{q}
\right)\,,\\
g_R^{q}&=
\frac{g_2}{\cos\theta_W}
Q_q\sin^2\theta_W\,.
\label{eq:gRq}
\end{align}
In the above expressions, $T_3^q$ represents the third component of weak
isospin associated with the quark field.
\section{Electric dipole moments}
\label{sec:EDM}
VSM contributes to Electric Dipole Moment (EDM). The observed experimental upper limit on the 
EDM at $90\%$ C.L. are given by:
\begin{gather}
|d_e|/e< 1.1\times 10^{-29}~\rm cm\,, \text{ ACME \cite{ACME:2018yjb}}\,\\
|d_\mu|/e<1.9\times 10^{-19}~\rm cm\,,\text{ Muon (g-2) \cite{Muong-2:2008ebm}}\,,\\
\left.\begin{aligned}  
{\rm Re}(d_\tau)/e=-0.62\pm 0.63\,\\
{\rm Im}(d_\tau)/e=-0.40\pm 0.32\,        
\end{aligned}\right\}\times 10^{-17}~\rm cm\,,~{\rm Belle}~\text{\cite{Belle:2021ybo}}\,,
\end{gather}
while the future sensitivity enhances the upper limit as indicated in \,.~\cite{Roussy:2022cmp, He:2025ewk,Lu:2025heu,Schmidt-Wellenburg:2023aga}. The bound on $|d_\tau|/e$ combining the limits from the real and imaginary parts turn out to be $(0.74 \pm 0.70) \times 10^{-17}$ cm.

For a VSM framework, assuming ($m_{\ell_{\alpha}}\ll \metap,m_{N_k}$), the EDM for a charged lepton $\ell_{\alpha}$ (with mass $m_{\ell_\alpha}$) can expressed as \cite{Abada:2018zra},
\begin{align}
d_\alpha=-\dfrac{e~m_{\ell_{\alpha}}}{(4\pi)^4m^2_{\eta^+}}\sum_{i,j=1}^3\sum_{\beta}\left[J^M_{ij\alpha\beta}\sqrt{x_ix_j}~I_M(x_i,x_j)+J^D_{ij\alpha\beta}~I_D(x_i,x_j)\right]\,,
\end{align}
where $x_i = m_{N_i}^2/m_{\eta^+}^2$ $(i=1,2)$, $J^{M}_{ij\alpha\beta} \equiv \mathrm{Im}(h_{i\alpha}h^*_{j\alpha} h_{i\beta}h^*_{j\beta})$ and $J^{D}_{ij\alpha\beta} \equiv \mathrm{Im}(h_{i\alpha}h^*_{j\alpha} h^*_{i\beta}h_{j\beta})$ denote the CP-violating phase factors arising from the two-loop calculation; for expressions of $I_{M,D}(x_1,x_2)$, see ref.~\cite{Abada:2018zra}. Although charged-lepton EDMs may provide sensitive probes of the Scotogenic model, our analysis lies in a parameter regime far from the exclusion region shown in ref.~\cite{Abada:2018zra}.
\begin{table}[htb!]
\centering
{\small
\begin{adjustbox}{max width=\textwidth}
\begin{tabular}{c c c c c c c }
\hline \hline 
 $\metaI~(\text{GeV})$ &
$\mntwo~(\text{TeV})$ &
$\Deltam~(\text{GeV})$ & $|h_{2\mu}|$
 &
$|d_e|/e$ (\text{cm}) & $|d_\mu|/e$ (\text{cm})& $|d_\tau|/e$  (\text{cm})\\
\hline
\hline
\multirow{3}*{$580$} & \multirow{3}*{$5.2$} & $0.01$ & $0.55$& $1.7\times 10^{-39}$ & $7.5\times 10^{-28}$ & $1.2\times 10^{-27}$ \\
 &  & $0.02$ & $0.41$ & $4.3\times 10^{-40}$ & $1.9\times 10^{-28}$ & $3.0\times 10^{-27}$ \\
 &  & $0.03$ & $0.34$ & $1.9\times 10^{-40}$ & $8.3\times 10^{-29}$ & $1.3\times 10^{-27}$ \\
 \hline
 \multirow{3}*{$1260$} & \multirow{3}*{$6.4$} & $0.01$ & $0.55$& $6.4\times 10^{-40}$ & $2.9\times 10^{-28}$ & $4.7\times10^{-27}$\\
 &  & $0.02$ & $0.37$ & $1.6\times 10^{-40}$ & $7.4\times 10^{-29}$ & $1.2\times 10^{-27}$ \\
 &  & $0.03$ & $0.30$ & $7.1\times 10^{-41}$ & $3.3\times 10^{-29}$ & $5.3\times 10^{-28}$ \\
 \hline \hline
\end{tabular}
\end{adjustbox}}
\caption{Electric dipole moments of the charged leptons corresponding to the same benchmark points and model parameter choices as those used in table~\ref{tab:LFV}.}
\label{tab:EDM}
\end{table}
For completeness, we show the EDM for two benchmark scenarios (same as table~\ref{tab:LFV}) that respect the exclusion limits on the electric dipole moments of the electron and muon in \cref{tab:EDM}. 
\newpage
\bibliographystyle{JHEP}
\bibliography{mustristan}

\providecommand{\href}[2]{#2}\begingroup\raggedright\begin{thebibliography}{100}

\bibitem{Zwicky:1933gu}
F.~Zwicky, \emph{{Die Rotverschiebung von extragalaktischen Nebeln}}, \href{https://doi.org/10.1007/s10714-008-0707-4}{\emph{Helv. Phys. Acta} {\bfseries 6} (1933) 110}.

\bibitem{Rubin:1970zza}
V.C.~Rubin and W.K.~Ford, Jr., \emph{{Rotation of the Andromeda Nebula from a Spectroscopic Survey of Emission Regions}}, \href{https://doi.org/10.1086/150317}{\emph{Astrophys. J.} {\bfseries 159} (1970) 379}.

\bibitem{Rubin:1980zd}
V.C.~Rubin, N.~Thonnard and W.K.~Ford, Jr., \emph{{Rotational properties of 21 SC galaxies with a large range of luminosities and radii, from NGC 4605 /R = 4kpc/ to UGC 2885 /R = 122 kpc/}}, \href{https://doi.org/10.1086/158003}{\emph{Astrophys. J.} {\bfseries 238} (1980) 471}.

\bibitem{Tyson:1990yt}
J.A.~Tyson, R.A.~Wenk and F.~Valdes, \emph{{Detection of systematic gravitational lens galaxy image alignments - Mapping dark matter in galaxy clusters}}, \href{https://doi.org/10.1086/185636}{\emph{Astrophys. J. Lett.} {\bfseries 349} (1990) L1}.

\bibitem{Bartelmann:1999yn}
M.~Bartelmann and P.~Schneider, \emph{{Weak gravitational lensing}}, \href{https://doi.org/10.1016/S0370-1573(00)00082-X}{\emph{Phys. Rept.} {\bfseries 340} (2001) 291} [\href{https://arxiv.org/abs/astro-ph/9912508}{{\ttfamily astro-ph/9912508}}].

\bibitem{Clowe:2006eq}
D.~Clowe, M.~Bradac, A.H.~Gonzalez, M.~Markevitch, S.W.~Randall, C.~Jones et~al., \emph{{A direct empirical proof of the existence of dark matter}}, \href{https://doi.org/10.1086/508162}{\emph{Astrophys. J. Lett.} {\bfseries 648} (2006) L109} [\href{https://arxiv.org/abs/astro-ph/0608407}{{\ttfamily astro-ph/0608407}}].

\bibitem{Planck:2018vyg}
{\scshape Planck} collaboration, \emph{{Planck 2018 results. VI. Cosmological parameters}}, \href{https://doi.org/10.1051/0004-6361/201833910}{\emph{Astron. Astrophys.} {\bfseries 641} (2020) A6} [\href{https://arxiv.org/abs/1807.06209}{{\ttfamily 1807.06209}}].

\bibitem{White:1993wm}
S.D.M.~White, J.F.~Navarro, A.E.~Evrard and C.S.~Frenk, \emph{{The Baryon content of galaxy clusters: A Challenge to cosmological orthodoxy}}, \href{https://doi.org/10.1038/366429a0}{\emph{Nature} {\bfseries 366} (1993) 429}.

\bibitem{Cyburt:2015mya}
R.H.~Cyburt, B.D.~Fields, K.A.~Olive and T.-H.~Yeh, \emph{{Big Bang Nucleosynthesis: 2015}}, \href{https://doi.org/10.1103/RevModPhys.88.015004}{\emph{Rev. Mod. Phys.} {\bfseries 88} (2016) 015004} [\href{https://arxiv.org/abs/1505.01076}{{\ttfamily 1505.01076}}].

\bibitem{Goodman:1984dc}
M.W.~Goodman and E.~Witten, \emph{{Detectability of Certain Dark Matter Candidates}}, \href{https://doi.org/10.1103/PhysRevD.31.3059}{\emph{Phys. Rev. D} {\bfseries 31} (1985) 3059}.

\bibitem{Bergstrom:1989jr}
L.~Bergstrom, \emph{{Radiative Processes in Dark Matter Photino Annihilation}}, \href{https://doi.org/10.1016/0370-2693(89)90585-6}{\emph{Phys. Lett. B} {\bfseries 225} (1989) 372}.

\bibitem{Ellis:1988qp}
J.R.~Ellis, R.A.~Flores, K.~Freese, S.~Ritz, D.~Seckel and J.~Silk, \emph{{Cosmic Ray Constraints on the Annihilations of Relic Particles in the Galactic Halo}}, \href{https://doi.org/10.1016/0370-2693(88)91385-8}{\emph{Phys. Lett. B} {\bfseries 214} (1988) 403}.

\bibitem{Rudaz:1987ry}
S.~Rudaz and F.W.~Stecker, \emph{{Cosmic Ray Anti-protons, Positrons and gamma-rays From Halo Dark Matter Annihilation}}, \href{https://doi.org/10.1086/165980}{\emph{Astrophys. J.} {\bfseries 325} (1988) 16}.

\bibitem{Fayet:1979sa}
P.~Fayet, \emph{{Relations Between the Masses of the Superpartners of Leptons and Quarks, the Goldstino Couplings and the Neutral Currents}}, \href{https://doi.org/10.1016/0370-2693(79)91229-2}{\emph{Phys. Lett. B} {\bfseries 84} (1979) 416}.

\bibitem{Fayet:1980rr}
P.~Fayet, \emph{{On the Search for a New Spin 1 Boson}}, \href{https://doi.org/10.1016/0550-3213(81)90122-X}{\emph{Nucl. Phys. B} {\bfseries 187} (1981) 184}.

\bibitem{Fayet:1982ky}
P.~Fayet, \emph{{Radiative Production of Gravitinos and Photinos in e+ e- Annihilation}}, \href{https://doi.org/10.1016/0370-2693(82)90583-4}{\emph{Phys. Lett. B} {\bfseries 117} (1982) 460}.

\bibitem{Goldberg:1983nd}
H.~Goldberg, \emph{{Constraint on the Photino Mass from Cosmology}}, \href{https://doi.org/10.1103/PhysRevLett.50.1419}{\emph{Phys. Rev. Lett.} {\bfseries 50} (1983) 1419}.

\bibitem{Haber:1984rc}
H.E.~Haber and G.L.~Kane, \emph{{The Search for Supersymmetry: Probing Physics Beyond the Standard Model}}, \href{https://doi.org/10.1016/0370-1573(85)90051-1}{\emph{Phys. Rept.} {\bfseries 117} (1985) 75}.

\bibitem{Jungman:1995df}
G.~Jungman, M.~Kamionkowski and K.~Griest, \emph{{Supersymmetric dark matter}}, \href{https://doi.org/10.1016/0370-1573(95)00058-5}{\emph{Phys. Rept.} {\bfseries 267} (1996) 195} [\href{https://arxiv.org/abs/hep-ph/9506380}{{\ttfamily hep-ph/9506380}}].

\bibitem{Sakharov:1967dj}
A.D.~Sakharov, \emph{{Violation of CP Invariance, C asymmetry, and baryon asymmetry of the universe}}, \href{https://doi.org/10.1070/PU1991v034n05ABEH002497}{\emph{Pisma Zh. Eksp. Teor. Fiz.} {\bfseries 5} (1967) 32}.

\bibitem{Trodden:1998ym}
M.~Trodden, \emph{{Electroweak baryogenesis}}, \href{https://doi.org/10.1103/RevModPhys.71.1463}{\emph{Rev. Mod. Phys.} {\bfseries 71} (1999) 1463} [\href{https://arxiv.org/abs/hep-ph/9803479}{{\ttfamily hep-ph/9803479}}].

\bibitem{Cohen:1993nk}
A.G.~Cohen, D.B.~Kaplan and A.E.~Nelson, \emph{{Progress in electroweak baryogenesis}}, \href{https://doi.org/10.1146/annurev.ns.43.120193.000331}{\emph{Ann. Rev. Nucl. Part. Sci.} {\bfseries 43} (1993) 27} [\href{https://arxiv.org/abs/hep-ph/9302210}{{\ttfamily hep-ph/9302210}}].

\bibitem{Affleck:1984fy}
I.~Affleck and M.~Dine, \emph{{A New Mechanism for Baryogenesis}}, \href{https://doi.org/10.1016/0550-3213(85)90021-5}{\emph{Nucl. Phys. B} {\bfseries 249} (1985) 361}.

\bibitem{Bodeker:2020ghk}
D.~Bodeker and W.~Buchmuller, \emph{{Baryogenesis from the weak scale to the grand unification scale}}, \href{https://doi.org/10.1103/RevModPhys.93.035004}{\emph{Rev. Mod. Phys.} {\bfseries 93} (2021) 035004} [\href{https://arxiv.org/abs/2009.07294}{{\ttfamily 2009.07294}}].

\bibitem{Riotto:1999yt}
A.~Riotto and M.~Trodden, \emph{{Recent progress in baryogenesis}}, \href{https://doi.org/10.1146/annurev.nucl.49.1.35}{\emph{Ann. Rev. Nucl. Part. Sci.} {\bfseries 49} (1999) 35} [\href{https://arxiv.org/abs/hep-ph/9901362}{{\ttfamily hep-ph/9901362}}].

\bibitem{Cui:2015eba}
Y.~Cui, \emph{{A Review of WIMP Baryogenesis Mechanisms}}, \href{https://doi.org/10.1142/S0217732315300281}{\emph{Mod. Phys. Lett. A} {\bfseries 30} (2015) 1530028} [\href{https://arxiv.org/abs/1510.04298}{{\ttfamily 1510.04298}}].

\bibitem{Morrissey:2012db}
D.E.~Morrissey and M.J.~Ramsey-Musolf, \emph{{Electroweak baryogenesis}}, \href{https://doi.org/10.1088/1367-2630/14/12/125003}{\emph{New J. Phys.} {\bfseries 14} (2012) 125003} [\href{https://arxiv.org/abs/1206.2942}{{\ttfamily 1206.2942}}].

\bibitem{Fukugita:1986hr}
M.~Fukugita and T.~Yanagida, \emph{{Baryogenesis Without Grand Unification}}, \href{https://doi.org/10.1016/0370-2693(86)91126-3}{\emph{Phys. Lett. B} {\bfseries 174} (1986) 45}.

\bibitem{Davidson:2008bu}
S.~Davidson, E.~Nardi and Y.~Nir, \emph{{Leptogenesis}}, \href{https://doi.org/10.1016/j.physrep.2008.06.002}{\emph{Phys. Rept.} {\bfseries 466} (2008) 105} [\href{https://arxiv.org/abs/0802.2962}{{\ttfamily 0802.2962}}].

\bibitem{Buchmuller:2005eh}
W.~Buchmuller, R.D.~Peccei and T.~Yanagida, \emph{{Leptogenesis as the origin of matter}}, \href{https://doi.org/10.1146/annurev.nucl.55.090704.151558}{\emph{Ann. Rev. Nucl. Part. Sci.} {\bfseries 55} (2005) 311} [\href{https://arxiv.org/abs/hep-ph/0502169}{{\ttfamily hep-ph/0502169}}].

\bibitem{Hugle:2018qbw}
T.~Hugle, M.~Platscher and K.~Schmitz, \emph{{Low-Scale Leptogenesis in the Scotogenic Neutrino Mass Model}}, \href{https://doi.org/10.1103/PhysRevD.98.023020}{\emph{Phys. Rev. D} {\bfseries 98} (2018) 023020} [\href{https://arxiv.org/abs/1804.09660}{{\ttfamily 1804.09660}}].

\bibitem{ParticleDataGroup:2024cfk}
{\scshape Particle Data Group} collaboration, \emph{{Review of particle physics}}, \href{https://doi.org/10.1103/PhysRevD.110.030001}{\emph{Phys. Rev. D} {\bfseries 110} (2024) 030001}.

\bibitem{Goldhaber:1958nb}
M.~Goldhaber, L.~Grodzins and A.W.~Sunyar, \emph{{Helicity of Neutrinos}}, \href{https://doi.org/10.1103/PhysRev.109.1015}{\emph{Phys. Rev.} {\bfseries 109} (1958) 1015}.

\bibitem{Wu:1957my}
C.S.~Wu, E.~Ambler, R.W.~Hayward, D.D.~Hoppes and R.P.~Hudson, \emph{{Experimental Test of Parity Conservation in $\beta$ Decay}}, \href{https://doi.org/10.1103/PhysRev.105.1413}{\emph{Phys. Rev.} {\bfseries 105} (1957) 1413}.

\bibitem{Feynman:1958ty}
R.P.~Feynman and M.~Gell-Mann, \emph{{Theory of Fermi interaction}}, \href{https://doi.org/10.1103/PhysRev.109.193}{\emph{Phys. Rev.} {\bfseries 109} (1958) 193}.

\bibitem{Sudarshan:1958vf}
E.C.G.~Sudarshan and R.e.~Marshak, \emph{{Chirality invariance and the universal Fermi interaction}}, \href{https://doi.org/10.1103/PhysRev.109.1860.2}{\emph{Phys. Rev.} {\bfseries 109} (1958) 1860}.

\bibitem{Gargamelle:1973}
F.J.e.a.~Hasert, \emph{Observation of neutrino-like interactions without muon or electron}, {\emph{Phys. Lett. B} {\bfseries 46} (1973) 138}.

\bibitem{ALEPH:2005ab}
{\scshape ALEPH, DELPHI, L3, OPAL, SLD, LEP Electroweak Working Group, SLD Electroweak Group, SLD Heavy Flavour Group} collaboration, \emph{{Precision electroweak measurements on the $Z$ resonance}}, \href{https://doi.org/10.1016/j.physrep.2005.12.006}{\emph{Phys. Rept.} {\bfseries 427} (2006) 257} [\href{https://arxiv.org/abs/hep-ex/0509008}{{\ttfamily hep-ex/0509008}}].

\bibitem{Davis:1968cp}
R.~Davis, Jr., D.S.~Harmer and K.C.~Hoffman, \emph{{Search for neutrinos from the sun}}, \href{https://doi.org/10.1103/PhysRevLett.20.1205}{\emph{Phys. Rev. Lett.} {\bfseries 20} (1968) 1205}.

\bibitem{SNO:2001kpb}
{\scshape SNO} collaboration, \emph{{Measurement of the rate of $\nu_e+d \to p+p+e^-$ interactions produced by $^8$B solar neutrinos at the Sudbury Neutrino Observatory}}, \href{https://doi.org/10.1103/PhysRevLett.87.071301}{\emph{Phys. Rev. Lett.} {\bfseries 87} (2001) 071301} [\href{https://arxiv.org/abs/nucl-ex/0106015}{{\ttfamily nucl-ex/0106015}}].

\bibitem{SNO:2011hxd}
{\scshape SNO} collaboration, \emph{{Combined Analysis of all Three Phases of Solar Neutrino Data from the Sudbury Neutrino Observatory}}, \href{https://doi.org/10.1103/PhysRevC.88.025501}{\emph{Phys. Rev. C} {\bfseries 88} (2013) 025501} [\href{https://arxiv.org/abs/1109.0763}{{\ttfamily 1109.0763}}].

\bibitem{Super-Kamiokande:1998kpq}
{\scshape Super-Kamiokande} collaboration, \emph{{Evidence for oscillation of atmospheric neutrinos}}, \href{https://doi.org/10.1103/PhysRevLett.81.1562}{\emph{Phys. Rev. Lett.} {\bfseries 81} (1998) 1562} [\href{https://arxiv.org/abs/hep-ex/9807003}{{\ttfamily hep-ex/9807003}}].

\bibitem{MACRO:1998ckv}
{\scshape MACRO} collaboration, \emph{{Measurement of the atmospheric neutrino induced upgoing muon flux using MACRO}}, \href{https://doi.org/10.1016/S0370-2693(98)00885-5}{\emph{Phys. Lett. B} {\bfseries 434} (1998) 451} [\href{https://arxiv.org/abs/hep-ex/9807005}{{\ttfamily hep-ex/9807005}}].

\bibitem{KamLAND:2002uet}
{\scshape KamLAND} collaboration, \emph{{First results from KamLAND: Evidence for reactor anti-neutrino disappearance}}, \href{https://doi.org/10.1103/PhysRevLett.90.021802}{\emph{Phys. Rev. Lett.} {\bfseries 90} (2003) 021802} [\href{https://arxiv.org/abs/hep-ex/0212021}{{\ttfamily hep-ex/0212021}}].

\bibitem{RENO:2012mkc}
{\scshape RENO} collaboration, \emph{{Observation of Reactor Electron Antineutrino Disappearance in the RENO Experiment}}, \href{https://doi.org/10.1103/PhysRevLett.108.191802}{\emph{Phys. Rev. Lett.} {\bfseries 108} (2012) 191802} [\href{https://arxiv.org/abs/1204.0626}{{\ttfamily 1204.0626}}].

\bibitem{DayaBay:2012fng}
{\scshape Daya Bay} collaboration, \emph{{Observation of electron-antineutrino disappearance at Daya Bay}}, \href{https://doi.org/10.1103/PhysRevLett.108.171803}{\emph{Phys. Rev. Lett.} {\bfseries 108} (2012) 171803} [\href{https://arxiv.org/abs/1203.1669}{{\ttfamily 1203.1669}}].

\bibitem{K2K:2002icj}
{\scshape K2K} collaboration, \emph{{Indications of neutrino oscillation in a 250 km long baseline experiment}}, \href{https://doi.org/10.1103/PhysRevLett.90.041801}{\emph{Phys. Rev. Lett.} {\bfseries 90} (2003) 041801} [\href{https://arxiv.org/abs/hep-ex/0212007}{{\ttfamily hep-ex/0212007}}].

\bibitem{NOvA:2019cyt}
{\scshape NOvA} collaboration, \emph{{First Measurement of Neutrino Oscillation Parameters using Neutrinos and Antineutrinos by NOvA}}, \href{https://doi.org/10.1103/PhysRevLett.123.151803}{\emph{Phys. Rev. Lett.} {\bfseries 123} (2019) 151803} [\href{https://arxiv.org/abs/1906.04907}{{\ttfamily 1906.04907}}].

\bibitem{T2K:2013ppw}
{\scshape T2K} collaboration, \emph{{Observation of Electron Neutrino Appearance in a Muon Neutrino Beam}}, \href{https://doi.org/10.1103/PhysRevLett.112.061802}{\emph{Phys. Rev. Lett.} {\bfseries 112} (2014) 061802} [\href{https://arxiv.org/abs/1311.4750}{{\ttfamily 1311.4750}}].

\bibitem{MINOS:2011neo}
{\scshape MINOS} collaboration, \emph{{Measurement of the Neutrino Mass Splitting and Flavor Mixing by MINOS}}, \href{https://doi.org/10.1103/PhysRevLett.106.181801}{\emph{Phys. Rev. Lett.} {\bfseries 106} (2011) 181801} [\href{https://arxiv.org/abs/1103.0340}{{\ttfamily 1103.0340}}].

\bibitem{Weinberg:1979sa}
S.~Weinberg, \emph{{Baryon and Lepton Nonconserving Processes}}, \href{https://doi.org/10.1103/PhysRevLett.43.1566}{\emph{Phys. Rev. Lett.} {\bfseries 43} (1979) 1566}.

\bibitem{Mohapatra:1979ia}
R.N.~Mohapatra and G.~Senjanovic, \emph{{Neutrino Mass and Spontaneous Parity Nonconservation}}, \href{https://doi.org/10.1103/PhysRevLett.44.912}{\emph{Phys. Rev. Lett.} {\bfseries 44} (1980) 912}.

\bibitem{Magg:1980ut}
M.~Magg and C.~Wetterich, \emph{{Neutrino Mass Problem and Gauge Hierarchy}}, \href{https://doi.org/10.1016/0370-2693(80)90825-4}{\emph{Phys. Lett. B} {\bfseries 94} (1980) 61}.

\bibitem{Foot:1988aq}
R.~Foot, H.~Lew, X.G.~He and G.C.~Joshi, \emph{{Seesaw Neutrino Masses Induced by a Triplet of Leptons}}, \href{https://doi.org/10.1007/BF01415558}{\emph{Z. Phys. C} {\bfseries 44} (1989) 441}.

\bibitem{Babu:1988ki}
K.S.~Babu, \emph{{Model of 'Calculable' Majorana Neutrino Masses}}, \href{https://doi.org/10.1016/0370-2693(88)91584-5}{\emph{Phys. Lett. B} {\bfseries 203} (1988) 132}.

\bibitem{Zee:1980ai}
A.~Zee, \emph{{A Theory of Lepton Number Violation, Neutrino Majorana Mass, and Oscillation}}, \href{https://doi.org/10.1016/0370-2693(80)90349-4}{\emph{Phys. Lett. B} {\bfseries 93} (1980) 389}.

\bibitem{Ma:2006km}
E.~Ma, \emph{{Verifiable radiative seesaw mechanism of neutrino mass and dark matter}}, \href{https://doi.org/10.1103/PhysRevD.73.077301}{\emph{Phys. Rev. D} {\bfseries 73} (2006) 077301} [\href{https://arxiv.org/abs/hep-ph/0601225}{{\ttfamily hep-ph/0601225}}].

\bibitem{Mohapatra:1986aw}
R.N.~Mohapatra, \emph{{Mechanism for Understanding Small Neutrino Mass in Superstring Theories}}, \href{https://doi.org/10.1103/PhysRevLett.56.561}{\emph{Phys. Rev. Lett.} {\bfseries 56} (1986) 561}.

\bibitem{Langacker:1998ut}
P.~Langacker, \emph{{A Mechanism for ordinary sterile neutrino mixing}}, \href{https://doi.org/10.1103/PhysRevD.58.093017}{\emph{Phys. Rev. D} {\bfseries 58} (1998) 093017} [\href{https://arxiv.org/abs/hep-ph/9805281}{{\ttfamily hep-ph/9805281}}].

\bibitem{Arkani-Hamed:1998wuz}
N.~Arkani-Hamed, S.~Dimopoulos, G.R.~Dvali and J.~March-Russell, \emph{{Neutrino masses from large extra dimensions}}, \href{https://doi.org/10.1103/PhysRevD.65.024032}{\emph{Phys. Rev. D} {\bfseries 65} (2001) 024032} [\href{https://arxiv.org/abs/hep-ph/9811448}{{\ttfamily hep-ph/9811448}}].

\bibitem{Altarelli:2010gt}
G.~Altarelli and F.~Feruglio, \emph{{Discrete Flavor Symmetries and Models of Neutrino Mixing}}, \href{https://doi.org/10.1103/RevModPhys.82.2701}{\emph{Rev. Mod. Phys.} {\bfseries 82} (2010) 2701} [\href{https://arxiv.org/abs/1002.0211}{{\ttfamily 1002.0211}}].

\bibitem{Mohapatra:2005wg}
R.N.~Mohapatra et~al., \emph{{Theory of Neutrinos: A White Paper}}, \href{https://doi.org/10.1088/0034-4885/70/11/R02}{\emph{Rept. Prog. Phys.} {\bfseries 70} (2007) 1757} [\href{https://arxiv.org/abs/hep-ph/0510213}{{\ttfamily hep-ph/0510213}}].

\bibitem{King:2003jb}
S.F.~King, \emph{{Neutrino mass models}}, \href{https://doi.org/10.1088/0034-4885/67/2/R01}{\emph{Rept. Prog. Phys.} {\bfseries 67} (2004) 107} [\href{https://arxiv.org/abs/hep-ph/0310204}{{\ttfamily hep-ph/0310204}}].

\bibitem{Ardila-Tafurth:2025qdf}
G.~Ardila-Tafurth, A.~Fl{\'o}rez, C.~Rodr{\'\i}guez, M.~Sarazin and {\'O}.~Zapata, \emph{{Feasibility to probe the dynamical scotogenic model at the LHC}},  \href{https://arxiv.org/abs/2512.17903}{{\ttfamily 2512.17903}}.

\bibitem{Barbieri:2006dq}
R.~Barbieri, L.J.~Hall and V.S.~Rychkov, \emph{{Improved naturalness with a heavy Higgs: An Alternative road to LHC physics}}, \href{https://doi.org/10.1103/PhysRevD.74.015007}{\emph{Phys. Rev. D} {\bfseries 74} (2006) 015007} [\href{https://arxiv.org/abs/hep-ph/0603188}{{\ttfamily hep-ph/0603188}}].

\bibitem{Hashemi:2015swh}
M.~Hashemi, M.~Krawczyk, S.~Najjari and A.F.~\.Zarnecki, \emph{{Production of Inert Scalars at the high energy $e^+ e^-$ colliders}}, \href{https://doi.org/10.1007/JHEP02(2016)187}{\emph{JHEP} {\bfseries 02} (2016) 187} [\href{https://arxiv.org/abs/1512.01175}{{\ttfamily 1512.01175}}].

\bibitem{Racker:2024fpn}
J.~Racker, \emph{{Low-scale leptogenesis in the scotogenic model: Spectator processes and benchmark points}}, \href{https://doi.org/10.1103/PhysRevD.111.L081301}{\emph{Phys. Rev. D} {\bfseries 111} (2025) L081301} [\href{https://arxiv.org/abs/2411.15120}{{\ttfamily 2411.15120}}].

\bibitem{Accettura:2023ked}
C.~Accettura et~al., \emph{{Towards a muon collider}}, \href{https://doi.org/10.1140/epjc/s10052-023-11889-x}{\emph{Eur. Phys. J. C} {\bfseries 83} (2023) 864} [\href{https://arxiv.org/abs/2303.08533}{{\ttfamily 2303.08533}}].

\bibitem{InternationalMuonCollider:2024jyv}
{\scshape International Muon Collider} collaboration, \emph{{Interim report for the International Muon Collider Collaboration (IMCC)}}, \href{https://doi.org/10.23731/CYRM-2024-002}{\emph{CERN Yellow Rep. Monogr.} {\bfseries 2/2024} (2024) 176} [\href{https://arxiv.org/abs/2407.12450}{{\ttfamily 2407.12450}}].

\bibitem{Accettura:2025xgt}
C.~Accettura et~al., \emph{{The Muon Collider}},  \href{https://arxiv.org/abs/2504.21417}{{\ttfamily 2504.21417}}.

\bibitem{LZ:2024zvo}
{\scshape LZ} collaboration, \emph{{Dark Matter Search Results from 4.2{\,}{\,}Tonne-Years of Exposure of the LUX-ZEPLIN (LZ) Experiment}}, \href{https://doi.org/10.1103/4dyc-z8zf}{\emph{Phys. Rev. Lett.} {\bfseries 135} (2025) 011802} [\href{https://arxiv.org/abs/2410.17036}{{\ttfamily 2410.17036}}].

\bibitem{Fermi-LAT:2025gei}
{\scshape Fermi-LAT, HAWC, H.E.S.S., MAGIC, VERITAS} collaboration, \emph{{Combined dark matter search towards dwarf spheroidal galaxies with Fermi-LAT, HAWC, H.E.S.S., MAGIC, and VERITAS}},  \href{https://arxiv.org/abs/2508.20229}{{\ttfamily 2508.20229}}.

\bibitem{XLZD:2024nsu}
{\scshape XLZD} collaboration, \emph{{The XLZD Design Book: towards the next-generation liquid xenon observatory for dark matter and neutrino physics}}, \href{https://doi.org/10.1140/epjc/s10052-025-14810-w}{\emph{Eur. Phys. J. C} {\bfseries 85} (2025) 1192} [\href{https://arxiv.org/abs/2410.17137}{{\ttfamily 2410.17137}}].

\bibitem{Swiezewska:2012eh}
B.~Swiezewska and M.~Krawczyk, \emph{{Diphoton rate in the inert doublet model with a 125 GeV Higgs boson}}, \href{https://doi.org/10.1103/PhysRevD.88.035019}{\emph{Phys. Rev. D} {\bfseries 88} (2013) 035019} [\href{https://arxiv.org/abs/1212.4100}{{\ttfamily 1212.4100}}].

\bibitem{Aiko:2023nqj}
M.~Aiko, J.~Braathen and S.~Kanemura, \emph{{Leading two-loop corrections to the Higgs di-photon decay in the inert doublet model}}, \href{https://doi.org/10.1140/epjc/s10052-025-14184-z}{\emph{Eur. Phys. J. C} {\bfseries 85} (2025) 489} [\href{https://arxiv.org/abs/2307.14976}{{\ttfamily 2307.14976}}].

\bibitem{Braathen:2024ckk}
J.~Braathen, M.~Gabelmann, T.~Robens and P.~Stylianou, \emph{{Probing the Inert Doublet Model via vector-boson fusion at a muon collider}}, \href{https://doi.org/10.1007/JHEP05(2025)055}{\emph{JHEP} {\bfseries 05} (2025) 055} [\href{https://arxiv.org/abs/2411.13729}{{\ttfamily 2411.13729}}].

\bibitem{Lozano:2025tst}
V.M.~Lozano, G.~Sanchez~Garcia and J.W.F.~Valle, \emph{{Collider signatures of fermionic scotogenic dark matter}},  \href{https://arxiv.org/abs/2502.05270}{{\ttfamily 2502.05270}}.

\bibitem{Abada:2018zra}
A.~Abada and T.~Toma, \emph{{Electric dipole moments in the minimal scotogenic model}}, \href{https://doi.org/10.1007/JHEP04(2018)030}{\emph{JHEP} {\bfseries 04} (2018) 030} [\href{https://arxiv.org/abs/1802.00007}{{\ttfamily 1802.00007}}].

\bibitem{SNO:2002tuh}
{\scshape SNO} collaboration, \emph{{Direct evidence for neutrino flavor transformation from neutral current interactions in the Sudbury Neutrino Observatory}}, \href{https://doi.org/10.1103/PhysRevLett.89.011301}{\emph{Phys. Rev. Lett.} {\bfseries 89} (2002) 011301} [\href{https://arxiv.org/abs/nucl-ex/0204008}{{\ttfamily nucl-ex/0204008}}].

\bibitem{Esteban:2020cvm}
I.~Esteban, M.C.~Gonzalez-Garcia, M.~Maltoni, T.~Schwetz and A.~Zhou, \emph{{The fate of hints: updated global analysis of three-flavor neutrino oscillations}}, \href{https://doi.org/10.1007/JHEP09(2020)178}{\emph{JHEP} {\bfseries 09} (2020) 178} [\href{https://arxiv.org/abs/2007.14792}{{\ttfamily 2007.14792}}].

\bibitem{Esteban:2018azc}
I.~Esteban, M.C.~Gonzalez-Garcia, A.~Hernandez-Cabezudo, M.~Maltoni and T.~Schwetz, \emph{{Global analysis of three-flavour neutrino oscillations: synergies and tensions in the determination of $\theta_{23}$, $\delta_{CP}$, and the mass ordering}}, \href{https://doi.org/10.1007/JHEP01(2019)106}{\emph{JHEP} {\bfseries 01} (2019) 106} [\href{https://arxiv.org/abs/1811.05487}{{\ttfamily 1811.05487}}].

\bibitem{Esteban:2024eli}
I.~Esteban, M.C.~Gonzalez-Garcia, M.~Maltoni, I.~Martinez-Soler, J.P.~Pinheiro and T.~Schwetz, \emph{{NuFit-6.0: updated global analysis of three-flavor neutrino oscillations}}, \href{https://doi.org/10.1007/JHEP12(2024)216}{\emph{JHEP} {\bfseries 12} (2024) 216} [\href{https://arxiv.org/abs/2410.05380}{{\ttfamily 2410.05380}}].

\bibitem{Casas:2001sr}
J.A.~Casas and A.~Ibarra, \emph{{Oscillating neutrinos and $\mu \to e, \gamma$}}, \href{https://doi.org/10.1016/S0550-3213(01)00475-8}{\emph{Nucl. Phys. B} {\bfseries 618} (2001) 171} [\href{https://arxiv.org/abs/hep-ph/0103065}{{\ttfamily hep-ph/0103065}}].

\bibitem{MEGII:2025gzr}
{\scshape MEG II} collaboration, \emph{{New limit on the ${\upmu ^+ \rightarrow e^+ \upgamma }$ decay with the MEG II experiment}}, \href{https://doi.org/10.1140/epjc/s10052-025-14906-3}{\emph{Eur. Phys. J. C} {\bfseries 85} (2025) 1177} [\href{https://arxiv.org/abs/2504.15711}{{\ttfamily 2504.15711}}].

\bibitem{MEGII:2023ltw}
{\scshape MEG II} collaboration, \emph{{A search for $\upmu ^+ \rightarrow \textrm{e}^+ \upgamma $ with the first dataset of the MEG~II experiment}}, \href{https://doi.org/10.1140/epjc/s10052-024-12416-2}{\emph{Eur. Phys. J. C} {\bfseries 84} (2024) 216} [\href{https://arxiv.org/abs/2310.12614}{{\ttfamily 2310.12614}}].

\bibitem{Toma:2013zsa}
T.~Toma and A.~Vicente, \emph{{Lepton Flavor Violation in the Scotogenic Model}}, \href{https://doi.org/10.1007/JHEP01(2014)160}{\emph{JHEP} {\bfseries 01} (2014) 160} [\href{https://arxiv.org/abs/1312.2840}{{\ttfamily 1312.2840}}].

\bibitem{Vicente:2014wga}
A.~Vicente and C.E.~Yaguna, \emph{{Probing the scotogenic model with lepton flavor violating processes}}, \href{https://doi.org/10.1007/JHEP02(2015)144}{\emph{JHEP} {\bfseries 02} (2015) 144} [\href{https://arxiv.org/abs/1412.2545}{{\ttfamily 1412.2545}}].

\bibitem{Valiente:2023blo}
P.E.~Valiente, \emph{{Neutrino Masses and Lepton Flavor Physics Beyond the Standard Model}}, Ph.D. thesis, U. Valencia (main), 2023.

\bibitem{Hundi:2022iva}
R.S.~Hundi, \emph{{Lepton flavor violating Z and Higgs decays in the scotogenic model}}, \href{https://doi.org/10.1140/epjc/s10052-022-10453-3}{\emph{Eur. Phys. J. C} {\bfseries 82} (2022) 505} [\href{https://arxiv.org/abs/2201.03779}{{\ttfamily 2201.03779}}].

\bibitem{Chen:2019nud}
C.-H.~Chen and T.~Nomura, \emph{{Influence of an inert charged Higgs boson on the muon $g-2$ and radiative neutrino masses in a scotogenic model}}, \href{https://doi.org/10.1103/PhysRevD.100.015024}{\emph{Phys. Rev. D} {\bfseries 100} (2019) 015024} [\href{https://arxiv.org/abs/1903.03380}{{\ttfamily 1903.03380}}].

\bibitem{SINDRUM:1987nra}
{\scshape SINDRUM} collaboration, \emph{{Search for the Decay $\mu^+ \to e^+ e^+ e^-$}}, \href{https://doi.org/10.1016/0550-3213(88)90462-2}{\emph{Nucl. Phys. B} {\bfseries 299} (1988) 1}.

\bibitem{Mu3e:2020gyw}
{\scshape Mu3e} collaboration, \emph{{Technical design of the phase I Mu3e experiment}}, \href{https://doi.org/10.1016/j.nima.2021.165679}{\emph{Nucl. Instrum. Meth. A} {\bfseries 1014} (2021) 165679} [\href{https://arxiv.org/abs/2009.11690}{{\ttfamily 2009.11690}}].

\bibitem{COMET:2025sdw}
{\scshape COMET, MEG, Mu2e, Mu3e} collaboration, \emph{{Charged Lepton Flavour Violations searches with muons: present and future}},  \href{https://arxiv.org/abs/2503.22461}{{\ttfamily 2503.22461}}.

\bibitem{Arganda:2007jw}
E.~Arganda, M.J.~Herrero and A.M.~Teixeira, \emph{{mu-e conversion in nuclei within the CMSSM seesaw: Universality versus non-universality}}, \href{https://doi.org/10.1088/1126-6708/2007/10/104}{\emph{JHEP} {\bfseries 10} (2007) 104} [\href{https://arxiv.org/abs/0707.2955}{{\ttfamily 0707.2955}}].

\bibitem{SINDRUMII:2006dvw}
{\scshape SINDRUM II} collaboration, \emph{{A Search for muon to electron conversion in muonic gold}}, \href{https://doi.org/10.1140/epjc/s2006-02582-x}{\emph{Eur. Phys. J. C} {\bfseries 47} (2006) 337}.

\bibitem{Mu2e:2022ggl}
{\scshape Mu2e} collaboration, \emph{{Mu2e Run I Sensitivity Projections for the Neutrinoless Conversion Search in Aluminum}}, \href{https://doi.org/10.3390/universe9010054}{\emph{Universe} {\bfseries 9} (2023) 54} [\href{https://arxiv.org/abs/2210.11380}{{\ttfamily 2210.11380}}].

\bibitem{Mu2e-II:2022blh}
{\scshape Mu2e-II} collaboration, \emph{{Mu2e-II: Muon to electron conversion with PIP-II}},  in \emph{{Snowmass 2021}}, 3, 2022 [\href{https://arxiv.org/abs/2203.07569}{{\ttfamily 2203.07569}}].

\bibitem{COMET:2018auw}
{\scshape COMET} collaboration, \emph{{COMET Phase-I Technical Design Report}}, \href{https://doi.org/10.1093/ptep/ptz125}{\emph{PTEP} {\bfseries 2020} (2020) 033C01} [\href{https://arxiv.org/abs/1812.09018}{{\ttfamily 1812.09018}}].

\bibitem{Moritsu:2022lem}
{\scshape COMET} collaboration, \emph{{Search for Muon-to-Electron Conversion with the COMET Experiment {\textdagger}}}, \href{https://doi.org/10.3390/universe8040196}{\emph{Universe} {\bfseries 8} (2022) 196} [\href{https://arxiv.org/abs/2203.06365}{{\ttfamily 2203.06365}}].

\bibitem{Baak:2014ora}
{\scshape Gfitter Group} collaboration, \emph{{The global electroweak fit at NNLO and prospects for the LHC and ILC}}, \href{https://doi.org/10.1140/epjc/s10052-014-3046-5}{\emph{Eur. Phys. J. C} {\bfseries 74} (2014) 3046} [\href{https://arxiv.org/abs/1407.3792}{{\ttfamily 1407.3792}}].

\bibitem{Hambye:2009pw}
T.~Hambye, F.S.~Ling, L.~Lopez~Honorez and J.~Rocher, \emph{{Scalar Multiplet Dark Matter}}, \href{https://doi.org/10.1007/JHEP05(2010)066}{\emph{JHEP} {\bfseries 07} (2009) 090} [\href{https://arxiv.org/abs/0903.4010}{{\ttfamily 0903.4010}}].

\bibitem{deBoer:2021pon}
T.~de~Boer, R.~Busse, A.~Kappes, M.~Klasen and S.~Zeinstra, \emph{{Indirect detection constraints on the scotogenic dark matter model}}, \href{https://doi.org/10.1088/1475-7516/2021/08/038}{\emph{JCAP} {\bfseries 08} (2021) 038} [\href{https://arxiv.org/abs/2105.04899}{{\ttfamily 2105.04899}}].

\bibitem{Lundstrom:2008ai}
E.~Lundstrom, M.~Gustafsson and J.~Edsjo, \emph{{The Inert Doublet Model and LEP II Limits}}, \href{https://doi.org/10.1103/PhysRevD.79.035013}{\emph{Phys. Rev. D} {\bfseries 79} (2009) 035013} [\href{https://arxiv.org/abs/0810.3924}{{\ttfamily 0810.3924}}].

\bibitem{ATLAS:2014zve}
{\scshape ATLAS} collaboration, \emph{{Search for direct production of charginos, neutralinos and sleptons in final states with two leptons and missing transverse momentum in $pp$ collisions at $\sqrt{s} =$ 8 TeV with the ATLAS detector}}, \href{https://doi.org/10.1007/JHEP05(2014)071}{\emph{JHEP} {\bfseries 05} (2014) 071} [\href{https://arxiv.org/abs/1403.5294}{{\ttfamily 1403.5294}}].

\bibitem{Pierce:2007ut}
A.~Pierce and J.~Thaler, \emph{{Natural Dark Matter from an Unnatural Higgs Boson and New Colored Particles at the TeV Scale}}, \href{https://doi.org/10.1088/1126-6708/2007/08/026}{\emph{JHEP} {\bfseries 08} (2007) 026} [\href{https://arxiv.org/abs/hep-ph/0703056}{{\ttfamily hep-ph/0703056}}].

\bibitem{Kalinowski:2018ylg}
J.~Kalinowski, W.~Kotlarski, T.~Robens, D.~Sokolowska and A.F.~Zarnecki, \emph{{Benchmarking the Inert Doublet Model for $e^+ e^-$ colliders}}, \href{https://doi.org/10.1007/JHEP12(2018)081}{\emph{JHEP} {\bfseries 12} (2018) 081} [\href{https://arxiv.org/abs/1809.07712}{{\ttfamily 1809.07712}}].

\bibitem{ALEPH:2002ftu}
{\scshape ALEPH} collaboration, \emph{{Search for charged Higgs bosons in $e^{+} e^{-}$ collisions at energies up to $\sqrt{s}$ = 209-GeV}}, \href{https://doi.org/10.1016/S0370-2693(02)02380-8}{\emph{Phys. Lett. B} {\bfseries 543} (2002) 1} [\href{https://arxiv.org/abs/hep-ex/0207054}{{\ttfamily hep-ex/0207054}}].

\bibitem{Banerjee:2025qkb}
A.~Banerjee, D.~Das, S.~Mukherjee and S.~Pandey, \emph{{Drell-Yan constraints on charged scalars: a weak isospin perspective}},  \href{https://arxiv.org/abs/2501.09796}{{\ttfamily 2501.09796}}.

\bibitem{CMS:2026cjr}
{\scshape CMS} collaboration, \emph{{Search for pair production of additional neutral scalars within the Inert Doublet Model in a final state with two electrons or two muons in proton-proton collisions at $\sqrt{s}$ = 13 TeV and 13.6 TeV}},  \href{https://arxiv.org/abs/2605.13614}{{\ttfamily 2605.13614}}.

\bibitem{CMS:2026rhc}
{\scshape CMS} collaboration, \emph{{Search for pair production of additional neutral scalars within the Inert Doublet Model in a final state with two electrons or two muons}}, .

\bibitem{ATLAS:2014otc}
{\scshape ATLAS} collaboration, \emph{{Search for charged Higgs bosons decaying via $H^{\pm} \rightarrow \tau^{\pm}\nu$ in fully hadronic final states using $pp$ collision data at $\sqrt{s} = 8$ TeV with the ATLAS detector}}, \href{https://doi.org/10.1007/JHEP03(2015)088}{\emph{JHEP} {\bfseries 03} (2015) 088} [\href{https://arxiv.org/abs/1412.6663}{{\ttfamily 1412.6663}}].

\bibitem{CMS:2015lsf}
{\scshape CMS} collaboration, \emph{{Search for a charged Higgs boson in pp collisions at $ \sqrt{s}=8 $ TeV}}, \href{https://doi.org/10.1007/JHEP11(2015)018}{\emph{JHEP} {\bfseries 11} (2015) 018} [\href{https://arxiv.org/abs/1508.07774}{{\ttfamily 1508.07774}}].

\bibitem{ATLAS:2018gfm}
{\scshape ATLAS} collaboration, \emph{{Search for charged Higgs bosons decaying via $H^{\pm} \to \tau^{\pm}\nu_{\tau}$ in the $\tau$+jets and $\tau$+lepton final states with 36 fb$^{-1}$ of $pp$ collision data recorded at $\sqrt{s} = 13$ TeV with the ATLAS experiment}}, \href{https://doi.org/10.1007/JHEP09(2018)139}{\emph{JHEP} {\bfseries 09} (2018) 139} [\href{https://arxiv.org/abs/1807.07915}{{\ttfamily 1807.07915}}].

\bibitem{Belyaev:2020wok}
A.~Belyaev, S.~Prestel, F.~Rojas-Abbate and J.~Zurita, \emph{{Probing dark matter with disappearing tracks at the LHC}}, \href{https://doi.org/10.1103/PhysRevD.103.095006}{\emph{Phys. Rev. D} {\bfseries 103} (2021) 095006} [\href{https://arxiv.org/abs/2008.08581}{{\ttfamily 2008.08581}}].

\bibitem{ATLAS:2017oal}
{\scshape ATLAS} collaboration, \emph{{Search for long-lived charginos based on a disappearing-track signature in pp collisions at $ \sqrt{s}=13 $ TeV with the ATLAS detector}}, \href{https://doi.org/10.1007/JHEP06(2018)022}{\emph{JHEP} {\bfseries 06} (2018) 022} [\href{https://arxiv.org/abs/1712.02118}{{\ttfamily 1712.02118}}].

\bibitem{ATLAS:2018jjf}
{\scshape ATLAS} collaboration, \emph{{ATLAS sensitivity to winos and higgsinos with a highly compressed mass spectrum at the HL-LHC}}, .

\bibitem{ATLAS:2019gqq}
{\scshape ATLAS} collaboration, \emph{{Search for heavy charged long-lived particles in the ATLAS detector in 36.1 fb$^{-1}$ of proton-proton collision data at $\sqrt{s} = 13$ TeV}}, \href{https://doi.org/10.1103/PhysRevD.99.092007}{\emph{Phys. Rev. D} {\bfseries 99} (2019) 092007} [\href{https://arxiv.org/abs/1902.01636}{{\ttfamily 1902.01636}}].

\bibitem{Belyaev:2018ext}
A.~Belyaev, T.R.~Fernandez Perez~Tomei, P.G.~Mercadante, C.S.~Moon, S.~Moretti, S.F.~Novaes et~al., \emph{{Advancing LHC probes of dark matter from the inert two-Higgs-doublet model with the monojet signal}}, \href{https://doi.org/10.1103/PhysRevD.99.015011}{\emph{Phys. Rev. D} {\bfseries 99} (2019) 015011} [\href{https://arxiv.org/abs/1809.00933}{{\ttfamily 1809.00933}}].

\bibitem{LopezHonorez:2006gr}
L.~Lopez~Honorez, E.~Nezri, J.F.~Oliver and M.H.G.~Tytgat, \emph{{The Inert Doublet Model: An Archetype for Dark Matter}}, \href{https://doi.org/10.1088/1475-7516/2007/02/028}{\emph{JCAP} {\bfseries 02} (2007) 028} [\href{https://arxiv.org/abs/hep-ph/0612275}{{\ttfamily hep-ph/0612275}}].

\bibitem{Alguero:2023zol}
G.~Alguero, G.~Belanger, F.~Boudjema, S.~Chakraborti, A.~Goudelis, S.~Kraml et~al., \emph{{micrOMEGAs 6.0: N-component dark matter}}, \href{https://doi.org/10.1016/j.cpc.2024.109133}{\emph{Comput. Phys. Commun.} {\bfseries 299} (2024) 109133} [\href{https://arxiv.org/abs/2312.14894}{{\ttfamily 2312.14894}}].

\bibitem{Belyaev:2012qa}
A.~Belyaev, N.D.~Christensen and A.~Pukhov, \emph{{CalcHEP 3.4 for collider physics within and beyond the Standard Model}}, \href{https://doi.org/10.1016/j.cpc.2013.01.014}{\emph{Comput. Phys. Commun.} {\bfseries 184} (2013) 1729} [\href{https://arxiv.org/abs/1207.6082}{{\ttfamily 1207.6082}}].

\bibitem{Billard:2013qya}
J.~Billard, L.~Strigari and E.~Figueroa-Feliciano, \emph{{Implication of neutrino backgrounds on the reach of next generation dark matter direct detection experiments}}, \href{https://doi.org/10.1103/PhysRevD.89.023524}{\emph{Phys. Rev. D} {\bfseries 89} (2014) 023524} [\href{https://arxiv.org/abs/1307.5458}{{\ttfamily 1307.5458}}].

\bibitem{OHare:2021utq}
C.A.J.~O'Hare, \emph{{New Definition of the Neutrino Floor for Direct Dark Matter Searches}}, \href{https://doi.org/10.1103/PhysRevLett.127.251802}{\emph{Phys. Rev. Lett.} {\bfseries 127} (2021) 251802} [\href{https://arxiv.org/abs/2109.03116}{{\ttfamily 2109.03116}}].

\bibitem{CTAConsortium:2017dvg}
{\scshape CTA Consortium} collaboration, B.S.~Acharya et~al., \emph{{Science with the Cherenkov Telescope Array}}, WSP (11, 2018), \href{https://doi.org/10.1142/10986}{10.1142/10986}, [\href{https://arxiv.org/abs/1709.07997}{{\ttfamily 1709.07997}}].

\bibitem{DOnofrio:2014rug}
M.~D'Onofrio, K.~Rummukainen and A.~Tranberg, \emph{{Sphaleron Rate in the Minimal Standard Model}}, \href{https://doi.org/10.1103/PhysRevLett.113.141602}{\emph{Phys. Rev. Lett.} {\bfseries 113} (2014) 141602} [\href{https://arxiv.org/abs/1404.3565}{{\ttfamily 1404.3565}}].

\bibitem{Mahanta:2019gfe}
D.~Mahanta and D.~Borah, \emph{{Fermion dark matter with $N_2$ leptogenesis in minimal scotogenic model}}, \href{https://doi.org/10.1088/1475-7516/2019/11/021}{\emph{JCAP} {\bfseries 11} (2019) 021} [\href{https://arxiv.org/abs/1906.03577}{{\ttfamily 1906.03577}}].

\bibitem{Harvey:1990qw}
J.A.~Harvey and M.S.~Turner, \emph{{Cosmological Baryon and Lepton Number in the Presence of Electroweak Fermion Number Violation}}, \href{https://doi.org/10.1103/PhysRevD.42.3344}{\emph{Phys. Rev. D} {\bfseries 42} (1990) 3344}.

\bibitem{MuonCollider:2022xlm}
{\scshape Muon Collider} collaboration, \emph{{The physics case of a 3 TeV muon collider stage}},  \href{https://arxiv.org/abs/2203.07261}{{\ttfamily 2203.07261}}.

\bibitem{Andreetto:2025mrd}
P.~Andreetto et~al., \emph{{Music: a multi-purpose detector concept for physics at the 10 TeV muon collider}}, \href{https://doi.org/10.1140/epjc/s10052-026-15654-8}{\emph{Eur. Phys. J. C} {\bfseries 86} (2026) 554} [\href{https://arxiv.org/abs/2511.23273}{{\ttfamily 2511.23273}}].

\bibitem{Hamada:2022mua}
Y.~Hamada, R.~Kitano, R.~Matsudo, H.~Takaura and M.~Yoshida, \emph{{$\mu$TRISTAN}}, \href{https://doi.org/10.1093/ptep/ptac059}{\emph{PTEP} {\bfseries 2022} (2022) 053B02} [\href{https://arxiv.org/abs/2201.06664}{{\ttfamily 2201.06664}}].

\bibitem{Kitano:2025xaj}
R.~Kitano, I.~Low, R.~Matsudo, S.~Okawa and S.~Roy, \emph{{Heavy Neutral Lepton at Same-Sign Muon Collider}},  \href{https://arxiv.org/abs/2510.18390}{{\ttfamily 2510.18390}}.

\bibitem{Alloul:2013bka}
A.~Alloul, N.D.~Christensen, C.~Degrande, C.~Duhr and B.~Fuks, \emph{{FeynRules 2.0 - A complete toolbox for tree-level phenomenology}}, \href{https://doi.org/10.1016/j.cpc.2014.04.012}{\emph{Comput. Phys. Commun.} {\bfseries 185} (2014) 2250} [\href{https://arxiv.org/abs/1310.1921}{{\ttfamily 1310.1921}}].

\bibitem{Alwall:2011uj}
J.~Alwall, M.~Herquet, F.~Maltoni, O.~Mattelaer and T.~Stelzer, \emph{{MadGraph 5 : Going Beyond}}, \href{https://doi.org/10.1007/JHEP06(2011)128}{\emph{JHEP} {\bfseries 06} (2011) 128} [\href{https://arxiv.org/abs/1106.0522}{{\ttfamily 1106.0522}}].

\bibitem{Bierlich:2022pfr}
C.~Bierlich et~al., \emph{{A comprehensive guide to the physics and usage of PYTHIA 8.3}}, \href{https://doi.org/10.21468/SciPostPhysCodeb.8}{\emph{SciPost Phys. Codeb.} {\bfseries 2022} (2022) 8} [\href{https://arxiv.org/abs/2203.11601}{{\ttfamily 2203.11601}}].

\bibitem{deFavereau:2013fsa}
{\scshape DELPHES 3} collaboration, \emph{{DELPHES 3, A modular framework for fast simulation of a generic collider experiment}}, \href{https://doi.org/10.1007/JHEP02(2014)057}{\emph{JHEP} {\bfseries 02} (2014) 057} [\href{https://arxiv.org/abs/1307.6346}{{\ttfamily 1307.6346}}].

\bibitem{Wilks:1938dza}
S.S.~Wilks, \emph{{The Large-Sample Distribution of the Likelihood Ratio for Testing Composite Hypotheses}}, \href{https://doi.org/10.1214/aoms/1177732360}{\emph{Annals Math. Statist.} {\bfseries 9} (1938) 60}.

\bibitem{Capdevilla:2021fmj}
R.~Capdevilla, F.~Meloni, R.~Simoniello and J.~Zurita, \emph{{Hunting wino and higgsino dark matter at the muon collider with disappearing tracks}}, \href{https://doi.org/10.1007/JHEP06(2021)133}{\emph{JHEP} {\bfseries 06} (2021) 133} [\href{https://arxiv.org/abs/2102.11292}{{\ttfamily 2102.11292}}].

\bibitem{MAIA:2025hzm}
{\scshape MAIA} collaboration, \emph{{MAIA: A new detector concept for a 10 TeV muon collider}},  \href{https://arxiv.org/abs/2502.00181}{{\ttfamily 2502.00181}}.

\bibitem{Littmann:2025ndn}
M.~Littmann, M.~Larson, B.~Rosser, T.~Flicker, K.~Huang, L.~Rozanov et~al., \emph{{Enabling searches for long-lived particles at a future 10 TeV Muon Collider}},  \href{https://arxiv.org/abs/2512.10097}{{\ttfamily 2512.10097}}.

\bibitem{Acanfora:2026aay}
F.~Acanfora, M.J.~Baker, T.~Martonhelyi and A.~Thamm, \emph{{Heavy Vector Triplets at a Muon Collider}},  \href{https://arxiv.org/abs/2605.13957}{{\ttfamily 2605.13957}}.

\bibitem{GEANT4:2002zbu}
{\scshape GEANT4} collaboration, \emph{{GEANT4 - A Simulation Toolkit}}, \href{https://doi.org/10.1016/S0168-9002(03)01368-8}{\emph{Nucl. Instrum. Meth. A} {\bfseries 506} (2003) 250}.

\bibitem{Antusch:2011nz}
S.~Antusch, P.~Di~Bari, D.A.~Jones and S.F.~King, \emph{{Leptogenesis in the Two Right-Handed Neutrino Model Revisited}}, \href{https://doi.org/10.1103/PhysRevD.86.023516}{\emph{Phys. Rev. D} {\bfseries 86} (2012) 023516} [\href{https://arxiv.org/abs/1107.6002}{{\ttfamily 1107.6002}}].

\bibitem{Kuno:1999jp}
Y.~Kuno and Y.~Okada, \emph{{Muon decay and physics beyond the standard model}}, \href{https://doi.org/10.1103/RevModPhys.73.151}{\emph{Rev. Mod. Phys.} {\bfseries 73} (2001) 151} [\href{https://arxiv.org/abs/hep-ph/9909265}{{\ttfamily hep-ph/9909265}}].

\bibitem{Kosmas:2001mv}
T.S.~Kosmas, S.~Kovalenko and I.~Schmidt, \emph{{Nuclear muon- e- conversion in strange quark sea}}, \href{https://doi.org/10.1016/S0370-2693(01)00657-8}{\emph{Phys. Lett. B} {\bfseries 511} (2001) 203} [\href{https://arxiv.org/abs/hep-ph/0102101}{{\ttfamily hep-ph/0102101}}].

\bibitem{ACME:2018yjb}
{\scshape ACME} collaboration, \emph{{Improved limit on the electric dipole moment of the electron}}, \href{https://doi.org/10.1038/s41586-018-0599-8}{\emph{Nature} {\bfseries 562} (2018) 355}.

\bibitem{Muong-2:2008ebm}
{\scshape Muon (g-2)} collaboration, \emph{{An Improved Limit on the Muon Electric Dipole Moment}}, \href{https://doi.org/10.1103/PhysRevD.80.052008}{\emph{Phys. Rev. D} {\bfseries 80} (2009) 052008} [\href{https://arxiv.org/abs/0811.1207}{{\ttfamily 0811.1207}}].

\bibitem{Belle:2021ybo}
{\scshape Belle} collaboration, \emph{{An improved search for the electric dipole moment of the $\tau$ lepton}}, \href{https://doi.org/10.1007/JHEP04(2022)110}{\emph{JHEP} {\bfseries 04} (2022) 110} [\href{https://arxiv.org/abs/2108.11543}{{\ttfamily 2108.11543}}].

\bibitem{Roussy:2022cmp}
T.S.~Roussy et~al., \emph{{An improved bound on the electron{\textquoteright}s electric dipole moment}}, \href{https://doi.org/10.1126/science.adg4084}{\emph{Science} {\bfseries 381} (2023) adg4084} [\href{https://arxiv.org/abs/2212.11841}{{\ttfamily 2212.11841}}].

\bibitem{He:2025ewk}
X.-G.~He, C.-W.~Liu, J.-P.~Ma, C.~Yang and Z.-Y.~Zou, \emph{Precise measurement of cp-violating $\tau$ edm through $e^{+}e^{-} \to \gamma^{*},\, \psi(2s) \to \tau^{+}\tau^{-}$}, \href{https://doi.org/10.1007/JHEP04(2025)001}{\emph{JHEP} {\bfseries 04} (2025) 001} [\href{https://arxiv.org/abs/2501.06687}{{\ttfamily 2501.06687}}].

\bibitem{Lu:2025heu}
P.-C.~Lu, Z.-G.~Si and H.~Zhang, \emph{{Probing the electromagnetic dipole moment of the ${\ensuremath{\tau}}$ lepton in the $e^+e^-{\textrightarrow}{\ensuremath{\gamma}}*/Z{\textrightarrow}{\ensuremath{\tau}}+{\ensuremath{\tau}}-$ reaction}}, \href{https://doi.org/10.1103/hvz6-b69m}{\emph{Phys. Rev. D} {\bfseries 112} (2025) 075039} [\href{https://arxiv.org/abs/2506.19557}{{\ttfamily 2506.19557}}].

\bibitem{Schmidt-Wellenburg:2023aga}
{\scshape muEDM} collaboration, \emph{{Preparations for a search of the muon EDM at PSI}}, \href{https://doi.org/10.1051/epjconf/202328901008}{\emph{EPJ Web Conf.} {\bfseries 289} (2023) 01008}.

\end{thebibliography}\endgroup
\end{document}